\documentclass[a4paper,11pt]{article}
\usepackage{jheppub}
\usepackage[utf8]{inputenc}
\usepackage{amsmath}
\usepackage{graphicx}
\usepackage{multirow}

\DeclareMathOperator{\diag}{diag}

\newcommand{\orcid}[1]{\href{https://orcid.org/#1}{#1}}
\newcommand{\eps}{\varepsilon}

\title{How to Identify Different New Neutrino Oscillation Physics Scenarios at DUNE}

\author[1]{Peter B.~Denton\note{\orcid{0000-0002-5209-872X}},}
\author[2]{Alessio Giarnetti\note{\orcid{0000-0001-8487-8045}},}
\author[2]{and Davide Meloni\note{\orcid{0000-0001-7680-6957}}}

\affiliation[1]{High Energy Theory Group, Physics Department, Brookhaven National Laboratory, Upton, NY 11973, USA}
\affiliation[2]{Dipartimento di Matematica e Fisica, Universit\`a di Roma Tre
Via della Vasca Navale 84, 00146 Rome, Italy}

\emailAdd{pdenton@bnl.gov}
\emailAdd{alessio.giarnetti@uniroma3.it}
\emailAdd{davide.meloni@uniroma3.it}

\makeatletter
\hypersetup{colorlinks=true,allcolors=[rgb]{1,0.56,0},pdftitle=\@title}
\makeatother

\abstract{
Next generation neutrino oscillation experiments are expected to measure the remaining oscillation parameters with very good precision.
They will have unprecedented capabilities to search for new physics that modify oscillations.
DUNE, with its broad band beam, good particle identification, and relatively high energies will provide an excellent environment to search for new physics.
If deviations from the standard three-flavor oscillation picture are seen however, it is crucial to know which new physics scenario is found so that it can be verified elsewhere and theoretically understood.
We investigate several benchmark new physics scenarios by looking at existing long-baseline accelerator neutrino data from NOvA and T2K and determine at what sensitivity DUNE can differentiate among them.
We consider sterile neutrinos and both vector and scalar non-standard neutrino interactions, all with new complex phases, the latter of which could conceivably provide absolute neutrino mass scale information.
We find that, in many interesting cases, DUNE will have good model discrimination.
We also perform a new fit to NOvA and T2K data with scalar NSI.}

\begin{document}

\maketitle

\section{Introduction}
Neutrino oscillation physics is on the advent of reaching the precision era.
Current long-baseline accelerator experiments NOvA \cite{Ayres:2007tu} and T2K \cite{Abe:2011ks} are making the first appearance measurements in accelerator neutrinos and the next-generation accelerator experiments HK \cite{Abe:2015zbg} and DUNE \cite{Acciarri:2015uup} will measure the appearance probabilities precisely providing a wealth of information about the three remaining oscillation unknowns: the octant of $\theta_{23}$, the atmospheric mass ordering, and the value of the complex phase $\delta$.

These powerful experiments will provide the strongest tests yet of the standard three-flavor oscillation hypothesis.
In the event there is new physics, however, it is important to check if that new physics can be robustly identified compared to alternatives.
To address this question, we use several benchmark new physics points, motivated by the slight tension in the NOvA and T2K data \cite{Denton:2020uda}, see also \cite{Chatterjee:2020kkm,Chatterjee:2020yak,Miranda:2019ynh,Forero:2021azc,deGouvea:2022kma,Majhi:2022wyp} and first test the level at which DUNE can identify them, and then the level at which they can be differentiated.
As scalar NSI has not yet been tested with T2K and NOvA data, we derive the first constraints on scalar NSI with existing long-baseline data.

Testing these benchmarks against no new physics and against each other will provide a fairly comprehensive overview of the capability of DUNE to correctly identify the new physics scenarios and the parameters within the scenario that are preferred, although partial degeneracies among different new physics scenarios for precise values of the parameters may weakened this sensitivity in some cases.
HK will also have sensitivity to many of these scenarios and their combination will be particularly powerful for disentangling things.
Nonetheless, as we will show in section \ref{sec:diff}, DUNE alone will provide very good model discrimination capabilities.

\section{New Physics Scenarios}
There are numerous new physics scenarios affecting neutrino oscillations \cite{Arguelles:2019xgp,Arguelles:2022xxa} including neutrino decay \cite{Berryman:2014qha,Picoreti:2015ika,SNO:2018pvg,Gonzalez-Garcia:2008mgl,Gomes:2014yua,Abrahao:2015rba,Pagliaroli:2016zab,Coloma:2017zpg,Choubey:2017dyu,Gago:2017zzy,Choubey:2017eyg,Denton:2018aml,deSalas:2018kri,Ascencio-Sosa:2018lbk,Choubey:2018cfz,Funcke:2019grs,Abdullahi:2020rge,Ghoshal:2020hyo,Porto-Silva:2020gma,Choubey:2020dhw,Picoreti:2021yct}, Lorentz invariance violation and CPT violation \cite{Kostelecky:2003cr,LSND:2005oop,MINOS:2008fnv,MINOS:2010kat,IceCube:2010fyu,Kostelecky:2011gq,MiniBooNE:2011pix,DoubleChooz:2012eiq,MINOS:2012ozn,Diaz:2013iba,Rebel:2013vc,Super-Kamiokande:2014exs,EXO-200:2016hbz,IceCube:2017qyp,T2K:2017ega,DayaBay:2018fsh}, background dark matter fields \cite{Reynoso:2016hjr,Berlin:2016woy,Krnjaic:2017zlz,Brdar:2017kbt,Liao:2018byh,Capozzi:2018bps,Huang:2018cwo,Farzan:2019yvo,Cline:2019seo,Dev:2020kgz,Huang:2021kam,Losada:2021bxx,Chun:2021ief,Dev:2022bae}, neutrino decoherence or wave-packet separation \cite{Liu:1997km,Chang:1998ea,Benatti:2000ph,Lisi:2000zt,Super-Kamiokande:2004orf,BalieiroGomes:2016ykp,BalieiroGomes:2018gtd,Stuttard:2021uyw,Hellmann:2021jyz}, unitary violation \cite{Blennow:2016jkn,Parke:2015goa,Denton:2021mso}, non-standard neutrino interactions (NSI) \cite{Wolfenstein:1977ue,deGouvea:2015ndi,Proceedings:2019qno,Denton:2020uda,Bakhti:2020fde,DUNE:2020fgq,Chatterjee:2021wac,Giarnetti:2021wur}, and sterile neutrinos \cite{Donini:2007yf,Dighe:2007uf,Berryman:2015nua,DUNE:2020fgq,Acero:2022wqg}.
We focus on the last two: NSI and sterile neutrinos.
NSI provides a general framework for quantifying modifications to the neutrino oscillation probability due to a new interaction with matter particles.
We further split NSIs into vector NSIs, which act like the regular matter effect \cite{Wolfenstein:1977ue} but possibly in a different basis or with a different dependence on the matter particles, and scalar NSI \cite{Ge:2018uhz}, which acts like a new mass term for neutrinos sourced by matter particles.
We focus on NSI with off-diagonal couplings as these are the parameters for which appearance measurements are particularly crucial to constrain.
Diagonal NSI are best constrained with solar and atmospheric neutrino oscillations and neutrino scattering.
Sterile neutrinos are well motivated extensions to the standard three-flavor neutrino oscillation picture due to theoretical arguments based on the fact that neutrinos have mass, as well as due to a host of confusing anomalies \cite{LSND:2001aii,T2K:2014xvp,Giunti:2010zu,Mention:2011rk,Barinov:2021asz,MiniBooNE:2020pnu,Denton:2021czb} which could be explained by new light sterile neutrinos in the $m_4\sim1$ eV range.
Sterile neutrino searches are also generally related to those involving unitary violation but are a bit more comprehensive in their ability to directly probe the new mass scale.
In the following subsections we review the formalism of each of these three cases as they apply for long-baseline neutrino oscillations.

\subsection{Vector Non-Standard Neutrino Interaction}
Since the introduction of the matter effects in neutrino oscillations, the possibility that neutrinos can undergo NSIs with matter has been widely studied.
Focusing only on neutral current vector NSI, which dominates over the axial-vector current assuming comparable coupling strengths, we can describe vector NSI using an effective theory approach.
The Lagrangian now includes the following terms: 
\begin{equation}
{\cal L}^{eff}_{\rm vector\ NSI} = -2\sqrt2 G_F\sum_{f,\alpha,\beta}
\varepsilon^f_{\alpha \beta} (\bar{\nu}_\alpha \gamma_\rho \nu_\beta)
( \bar {f} \gamma^\rho f ) \,,
\label{eq:efflag}
\end{equation}
where $G_F$ is the Fermi constant, $\varepsilon_{\alpha \beta}^f$
is the parameter which describes the strength of the NSI, $f$ is a
first generation SM charged-fermion ($e$, $u$, or $d$) and $\alpha$ and $\beta$ denote the neutrino flavors $e$, $\mu$ or $\tau$.
The $\varepsilon$ parameter can be related to the parameters in a simplified model or even a UV complete scenarios.
Since these details do not affect oscillations within a single experiment, we focus only on the $\varepsilon$ effective parameter.
Notice that in this subsection we will only consider interactions mediated by vector particles.

The presence of such interactions modifies the neutrino oscillation Hamiltonian to
\begin{align} 
H
 =\frac{1}{2E} \left[ U M^2 U^\dagger +
                  a \left( \begin{array}{ccc}
            1 + \varepsilon_{ee}     & \varepsilon_{e\mu} & \varepsilon_{e\tau} \\
            \varepsilon_{e \mu }^*  & \varepsilon_{\mu\mu}  & \varepsilon_{\mu\tau} \\
            \varepsilon_{e \tau}^* & \varepsilon_{\mu \tau }^* & \varepsilon_{\tau\tau} 
                   \end{array} 
                   \right) \right]\,,
\label{eq:matter}
\end{align}
where $U$ is the PMNS matrix \cite{Pontecorvo:1957cp,Maki:1962mu}, $M^2=\diag(0,\Delta m_{21}^2, \Delta m_{31}^2)$, $a=2\sqrt{2}G_F N_e E$, and $N_e$ is the electron number density.
Due to the hermiticity of the Hamiltonian matrix, the diagonal NSI couplings $\varepsilon_{\alpha\alpha}$ must be real, while the non-diagonal ones are in general complex and can be written as $\varepsilon_{\alpha\beta}=|\varepsilon_{\alpha\beta}|e^{i \phi_{\alpha\beta}}$. Since we can subtract a matrix proportional to the identity without changing the oscillation probabilities, only two of the diagonal NSI parameters are independent.
The Hamiltonian level NSI parameters relevant for neutrino oscillations, those without superscripts $\varepsilon_{\alpha\beta}$, are related to the Lagrangian level NSI terms via
\begin{equation}
\varepsilon_{\alpha\beta}=\sum_{f\in\{e,u,d\}}\frac{N_f}{N_e}\varepsilon_{\alpha\beta}^f\,,
\end{equation}
where $N_f$ is the number density of fermion $f$.
When looking at NSI in the Sun and the Earth simultaneously one must consider the Lagrangian level NSI to accurately translate between them.
Here we are only considering experiments in the Earth's crust so we can safely work with the Hamiltonian level parameters.
Many studies of vector NSI exist in the literature, see e.g.~\cite{deGouvea:2015ndi,Coloma:2015kiu,Denton:2020uda,Chatterjee:2020kkm,Esteban:2020cvm}.

Several analyses of oscillation data\footnote{Scattering data is also sensitive to NSI \cite{Coloma:2017egw,Liao:2017uzy,Farzan:2018gtr,Denton:2018xmq,Denton:2020hop,Denton:2022nol}, although these data sets have a non-trivial dependence on the mediator mass, while oscillation data is essentially \cite{Joshipura:2003jh,Grifols:2003gy,Gonzalez-Garcia:2006vic,Bandyopadhyay:2006uh,Samanta:2010zh,Davoudiasl:2011sz,Wise:2018rnb,Smirnov:2019cae,Coloma:2020gfv} independent of it.} have been considered under various assumptions.
A recent global analysis of oscillation data in the context of NSIs has estimated the constraints on the NSI parameters in the context of both LMA (Large Mixing Angle solution of the solar neutrino problem) and LMA-Dark results are shown, with the difference mainly affecting $\varepsilon_{ee}$.
The LMA-Dark solution \cite{Miranda:2004nb,Escrihuela:2009up,Gonzalez-Garcia:2013usa,Bakhti:2014pva,Coloma:2016gei,Farzan:2017xzy,Coloma:2017egw,Denton:2021vtf,Denton:2022nol} is the solution with $\varepsilon_{ee}\simeq-2$ and the opposite sign\footnote{We take the definition of the three mass eigenstates as $|U_{e1}|>|U_{e2}|>|U_{e3}|$.
Thus $\theta_{12}<45^\circ$ by definition and the sign of $\Delta m^2_{21}$ has been measured experimentally with solar neutrinos.
Some define the mass eigenstates by $m_1<m_2$, $|U_{e1}|>|U_{e3}|$, and $|U_{e2}|>|U_{e3}|$.
In this case $\Delta m^2_{21}>0$ by definition and the octant of $\theta_{12}$ is to be determined experimentally.
See \cite{Denton:2020exu,Denton:2021vtf}.} on $\Delta m^2_{31}$, $\Delta m^2_{21}$, and $\delta$.
For a recent discussion of LMA-Dark in the context of the latest reactor constraints see \cite{Denton:2022nol}.
We note that while the allowed values in the global analysis \cite{Esteban:2018ppq} they find might seem to disfavor some of the values preferred in recent analyses long-baseline data \cite{Denton:2020uda,Chatterjee:2020kkm} used in this paper (see table \ref{tab:nova t2k vector}), it is easy to see that the constraints on real NSI and NSI with a large complex component can be quite different.

It might appear that charged lepton flavor violating probes would always be stronger than those from oscillations, but numerous UV complete models with large $\varepsilon_{\alpha\beta}\gtrsim0.1$ exist in the literature where oscillations provide the strongest probes \cite{Forero:2016ghr,Denton:2018dqq,Dey:2018yht,Babu:2017olk,Farzan:2016wym,Farzan:2015hkd,Farzan:2015doa,Babu:2019mfe,Proceedings:2019qno}.
All of these models can be recast into the language of NSI which is exactly what makes NSI such an attractive BSM scenario to investigate.

In order to gain a good understanding of the impact of vector NSI on oscillation experimental data, we derive approximate expressions for the vector NSI contribution to neutrino oscillations in matter in appendix \ref{sec:approximate} by performing a perturbative expansion in various parameters known to be small.

\subsection{Scalar Non-Standard Neutrino Interaction}

In addition to a vector mediator, one can consider different Lorentz structure for the underlying theory behind a new neutrino interaction.
Scalar NSI has been investigated in the context of some neutrino oscillation experiments as well as early universe constraints \cite{Ge:2018uhz,Babu:2019iml,Smirnov:2019cae,Venzor:2020ova,Medhi:2021wxj,Medhi:2022qmu,Sarkar:2022ujy,Dutta:2022fdt}.
All previous studies, to our knowledge, focused on the diagonal scalar NSI parameters; instead, we focus here on the off-diagonal parameters.
Early universe constraints and fifth-force probes may be stronger than terrestrial probes in many cases, although not necessarily all, depending primarily on the mediator mass \cite{Smirnov:2019cae}.
Given the highly disparate environments between the early universe and terrestrial oscillations for which an UV complete model may behave differently, in addition to some hints for a new interaction in early universe data \cite{Kreisch:2019yzn,Barenboim:2019tux}, we consider this scenario in DUNE data nonetheless.
That said, we do caution the reader to be aware of important non-oscillation constraints on scalar NSI.

The effective Lagrangian for scalar NSI is:
\begin{equation}
{\cal L}^{eff}_{\rm scalar\ NSI} = \frac{y_f y_{\alpha\beta}}{m_\phi^2}(\bar{\nu}_\alpha  \nu_\beta)
( \bar {f}  f ) \,,
\label{eq:efflag_sc}
\end{equation}
where the $y$'s are the Yukawa couplings to matter fermions and neutrinos and $m_\phi$ is the mass of the scalar mediator.
We note that this can no longer be considered as a matter potential, but it can be seen as a Yukawa interaction term for Dirac neutrinos that induces a correction to the mass term that depends on the density of fermions sourcing the new interactions.
After the spin summation of the environmental fermion, the Dirac equation for neutrinos becomes:
\begin{equation}
\bar{\nu}_\beta\left[i\partial_{\mu}\gamma^\mu+\left(M_{\beta\alpha}+\frac{\sum_f N_f y_f y_{\alpha\beta}}{m_\phi^2}\right)\right]\nu_\alpha=0\,,
\end{equation}
where $M_{\beta\alpha}$ is the Dirac mass matrix of the neutrinos and $N_f$ is the number density of fermion $f$.
The effect of such scalar NSI is thus a modification of the neutrino mass matrix by a factor of $\delta M$.
As a consequence, the Hamiltonian governing neutrino oscillations is modified from the diagonal $M^2$ term to $(M+\delta M)(M+\delta M)^\dagger$.
We then parameterize the correction term $\delta M$ as:
\begin{equation}
\delta M=\sqrt{|\Delta m_{31}^2|}
\begin{pmatrix}
\eta_{ee} & \eta_{e\mu} & \eta_{e\tau} \\
\eta_{e\mu}^* & \eta_{\mu\mu} & \eta_{\mu\tau} \\
\eta_{e\tau}^* & \eta_{\mu\tau}^* & \eta_{\tau\tau}
\end{pmatrix}\,,
\end{equation}
where we have chosen to scale the size of $\delta M$ relative to $\sqrt{|\Delta m^2_{31}|}$ to make the parameters of the model, $\eta_{\alpha\beta}$, dimensionless\footnote{Note that this $\eta_{\alpha\beta}$ parameter is unrelated to the parameter sometimes used in the context of unitarity violation, see e.g.~\cite{Fernandez-Martinez:2007iaa}.}.
We have also chosen to make $\delta M$ Hermitian although it need not be (see e.g.~\cite{Sarkar:2022ujy}) depending on if the scalar mediator is real or complex; we encourage further research into the non-Hermitian case.
In addition, while we have used the Dirac equation for Dirac neutrinos, the effect is the same for Majorana neutrinos in the ultrarelativistic case, see e.g.~\cite{Babu:2020ivd}.

Unlike in the vector case parameterized by the dimensionless $\varepsilon_{\alpha\beta}$ parameters, these dimensionless scalar NSI parameters, $\eta_{\alpha\beta}$, are proportional to the matter density since they depend on $N_f$.
As in the vector case, the diagonal parameters are real while the off-diagonal elements are complex, $\eta_{\alpha\beta}=|\eta_{\alpha\beta}|e^{i\phi_{\alpha\beta}}$.
To be explicit, we can relate these $\eta_{\alpha\beta}$ parameters to the parameters of the underlying theory as
\begin{equation}
\eta_{\alpha\beta}=\frac1{m_\phi^2\sqrt{|\Delta m^2_{31}|}}\sum_fN_fy_fy_{\alpha\beta}\,.
\end{equation}

Since the dimensionless parameters for scalar NSI depend on the density, unlike for vector NSI, the density at which they are calculated must be considered.
For clarity, we will consider all values of $\eta_{\alpha\beta}$ presented to be at 3 g/cc as a benchmark, and then the corresponding value used in NOvA, T2K, or DUNE will be appropriately rescaled to the density of that experiment.
The effect of this rescaling is small for long-baseline experiments, but would be quite significant if atmospheric or solar neutrinos were also considered.

One of the main feature of such a model is that in the neutrino oscillation Hamiltonian we cannot subtract an identity matrix proportional to the lightest neutrino mass.
This procedure, in the standard oscillation, allows us to write all the probabilities only in terms of mass splittings $\Delta m_{ij}^2$.
In the scalar NSI case, on the other hand, the absolute neutrino mass (parameterized as the lightest neutrino mass which $m_1$ in the normal ordering (NO) and $m_3$ in the inverted ordering (IO)) survives, and appears in the probabilities. Thus, in presence of such interactions, the oscillation probabilities are, in principle, sensitive to the neutrino mass scale.
We find that the impact, however, is quite small for realistic parameters.

Using the same expansion procedure described in the previous section and expanding up to the second order also in the parameter $\eta_{ij}$ it is possible to obtain approximate expressions also for the scalar NSI case see Ref.~\cite{Medhi:2021wxj}. In order to avoid cumbersome expressions, we did not show here the effect of the solar mass splitting and of the lightest neutrino mass.
See appendix \ref{sec:approximate}.

\subsection{Sterile Neutrino}
Sterile neutrinos are a simple, phenomenologically rich, and theoretically and experimentally motivated extension to the standard three-flavor neutrino scenario.
Since neutrinos have mass, there are additional particles and sterile neutrinos are present in many of the explanations.
In addition, there are numerous hints of various significances and robustness that indicates that new light ($m_4\lesssim10$ eV) neutrinos may exist \cite{LSND:2001aii,T2K:2014xvp,Giunti:2010zu,Mention:2011rk,Barinov:2021asz,MiniBooNE:2020pnu,Denton:2021czb} although strong constraints also exist \cite{MINOS:2017cae,IceCube:2020phf,Hagstotz:2020ukm}, see refs.~\cite{Acero:2022wqg,Dentler:2018sju,Diaz:2019fwt,Boser:2019rta,Machado:2019oxb,Berryman:2020agd,Dasgupta:2021ies} for recent reviews.
Sterile neutrinos also play a role in long-baseline accelerator neutrino experiments, although typically at somewhat lighter masses than the above hints $\sim1$ eV \cite{Donini:2007yf,Dighe:2007uf,Meloni:2010zr,Bhattacharya:2011ee,Hollander:2014iha,Klop:2014ima,Gandhi:2015xza,Palazzo:2015gja,Berryman:2015nua,Agarwalla:2016mrc,Agarwalla:2016xxa,Dutta:2016glq,Kelly:2017kch,Coloma:2017ptb,Ghosh:2017atj,Choubey:2017cba,Choubey:2017ppj,Agarwalla:2018nlx,Gupta:2018qsv,KumarAgarwalla:2019blx,Majhi:2019hdj,Reyimuaji:2019wbn,Ghosh:2019zvl,Chatterjee:2020yak,Penedo:2022etl}.
We focus on the scenario with a single light sterile neutrino which modifies the neutrino oscillation Hamiltonian to
\begin{equation}
H=\frac1{2E}\left[U_4
\begin{pmatrix}
0\\&\Delta m^2_{21}\\&&\Delta m^2_{31}\\&&&\Delta m^2_{41}
\end{pmatrix}
U_4^\dagger+a
\begin{pmatrix}
1\\&0\\&&0\\&&&\frac12\frac{N_n}{N_e}
\end{pmatrix}\right]\,,
\label{eq:Hsterile}
\end{equation}
where we have chosen to parameterize the mixing matrix as
\begin{equation}
U_4\equiv R_{34}(\theta_{34})R_{24}(\theta_{24})R_{14}(\theta_{14})U_{23}(\theta_{23},\delta_{23})U_{13}(\theta_{13},\delta_{13})U_{12}(\theta_{12},\delta_{12})\,,
\end{equation}
the relevant $2\times2$ submatrix of $U_{ij}$ is
\begin{equation}
\begin{pmatrix}
c_{ij}&s_{ij}e^{-i\delta_{ij}}\\
-s_{ij}e^{i\delta_{ij}}&c_{ij}
\end{pmatrix}\,,
\end{equation}
and $R_{ij}(\theta_{ij})=U_{ij}(\theta_{ij},0)$.
In the three-flavor limit where $\theta_{i4}=0$ we get that the regular complex phase is given by $\delta=\delta_{13}-\delta_{12}-\delta_{23}$ \cite{Rodejohann:2011vc,Denton:2021vtf}.
Note that in Eq.~\ref{eq:Hsterile} we have subtracted off the neutral current potential leading to an apparent potential for sterile neutrinos.

Similar to the two NSI cases, we again perform a perturbative calculation of the sterile neutrino contribution to the oscillation probability in appendix \ref{sec:approximate}.

\section{Benchmark Scenarios}
\label{sec:bench}
In order to determine the level at which DUNE can differentiate different new physics scenarios, we use several benchmark points in the parameter spaces of each scenario.
These points are motivated by existing long-baseline data from NOvA and T2K.
This serves two useful purposes.
First, if the tension between NOvA and T2K is due to new physics, then these are the parameters that DUNE should be looking at.
Second, even if this is not the first hint of new physics, it represents a good estimate of the expected upper limit on new physics parameters from the current generation of long-baseline accelerator neutrino experiments, generally consistent with the constraints from IceCube for $\varepsilon_{e\mu}$ and $\varepsilon_{e\tau}$ \cite{IceCubeCollaboration:2021euf}.

While different benchmark points might yield somewhat different results in sections \ref{sec:results} and \ref{sec:diff} below, we have checked that the results do not drastically change as these benchmarks are varied.

\subsection{Vector NSI Motivated by NOvA and T2K}
The first set of benchmark points we consider are from Ref.~\cite{Denton:2020uda} which investigated the possibility that NOvA and T2K data could be described by off-diagonal complex vector NSI.
The fit was performed to both NOvA and T2K disappearance and appearance data in both neutrino and anti-neutrino modes and wrong sign lepton contributions were included.
In order to fully constrain all six regular oscillation parameters plus the new physics parameters, data from Daya Bay were used for $\theta_{13}$ and $\Delta m^2_{31}$ and from KamLAND for $\theta_{12}$ and $\Delta m^2_{21}$.
Such data sets were chosen since the effect of NSI would not modify these external constraints.
This analysis finds results generally consistent with other similar analyses in the literature \cite{Kelly:2020fkv,Chatterjee:2020kkm,Esteban:2020cvm}.
The best fit values can be found in the appendix in table \ref{tab:nova t2k vector}.

\subsection{Scalar NSI Motivated by NOvA and T2K}
We reanalyze the NOvA and T2K data using the same assumptions as in the previous section, but now in the context of scalar NSI.
The best fit points are shown in table \ref{tab:nova t2k scalar} and the preferred regions are shown in Fig.~\ref{fig:nova t2k scalar} in appendix \ref{sec:nova t2k scalar}.
The physics impact of this analysis is also discussed in appendix \ref{sec:nova t2k scalar}.

\begin{figure}
\centering
\includegraphics[width=0.49\textwidth]{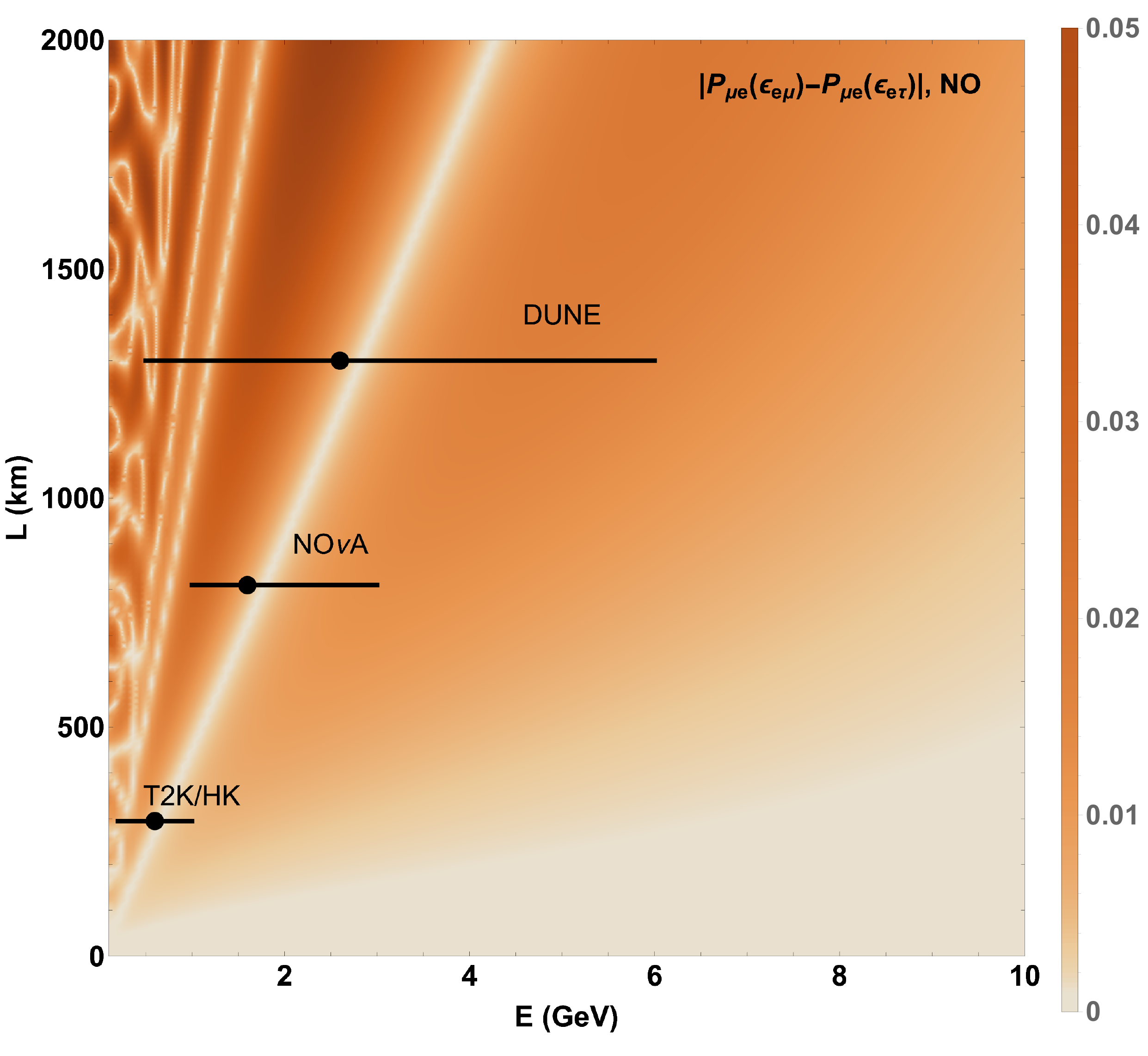}
\includegraphics[width=0.49\textwidth]{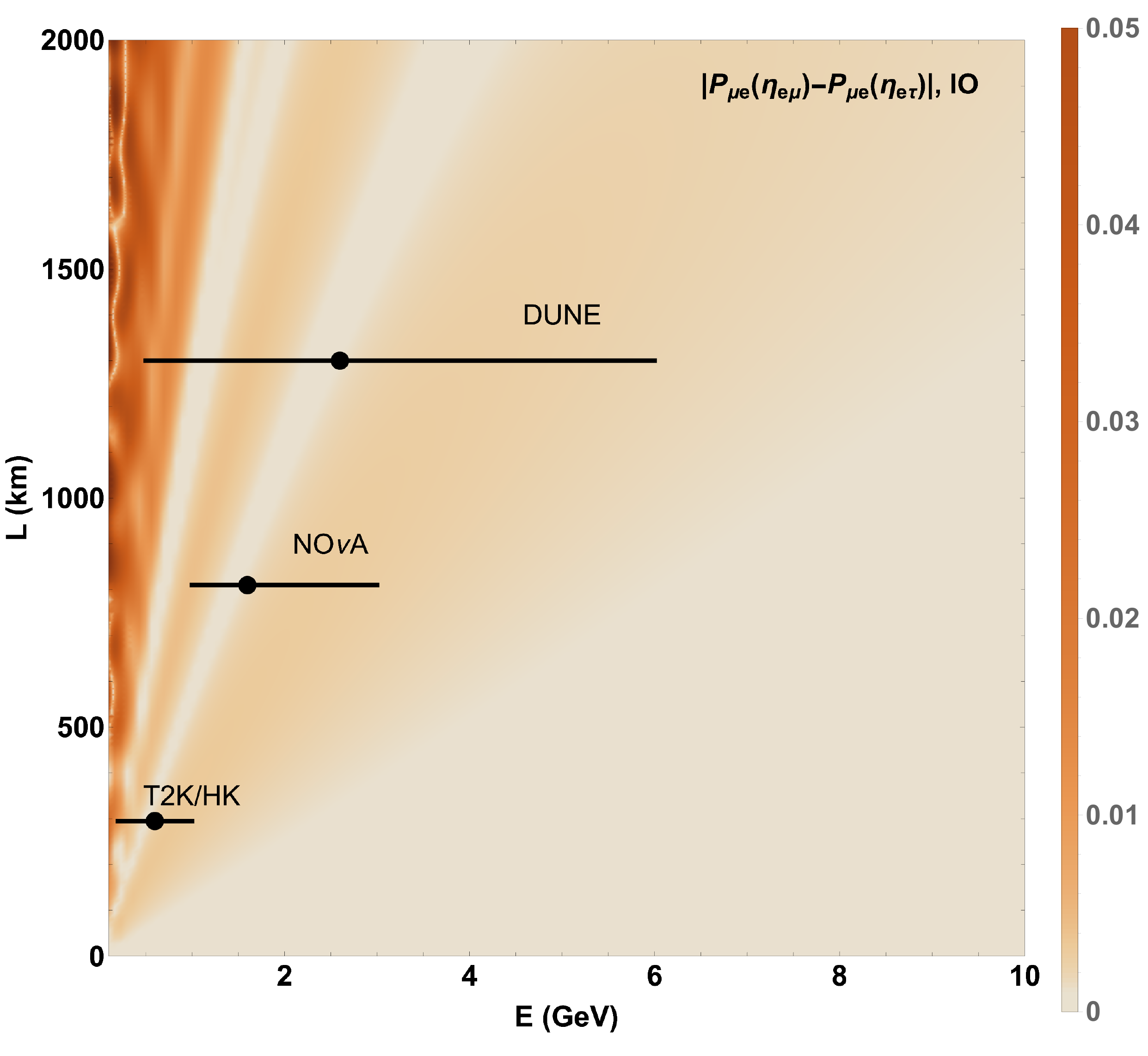}
\caption{The difference in neutrino appearance probabilities for two benchmark cases (see table \ref{tab:nova t2k vector}) as a function of baseline and neutrino energy.
On the left the difference is between the probabilities with vector NSI and $\varepsilon_{e\mu}$ and $\varepsilon_{e\tau}$ in normal ordering and the right is between probabilities with scalar NSI and $\eta_{e\mu}$ and $\eta_{e\tau}$ in inverted ordering.
The regions probed by the different long-baseline experiments are indicated.
The density is taken to be that for DUNE, 2.848 g/cm$^3$, throughout.}
\label{fig:oscillogram}
\end{figure}

To visually see the different effects of vector and scalar NSI at different long-baseline experiments, we show the difference in the appearance probabilities between to two benchmark NSI cases in Fig.~\ref{fig:oscillogram}, in particular $|P_{\mu e}(\varepsilon_{e\mu})-P_{\mu e}(\varepsilon_{e\tau})|$ in NO and $|P_{\mu e}(\eta_{e\mu})-P_{\mu e}(\eta_{e\tau})|$ in IO. These particular benchmark cases have been chosen being among the most significant ones in the  T2K-NOvA fit (see appendix \ref{fitAppendix}). We checked that results obtained using other scenarios are similar.
It is possible to notice that DUNE's broadband beam allows to catch more features of the probabilities than the other long-baseline experiments, and that the strength of the effects tend to be somewhat higher at larger baselines and energies, while is very mild at the first oscillation maximum.

\subsection{Sterile Neutrino Motivated by NOvA and T2K}
Light sterile neutrinos $m_4\lesssim10$ eV are interesting due to numerous anomalies.
While the slight tension in the NOvA and T2K data isn't substantially improved by the addition of a light sterile neutrino, one can nonetheless perform a fit to the experiments in the context of a 3+1 scenario.
In Ref.~\cite{Chatterjee:2020yak} they found that the benchmark point of $\theta_{14}=\theta_{24}=8^\circ$ and $\theta_{34}=0$ with $\Delta m^2_{41}=1$ eV$^2$ provides modest improvement to the NOvA and T2K data over the three-flavor oscillation hypothesis.
We note that long-baseline experiments are not particularly sensitive to the value of $\Delta m^2_{41}$ as long as $\Delta m^2_{41}\gg|\Delta m^2_{31}|$.
Meanwhile, these values of the two non-zero sterile mixing angles, $\theta_{14}$ and $\theta_{24}$ are near the existing limits from solar data \cite{Goldhagen:2021kxe} and long-baseline disappearance data \cite{MINOS:2017cae} respectively.
As they are not ruled out from other experiments, we then consider two benchmark points, one for each mass ordering.
While the NOvA and T2K data are somewhat constraining for $\delta_{13}$, the usual CP phase, they are not that constraining for the other two new phases, notably $\delta_{23}$.
Some information does exist, however, for $\delta_{12}$.
The benchmark points that provide the best fit to NOvA and T2K data in a sterile neutrino picture are shown in table \ref{tab:nova t2k sterile}.

\begin{table}
\centering
\begin{tabular}{c||c|c|c|c|c}
MO & $\Delta m^2_{41}$ [eV$^2$] & $\theta_{14}$ [$^\circ$] & $\theta_{24}$ [$^\circ$] & $\delta_{13}/\pi$ & $\delta_{12}/\pi$ \\\hline
NO & 1 & 8 & 8 & 1.9 & 0.7\\
IO & 1 & 8 & 8 & 0 & 0.5
\end{tabular}
\caption{Benchmark sterile neutrino parameters from NOvA and T2K data from Ref.~\cite{Chatterjee:2020yak}.
Sterile parameters not shown are zero and standard oscillation parameters not shown are taken to the standard values shown in table \ref{tab:bf}.}
\label{tab:nova t2k sterile}
\end{table}

\section{DUNE Analysis Details}

The DUNE (Deep Underground Neutrino Experiment) \cite{DUNE:2020ypp} is a next generation long baseline accelerator experiment that is under construction in the US. The near site, located at Fermilab, will host the Long Baseline Neutrino Facility (LBNF) and the Near Detectors complex. The Far Detector, a 40kt fiducial volume Liquid Argon Time Projection Chamber (LAr-TPC), on the other hand will be located at the Sanford Underground Research Facility (SURF) in South Dakota, 1285 km away from Fermilab. According to the most recent Technical Design Report (TDR), DUNE will use a 120 GeV proton beam of 1.2 MW power, which will deliver $1.1\, \times\, 10^{21}$ protons on target (POT) per year. DUNE will use an intense, on-axis, broad-band muon neutrino beam, peaked at 2.5 GeV (in order to sit around the first atmospheric oscillation maximum), with a small intrinsic $\nu_e$ contamination. Such a beam can also operate in antineutrino mode, providing an anti-muon neutrino flux. 

For the simulation of the DUNE experiment, we use the GLoBES software \cite{Huber:2004ka}. Considering the latest version of the DUNE GLoBES files \cite{DUNE:2021cuw}, we take into account an exposure of 6.5 years in neutrino mode and 6.5 years in antineutrino mode (total exposure of 312 kt-MW-years for each mode).
Accelerator upgrades to 2.4 MW would increase the statistics further or get to the quoted exposure faster, while realistic far detector staging would slow down the statistics somewhat.
The two oscillation channels we consider are the $\nu_\mu$ disappearance and the $\nu_e$ appearance. For the former the DUNE collaboration suggests a systematic uncertainty of 5\%; backgrounds to this channel are misidentified $\nu_e$ and Neutral Current (NC) events. For the latter, the systematic uncertainty is 2\% and the backgrounds consists on misidentified $\nu_\tau$ and NC events. Energy resolution and efficiency functions, as well as smearing matrices are given by the collaboration. We consider the Earth matter density to be 2.848 g/cm$^3$.

In order to study the DUNE performances in the measurements of the new physics parameters, we derive the statistical significance of our results using a $\Delta \chi^2$ function calculation with priors based on the pull method \cite{Fogli:2002pt}. As true values for the standard oscillation parameters, except for the CP-violating phase $\delta$ whose value depends on the given benchmark scenario, we consider the ones from the global fit without SuperKamiokande atmospheric data \cite{Esteban:2020cvm}, summarized in table \ref{tab:param}. In our simulations we also marginalize over the cited parameters using Gaussian priors given by their $1\sigma$ uncertainties shown in the above mentioned table; the mass ordering is considered to be known. It is worth to notice that other global fits \cite{deSalas:2020pgw,Capozzi:2021fjo} obtained slightly different results for the oscillation parameters. However, we checked in our simulations that the results are not significantly affected by the choice of the sets of parameters and their uncertainties. Moreover, the inclusion of the atmospheric data in the fit, can change the preference of the $\theta_{23}$ octant from the upper one to the lower one in NO. We also checked that the performances of DUNE in constraining our benchmark scenarios are not drastically affected by the atmospheric angle octant.
This is because DUNE will have measurements of $\theta_{23}$, $\Delta m^2_{31}$, and $\delta$ that will all be considerably better than currently available, and is significantly less sensitive to the other three oscillation parameters.

\begin{table}
\centering
\begin{tabular}{c||c|c|c|c|c}
MO & $\theta_{12}$ ($^\circ$) & $\theta_{23}$ ($^\circ$) & $\theta_{13}$ ($^\circ$) & $\Delta m_{21}^2$ [$10^{-5}$ eV$^2$]  & $\Delta m_{3l}^2$ [$10^{-3}$ eV$^2$]    \\ \hline
NO & $33.44^{+0.77}_{-0.74}$                  & $49.2^{+1.0}_{-1.3}$                   & $8.57^{+0.13}_{-0.12}$                   & $7.42^{+0.21}_{-0.20}$ & $2.515^{+0.028}_{-0.028}$  \\ \hline
IO & $33.45^{+0.77}_{-0.74}$                  & $49.5^{+1.0}_{-1.2}$                   & $8.60^{+0.12}_{-0.12}$                   & $7.42^{+0.21}_{-0.20}$ & $-2.498^{+0.028}_{-0.029}$
\end{tabular}
\caption{\label{tab:param} Preferred values from a recent global fit to oscillation data \cite{Esteban:2020cvm}.
$\Delta m^2_{3l}$ is $\Delta m^2_{31}$ when positive and $\Delta m^2_{32}$ when negative.}
\label{tab:bf}
\end{table}

While DUNE will have a state-of-the-art near detector facility \cite{DUNE:2021tad} which will provide important information about sterile neutrinos depending on the $\Delta m^2_{41}$ values, we conservatively consider only the far detector in this study.

\section{Single Scenario Results}
\label{sec:results}
In this section we present our results of DUNE's sensitivity to just a single new physics scenario.
This is done to show how well DUNE can identify the benchmarks we are considering when compared against the standard oscillation picture.
It also allows us to present the first results of scalar NSI at NOvA and T2K.
We find that, in general, DUNE has very good sensitivity to discover new physics at the level indicated by current long-baseline oscillation experiments, as expected.

\subsection{Vector NSI}
\begin{figure}
\begin{center}
\includegraphics[width=0.32\textwidth]{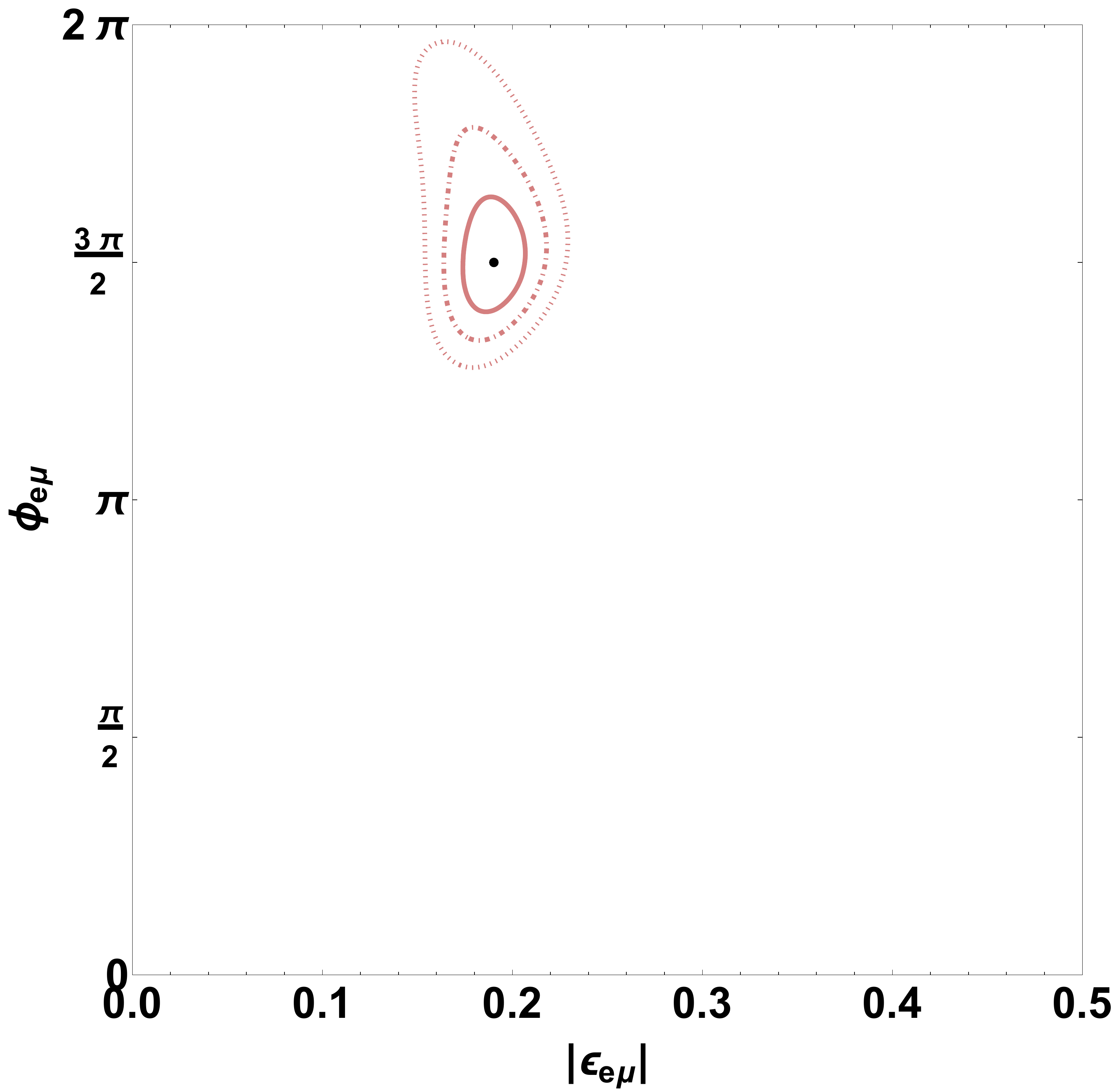}
\includegraphics[width=0.32\textwidth]{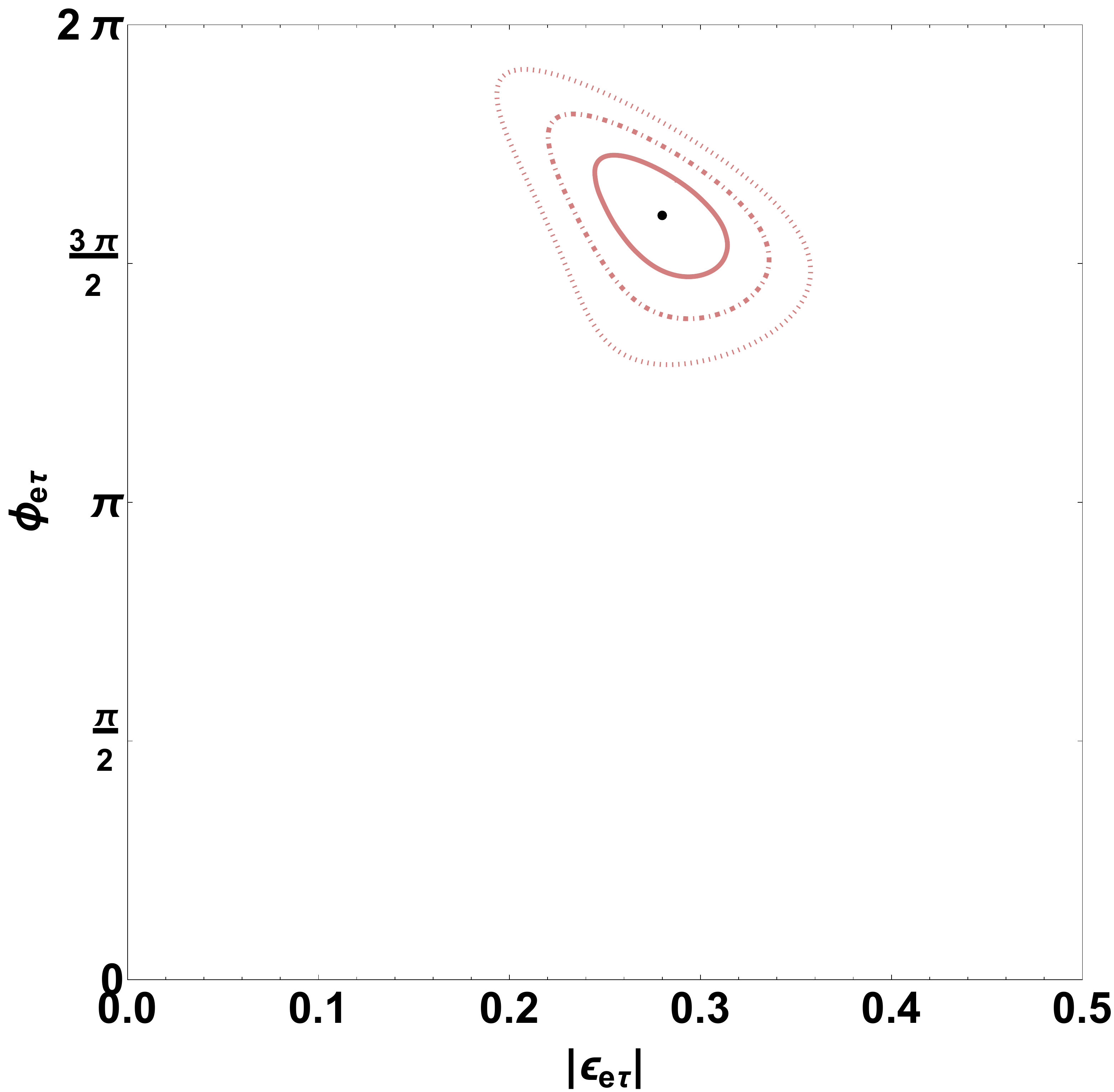}
\includegraphics[width=0.32\textwidth]{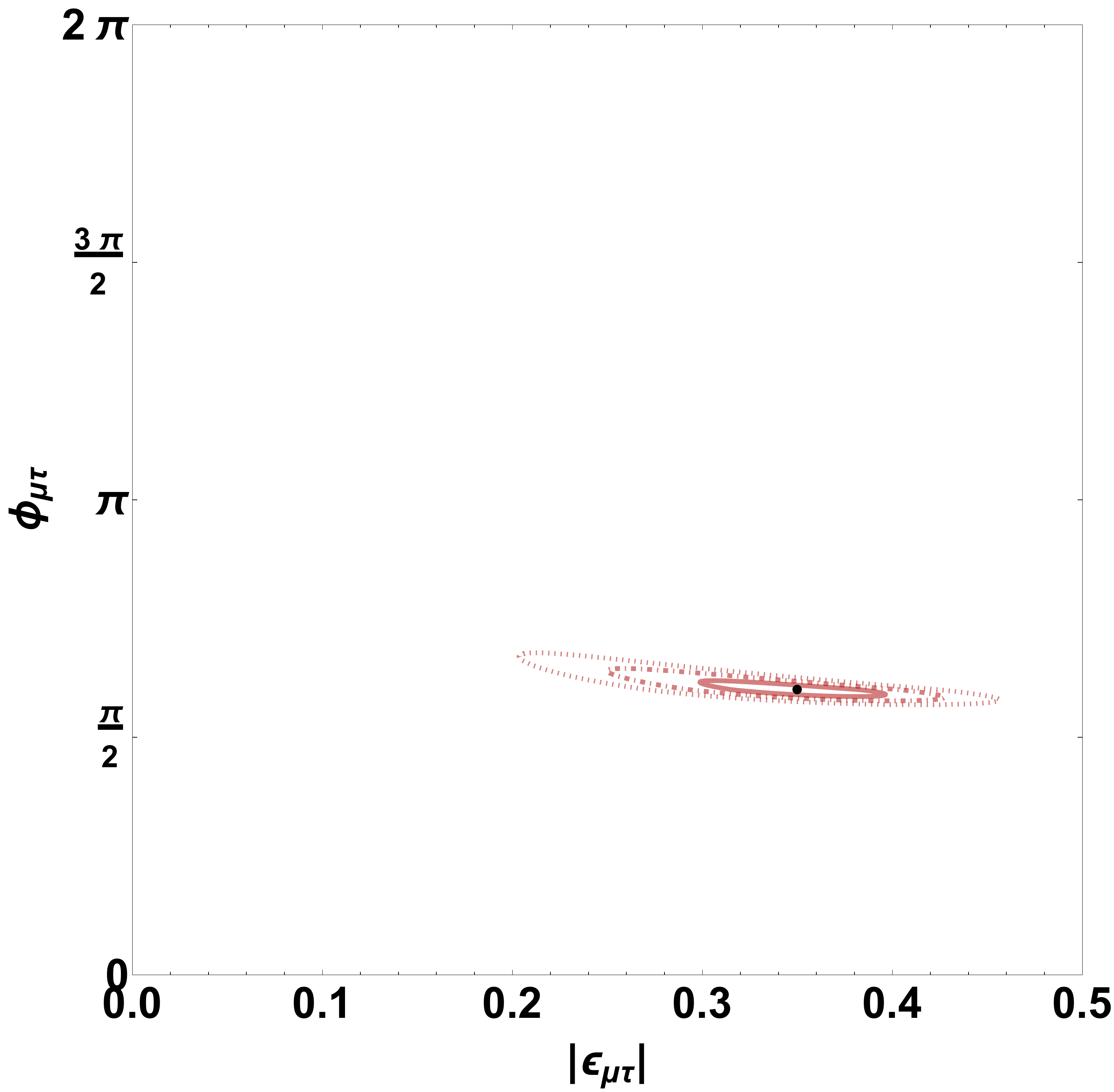}
\includegraphics[width=0.32\textwidth]{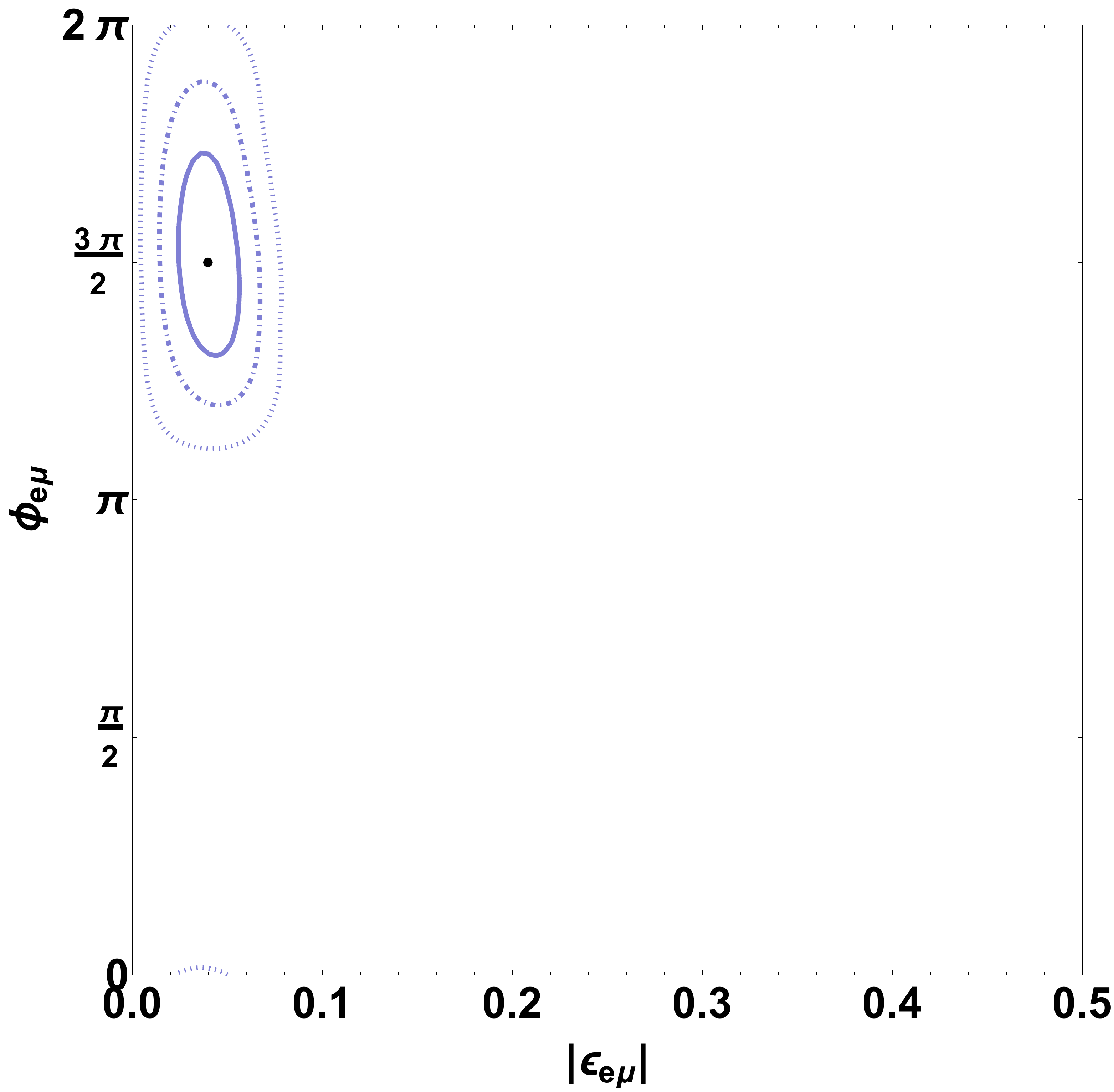}
\includegraphics[width=0.32\textwidth]{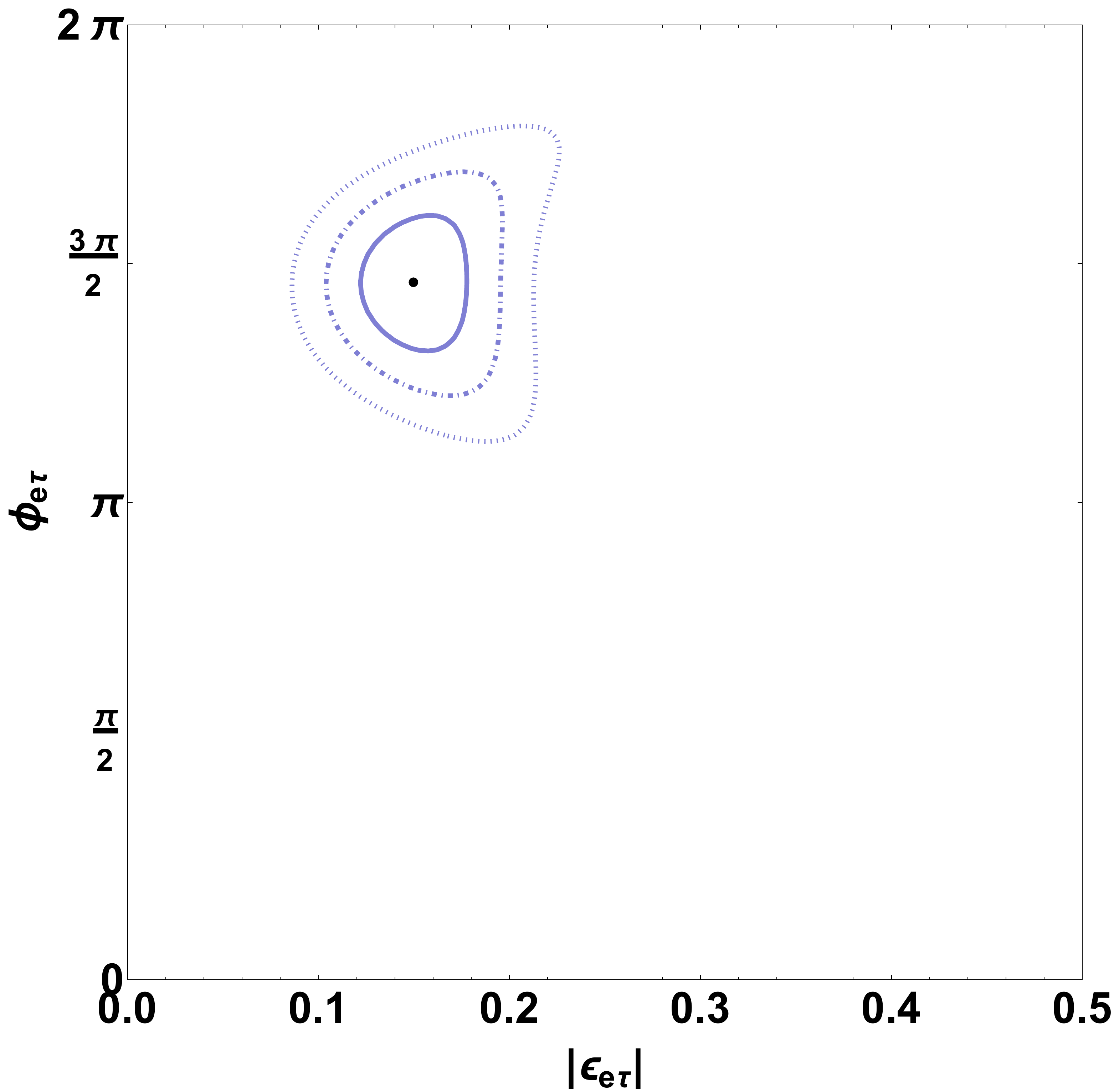}
\includegraphics[width=0.32\textwidth]{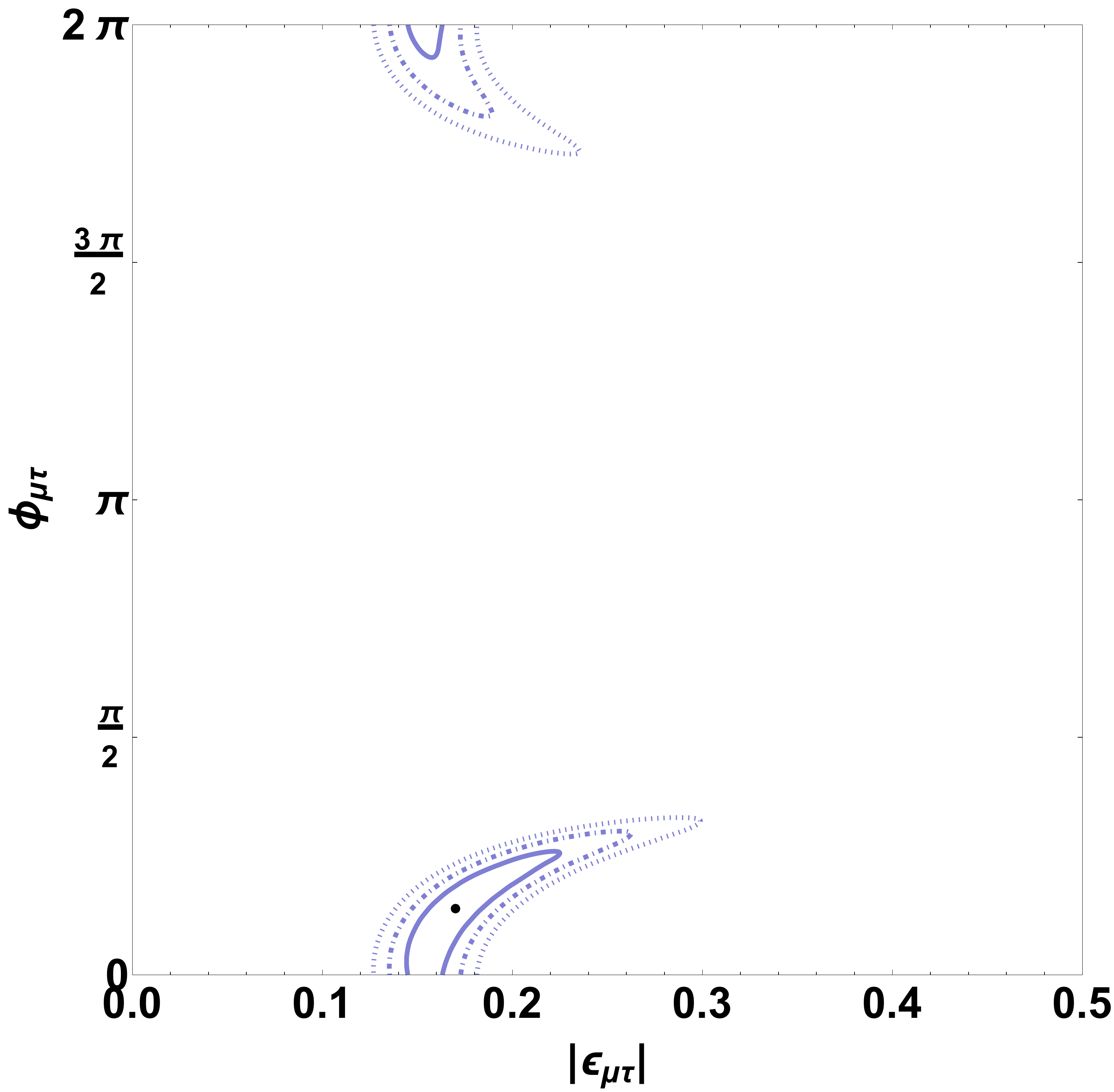}
\caption{68\% (solid lines), 95\% (dashed lines) and 99\% (dotted lines) contours in DUNE in the vector NSI $(|\varepsilon_{ij}|-\phi_{ij})$-planes when data are generated using vector NSI best fits considering one parameter at a time.
The top (bottom) panels with red (blue) contours have been obtained using NO (IO) best fits.} 
\label{fig:vector}
\end{center}
\end{figure}

We discuss here the performances of the DUNE Far Detector in constraining the NSI parameters in the benchmark scenarios discussed in section \ref{sec:bench}.
Fig.~\ref{fig:vector} shows the DUNE allowed regions at 68, 90, and 99\% CL in the ($\varepsilon_{\alpha\beta}-\phi_{\alpha\beta}$) plane. The top (bottom) panels show the results using the NO (IO) hypothesis.
In order to obtain the contours, we consider as true values for the two mass splittings and the mixing angles the ones summarized in table \ref{tab:param} and the values for $\varepsilon_{\alpha\beta}$, $\phi_{\alpha\beta}$ and $\delta$ from the NOvA-T2K fit. For the fit we marginalize over the oscillation parameters with pull terms. All the NSI parameters that do not appear in each plot are fixed to 0 both in the theory and in the fit.

For the vector NSI case, we observe that the most interesting results are obtained when the mass ordering is normal, since the $\Delta\chi^2$-s in respect to the standard model are 4.44 and 3.65 when the fits are performed in NO considering non-vanishing $\varepsilon_{e\mu}$ and $\varepsilon_{e\tau}$, respectively (see table \ref{tab:nova t2k vector} taken from \cite{Denton:2020uda}); however, for the sake of completeness, we show the results considering also the IO scenarios.

It is clear that DUNE is expected to be able to measure all the above mentioned vector NSI parameter with a good precision. Indeed, when $\varepsilon_{e\mu}$ is considered in NO, in the 2-dimensional plane (2 degrees of freedom), the 68\% CL allowed region includes the intervals [0.16,0.21] for $\varepsilon_{e\mu}$ and [1.3,1.8]$\pi$ for $\phi_{e\mu}$. This means that in the first benchmark scenario, DUNE would be able to determine both NSI parameters with a precision of roughly 10\% at the given benchmarks. In the IO case, since the best fit value for $\varepsilon_{e\mu}$ is smaller (0.04), the allowed region is bigger, but still excludes at 99\% CL the standard model.

In the middle panels, namely when the benchmark scenario presents non-zero values for $\varepsilon_{e\tau}$ and $\phi_{e\tau}$, it is possible to observe that the DUNE performances are similar to the previous case. In particular, the allowed regions includes the intervals [0.22,0.34] for $\varepsilon_{e\tau}$ and [1.4,1.8]$\pi$ for $\phi_{e\tau}$ which correspond to a precision of roughly 20\% for the magnitude of the parameter and 10\% for its phase for the NO case. When we consider the IO scenario, the precision on the parameters remains basically the same, even though the best fit value for the NSI coupling magnitude is reduced by almost a factor of 2.

The results are very different when $\varepsilon_{\mu\tau}$ is considered. In this case, the best fit value for $\phi_{\mu\tau}$ from NOvA and T2K is very close to $\pi/2$ in the NO case. Since the NSI correction to the $\nu_\mu$ disappearance probability depends at the leading order on the combination $\varepsilon_{\mu\tau}\cos\phi_{\mu\tau}$, see Eq.~\ref{eq:approx vector dis}, DUNE is expected to be very sensitive to small variation of the phase around $3/2\pi$, but is not adequate to constrain the magnitude $|\varepsilon_{\mu\tau}|$ with the same precision reached for the other parameters. When we consider the IO hypothesis, the situation is the opposite, since the phase best fit is close to zero: the magnitude is tightly constrained while the phase can vary in a relatively large interval.
We note also that atmospheric data from SK and IceCube is quite strong for this parameter \cite{Super-Kamiokande:2011dam,IceCube:2022ubv}.

\begin{figure}
\begin{center}
\includegraphics[width=0.32\textwidth]{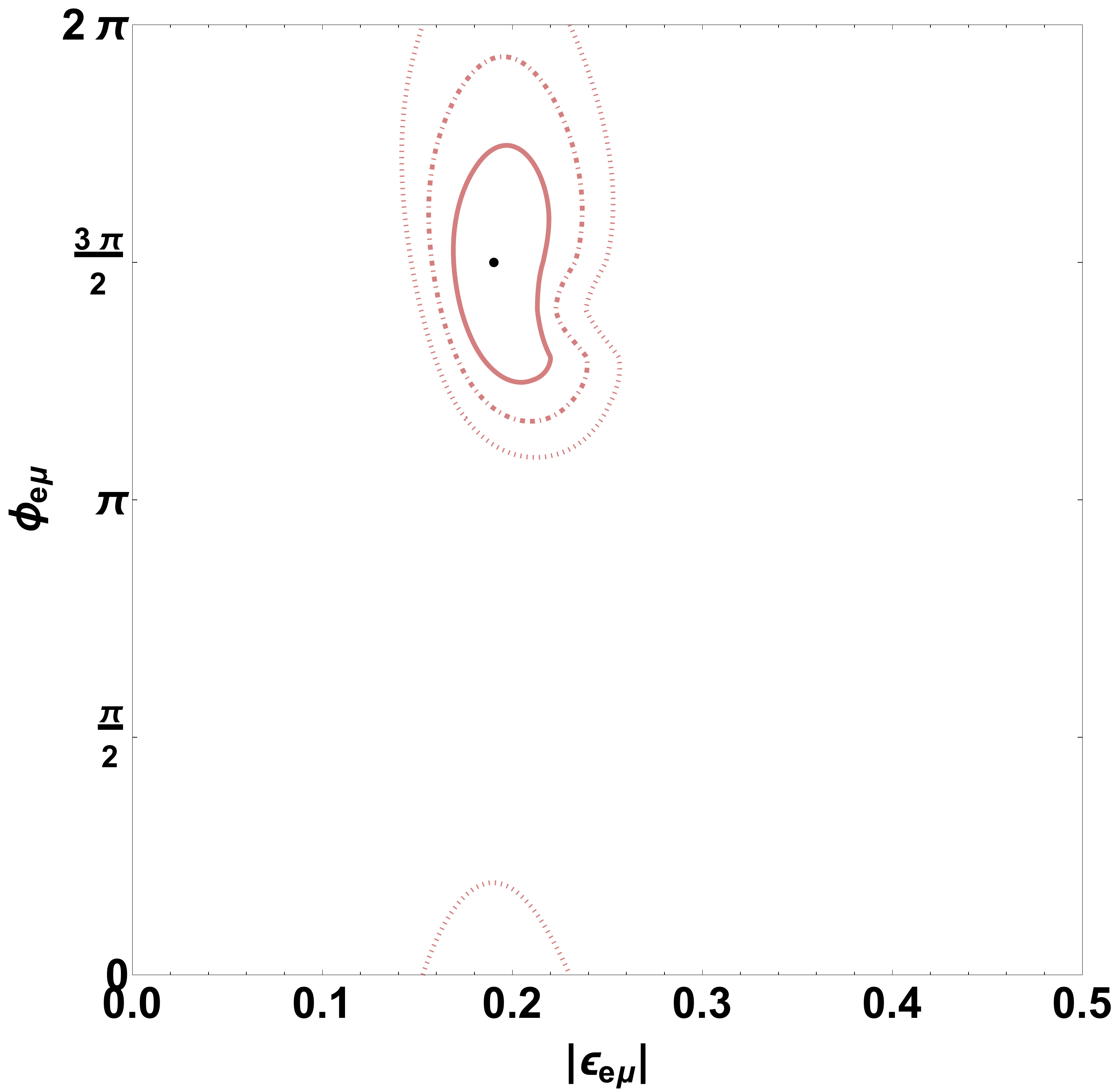}
\includegraphics[width=0.32\textwidth]{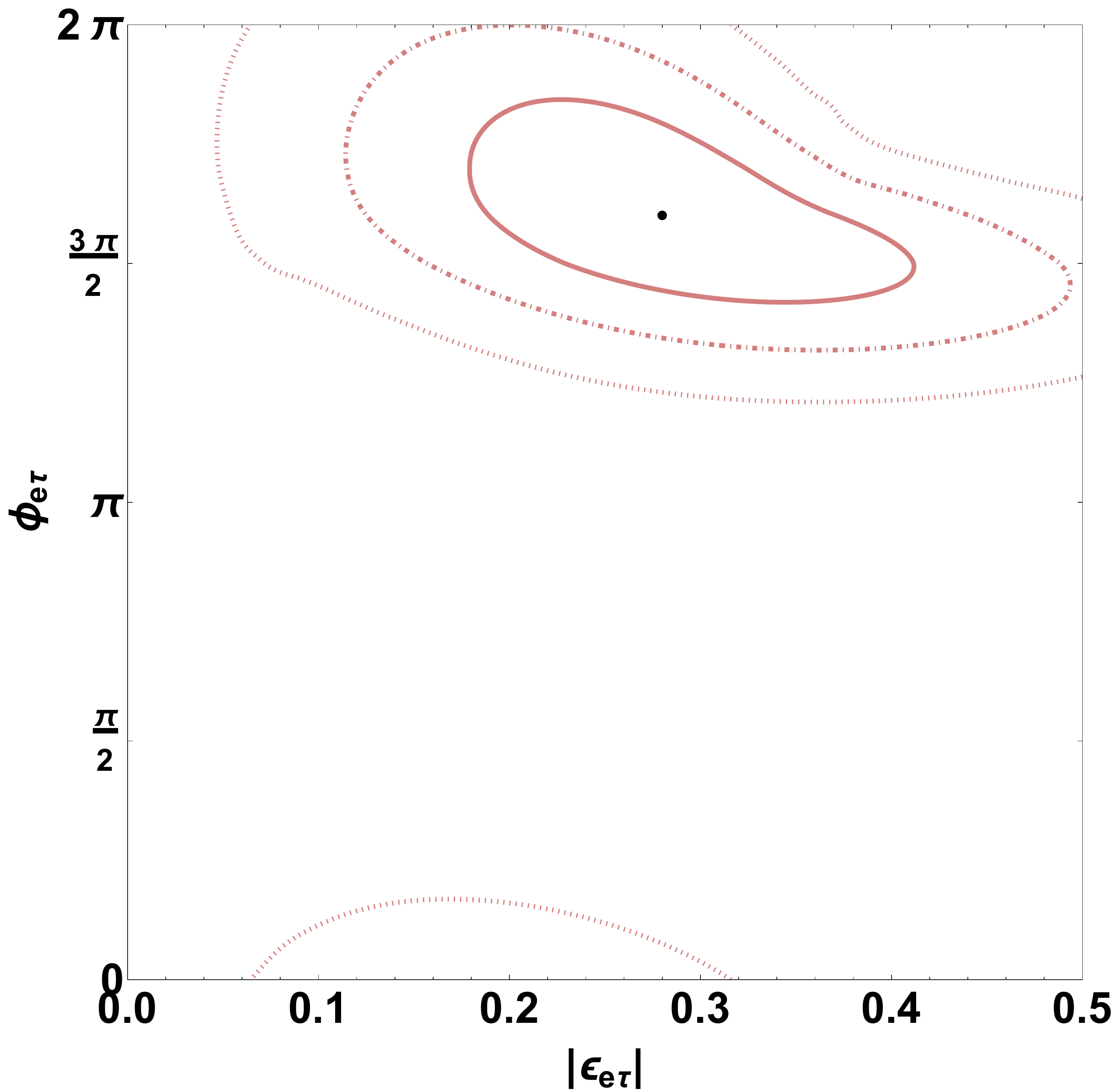}
\includegraphics[width=0.32\textwidth]{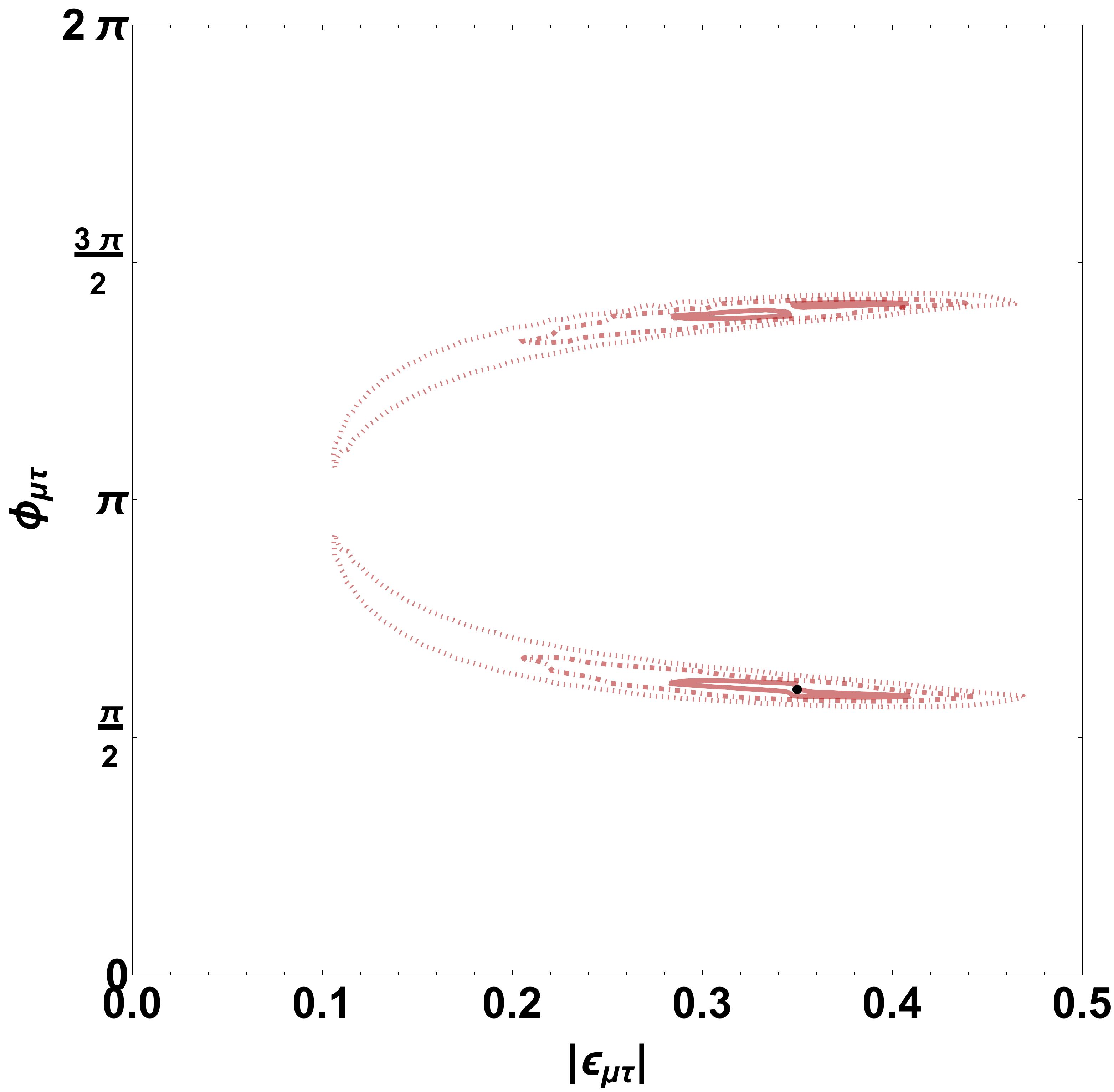}
\includegraphics[width=0.32\textwidth]{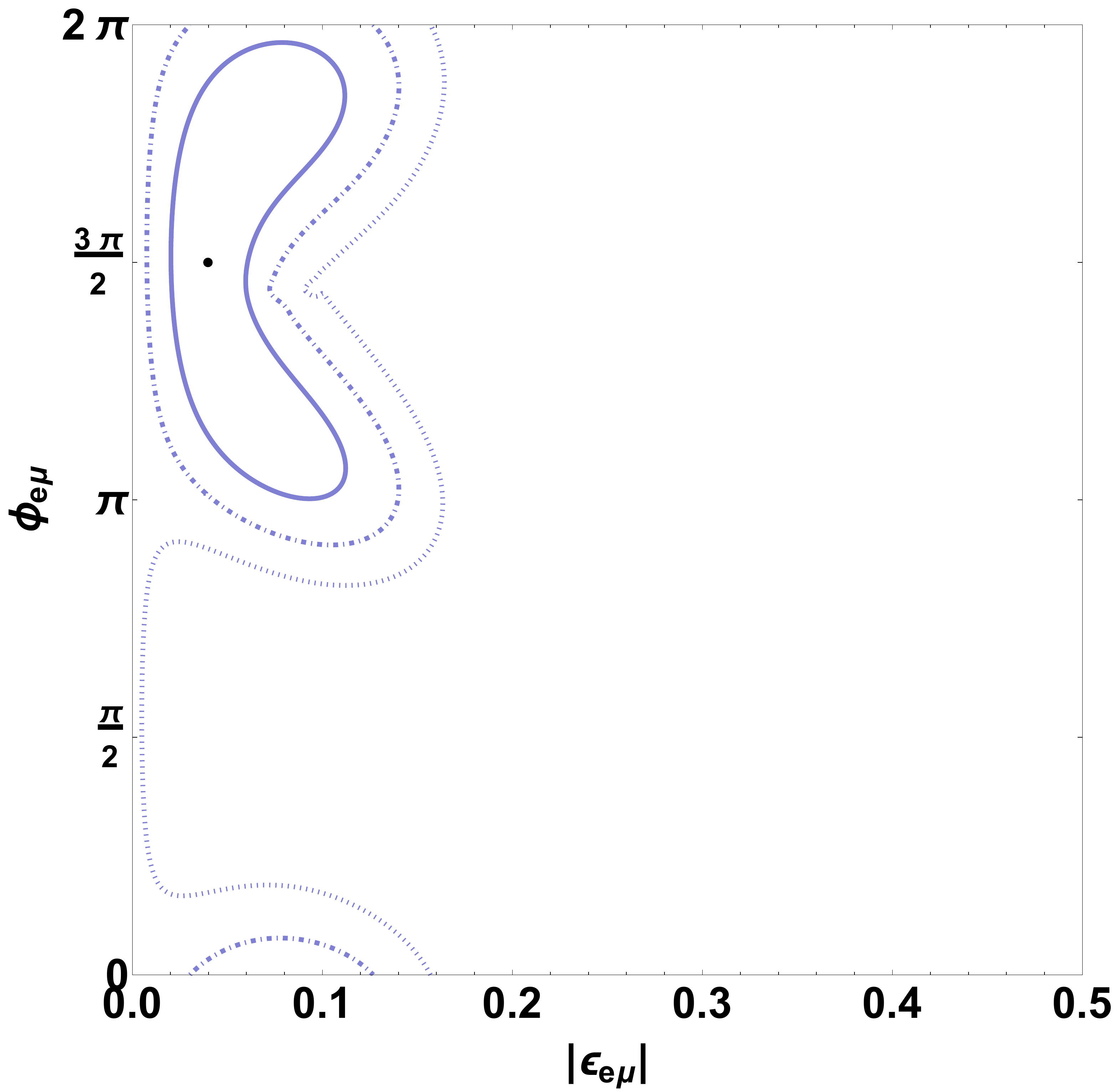}
\includegraphics[width=0.32\textwidth]{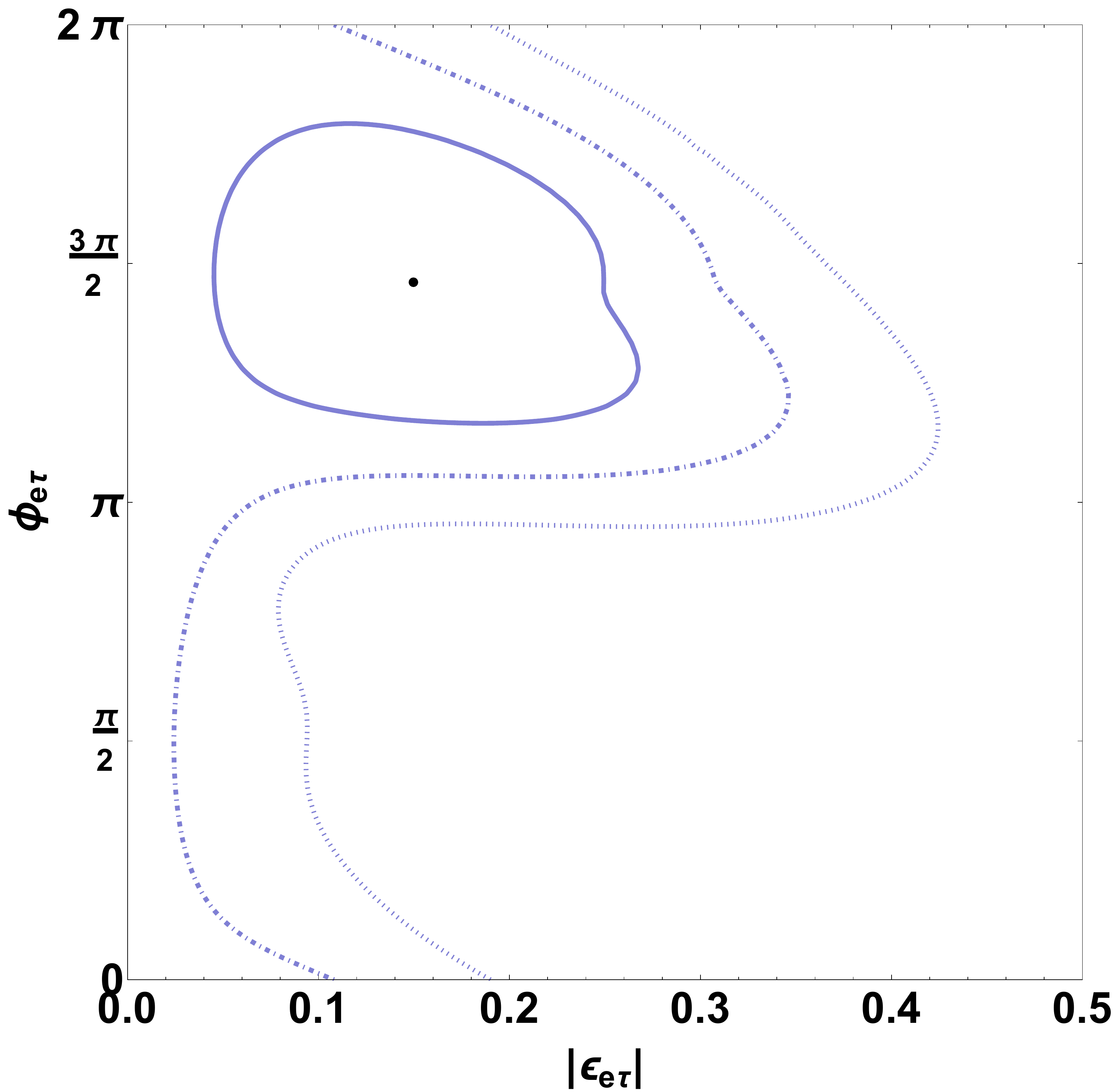}
\includegraphics[width=0.32\textwidth]{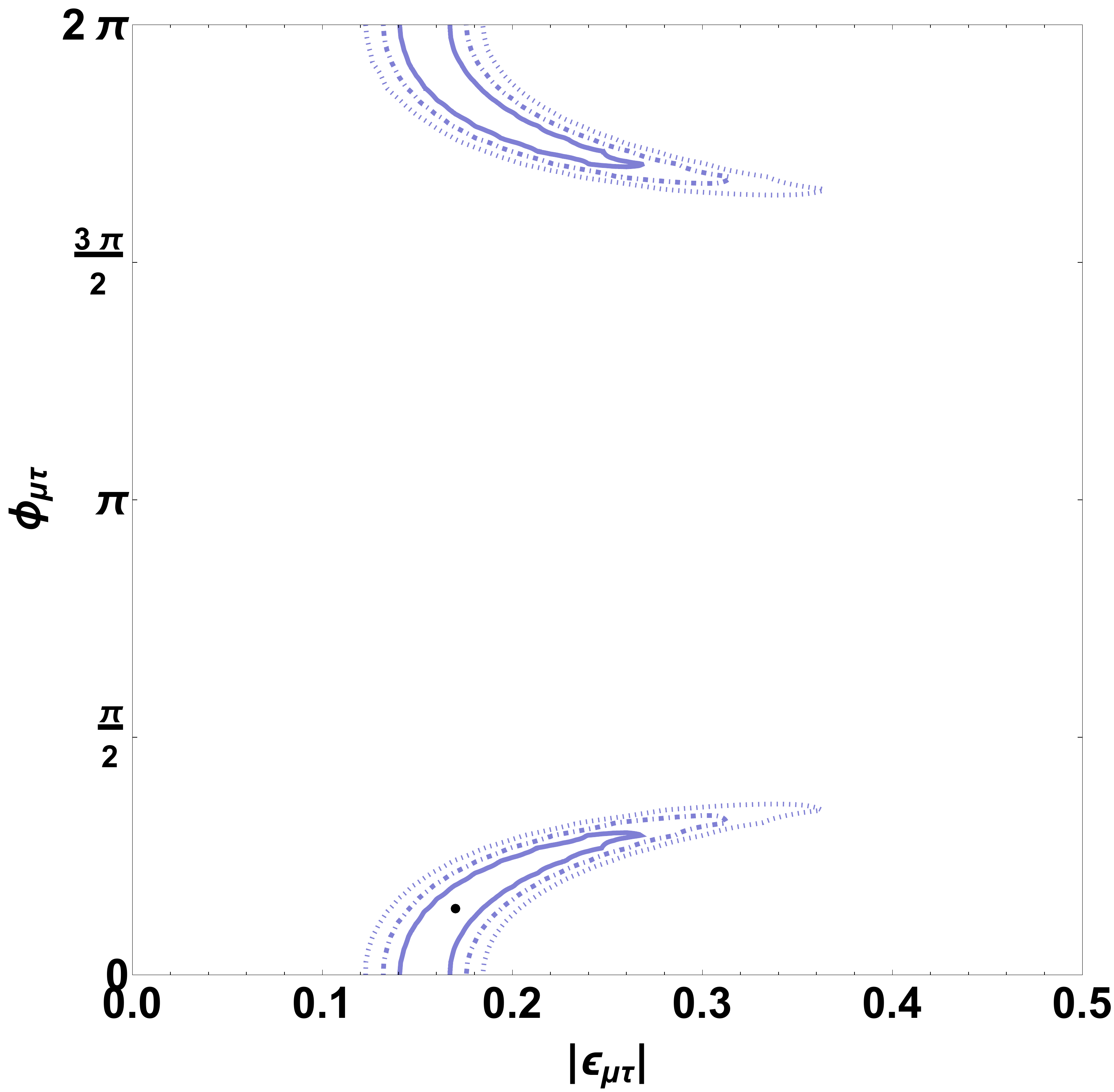}
\caption{68\% (solid lines), 95\% (dashed lines) and 99\% (dotted lines) contours in DUNE in the vector NSI $(|\varepsilon_{ij}|-\phi_{ij})$-planes when data are generated using NSI best fits marginalizing over all the NSI parameters. The three top (bottom) plots with red (blue) contours have been obtained using NO (IO) best fits.
} 
\label{fig:marg}
\end{center}
\end{figure}

Then, instead of considering just a single NSI parameter at a time, we marginalize over all NSI parameters (magnitudes and phases) with select input priors: $|\varepsilon_{\mu\tau}|<0.02$ (when undisplayed) \cite{IceCubeCollaboration:2021euf} and $\varepsilon_{ee}<0.3$ \cite{Esteban:2019lfo}.
The results are shown in Fig.~\ref{fig:marg}. Compared to the single parameter case the contours are obviously enlarged, however DUNE is still be able to exclude large portions of the parameters spaces in the studied benchmark scenarios and easily discover new physics.
The most interesting features that can be observed are the following:
\begin{itemize}
    \item In both the NO and IO cases, when we consider $\varepsilon_{\mu\tau}$, the $\cos\phi_{\mu\tau}$ degeneracy appears.
    \item In the IO case, now the SM is allowed at 99\% (95\%) CL in the $\varepsilon_{e\mu}$ ($\varepsilon_{e\tau}$) scenarios due partially to slightly smaller benchmark parameters.
\end{itemize}
We note that the degeneracy is $\varepsilon_{\mu\tau}$ appears only in the case where all NSI parameters are scanned over because $\varepsilon_{\mu\tau}$ does appear in the appearance probability at a subleading level which is enough to break the degeneracy between $\sin\phi_{\mu\tau}\leftrightarrow-\sin\phi_{\mu\tau}$.
When all NSI parameters are considered, however, there is more than enough freedom to cancel out the effect in appearance mode without significantly affect the disappearance measurement.
See also \cite{Masud:2018pig} for more discussion of some of these degeneracies.

Other studies have also investigated complex off-diagonal vector NSI at DUNE, often using this benchmark approach \cite{deGouvea:2015ndi,Bakhti:2020fde,DUNE:2020fgq,Chatterjee:2021wac}.
As they have used different benchmarks, either the SM or particular NSI values, it is not possibly to directly compare them, but qualitatively they find comparable sensitivity to vector NSI at the $|\varepsilon_{\alpha\beta}|\sim0.1$ level with a sizable dependence on the complex NSI phases and a stronger sensitivity and more precision for $\varepsilon_{\mu\tau}$ than either of $\varepsilon_{e\beta}$.

\subsection{Scalar NSI}
\begin{figure}
\begin{center}
\includegraphics[width=0.32\textwidth]{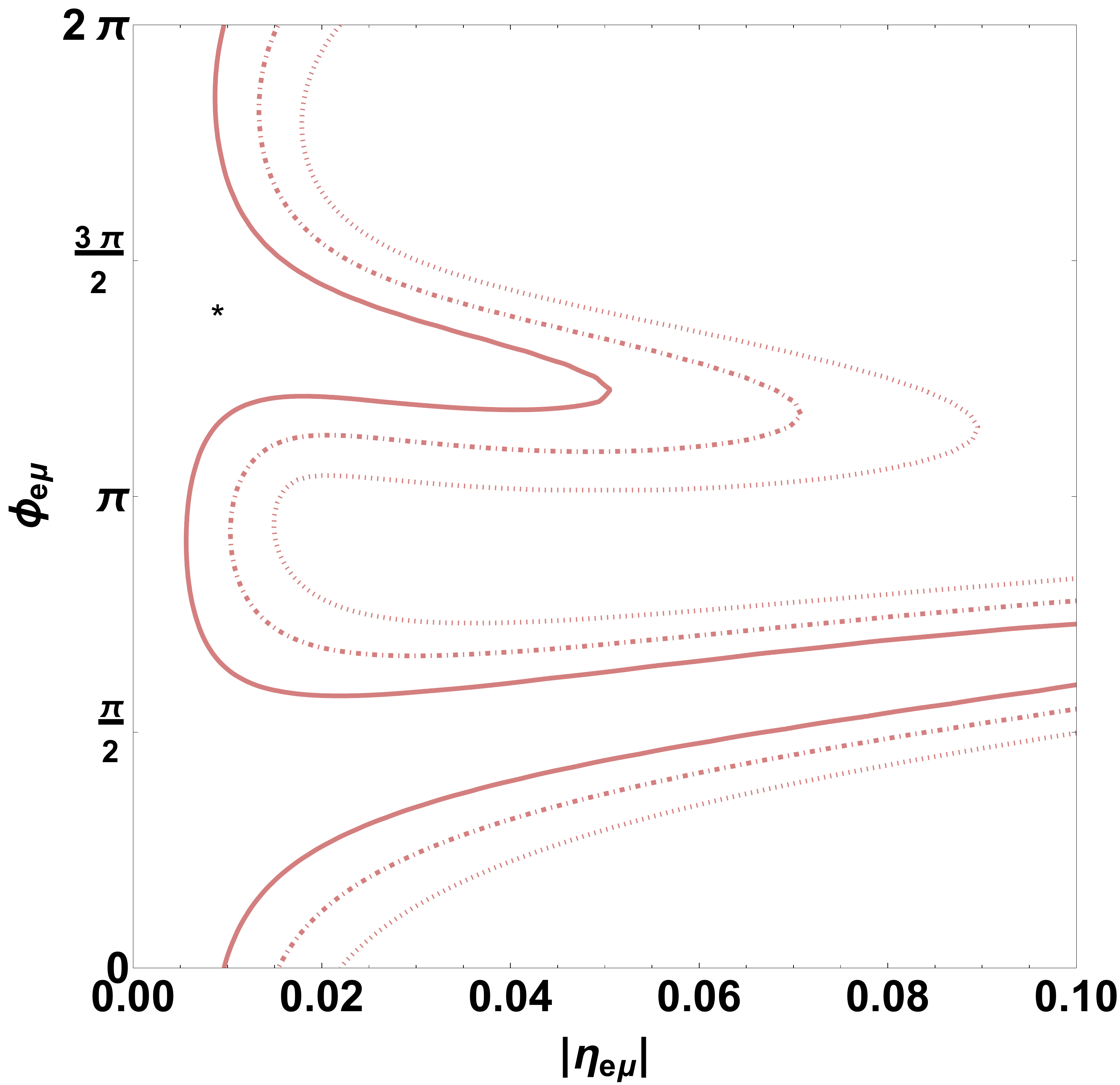}
\includegraphics[width=0.32\textwidth]{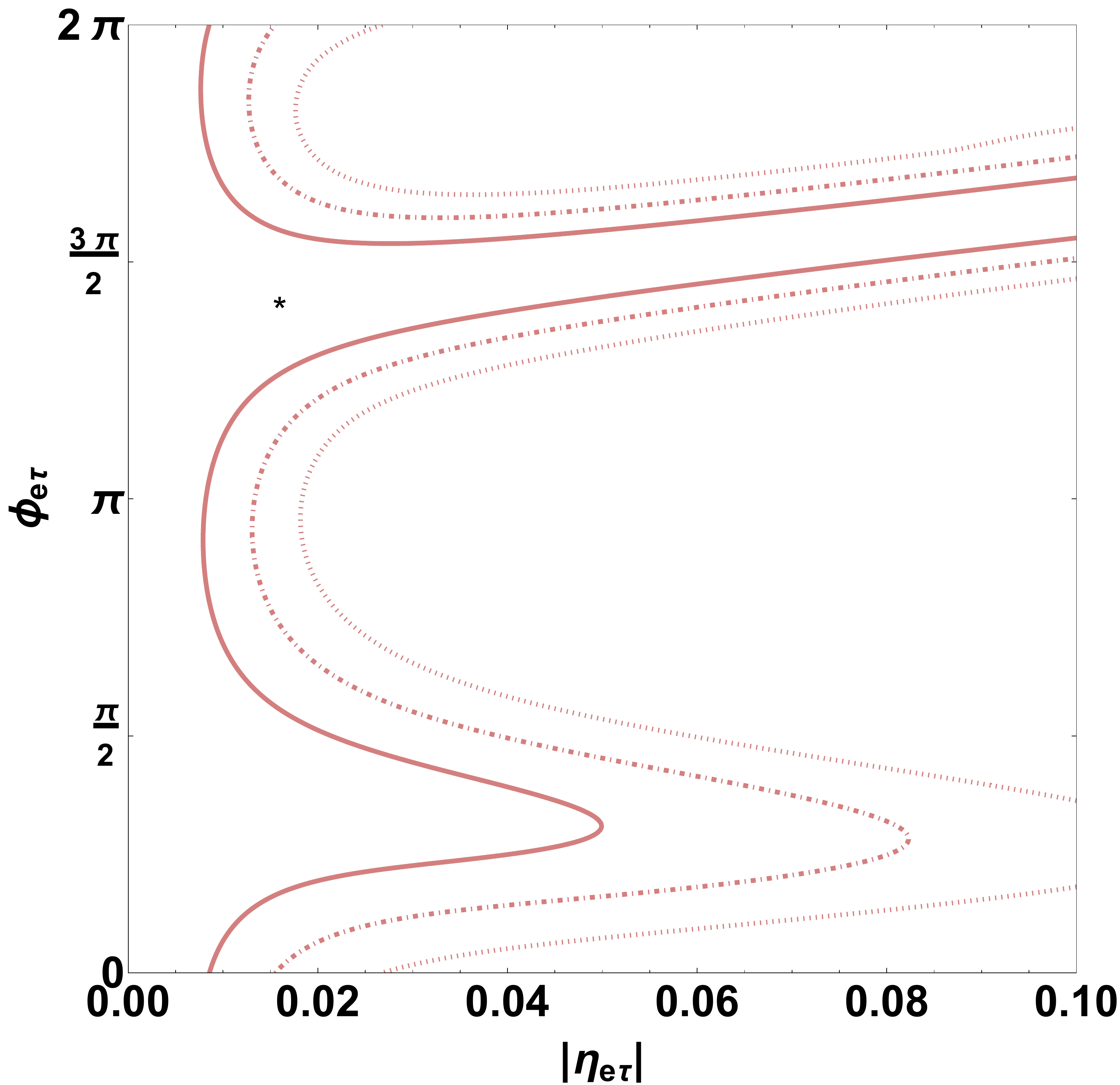}
\includegraphics[width=0.32\textwidth]{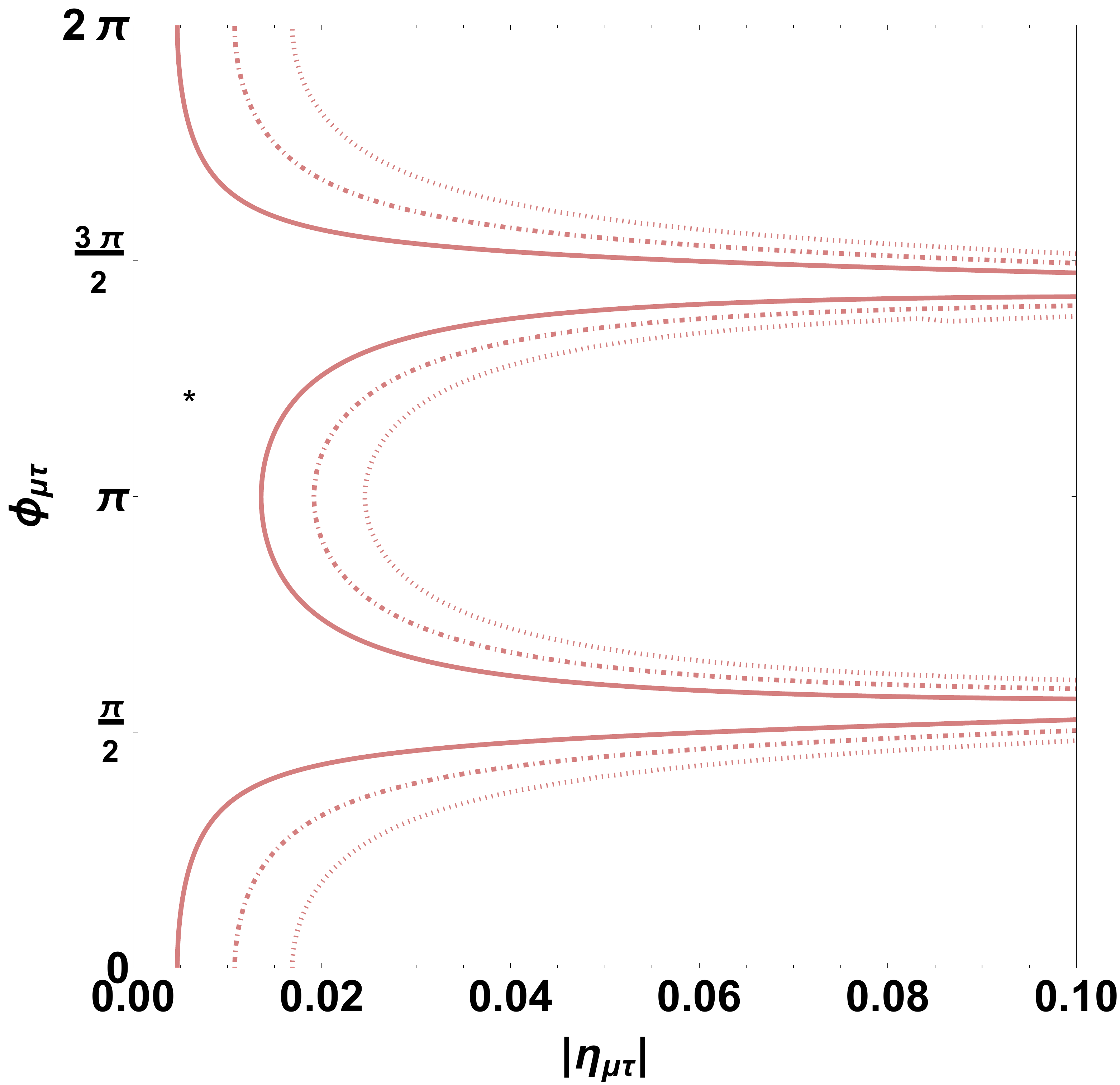}
\includegraphics[width=0.32\textwidth]{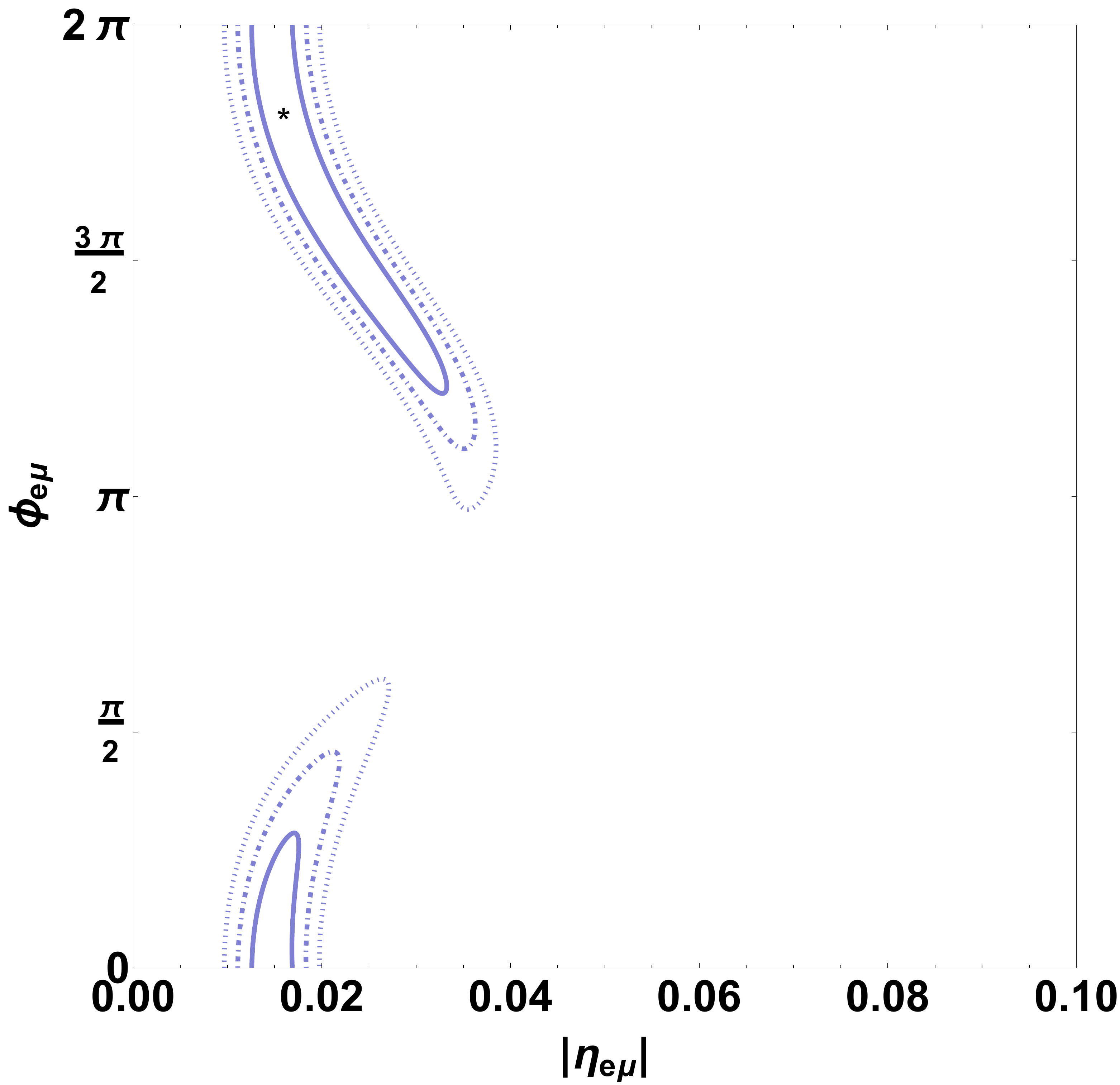}
\includegraphics[width=0.32\textwidth]{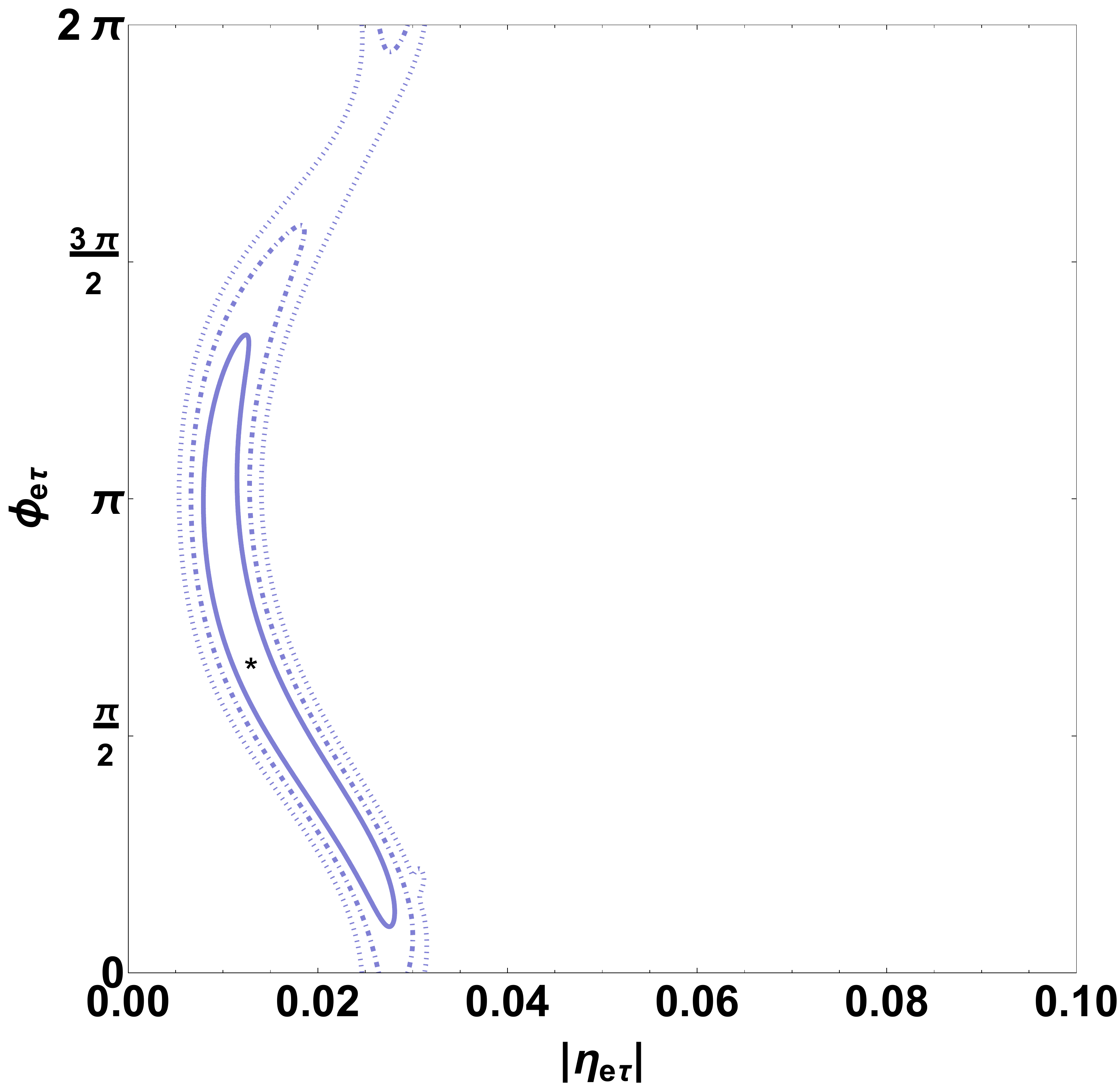}
\includegraphics[width=0.32\textwidth]{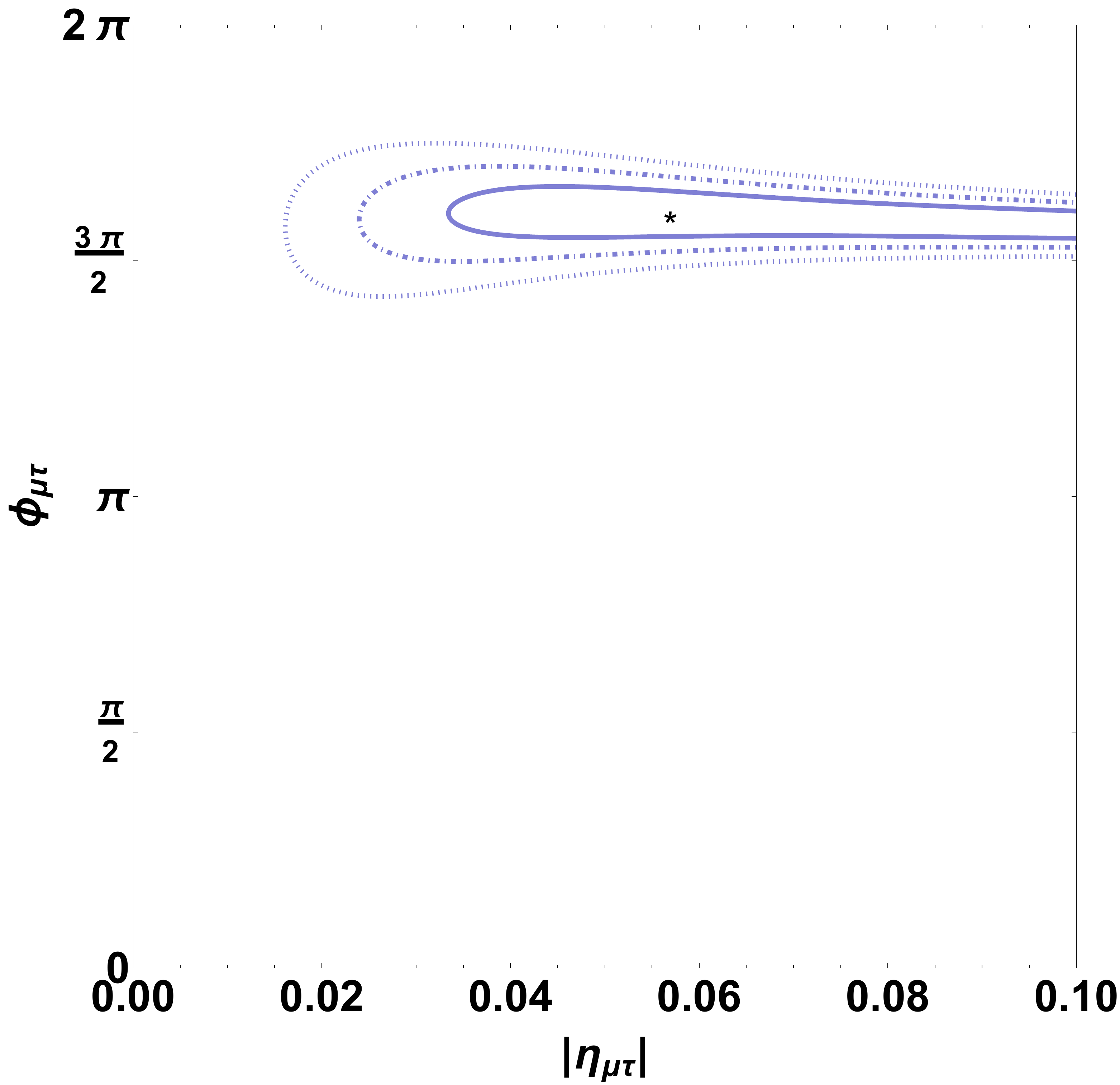}
\caption{68\% (solid lines), 95\% (dashed lines) and 99\% (dotted lines) contours in DUNE in the scalar NSI $(|\eta_{ij}|-\phi_{ij})$-planes where $\eta_{ij}$ is the value of $\eta_{ij}$ at $\rho=3$ g/cm$^3$.
The true data are generated using scalar NSI best fits (with $m_{\rm lightest}=0$) considering one NSI parameter at a time.
The three top (bottom) plots with red (blue) contours have been obtained using NO (IO) best fits.}
\label{fig:scalar}
\end{center}
\end{figure}

For the sensitivities performed using the off-diagonal scalar NSI benchmark scenarios, the first study of its kind, we use the same approach described in the previous subsection. Differently from the previous case, the results with more statistical significance in the NOvA-T2K fit are the IO ones.
In Fig.~\ref{fig:scalar} we show our results in the ($\eta_{\alpha\beta}-\phi_{\alpha\beta}$) planes in the case in which the lightest neutrino mass is zero.

When the mass ordering is inverted, DUNE will to constrain scalar NSI at the following levels at 68\% CL: $|\eta_{e\mu}|\in [0.012,0.035]$ and $|\eta_{e\tau}|\in[0.008,0.012]$. On the other hand, DUNE cannot set remarkable bounds on the phases, being able to exclude at 68\% CL only one third of the possible values of $\phi_{e\mu}$ (from $0.44\pi$ to $1.1\pi$) and one sixth of the possible values of $\phi_{e\tau}$ (from $1.6\pi$ to  $1.9\pi$). For the non-zero $\eta_{\mu\tau}$ scenario, in which the scalar NSI coupling best fit is bigger, we have a different situation. Indeed, DUNE is expected in this case to bound with good precision the phase, but is very unconstraining in the magnitude.
In the NO case, the NOvA-T2K results are characterized by very small best values and small $\Delta\chi^2$-s. When we perform sensitivity scans with DUNE, we can observe that this experiment is not able to distinguish the new physics scenarios from the Standard Model not even at 68\% level. Moreover, the magnitudes of the three off-diagonal scalar NSI parameters cannot be bounded from above and the phases are unconstrained by DUNE.

We checked that when a full marginalization is performed, DUNE is not able to exclude remarkable portions of the parameters space at a good confidence level. Indeed, the scalar NSI scenarios produce very similar phenomenology at DUNE (see section \ref{sec:diff}); thus, when we allow all the parameters to vary in the fit, the effects of the $\eta$-s can be reduced by the presence of other non-zero parameters.

\begin{figure}
\begin{center}
\includegraphics[width=0.32\textwidth]{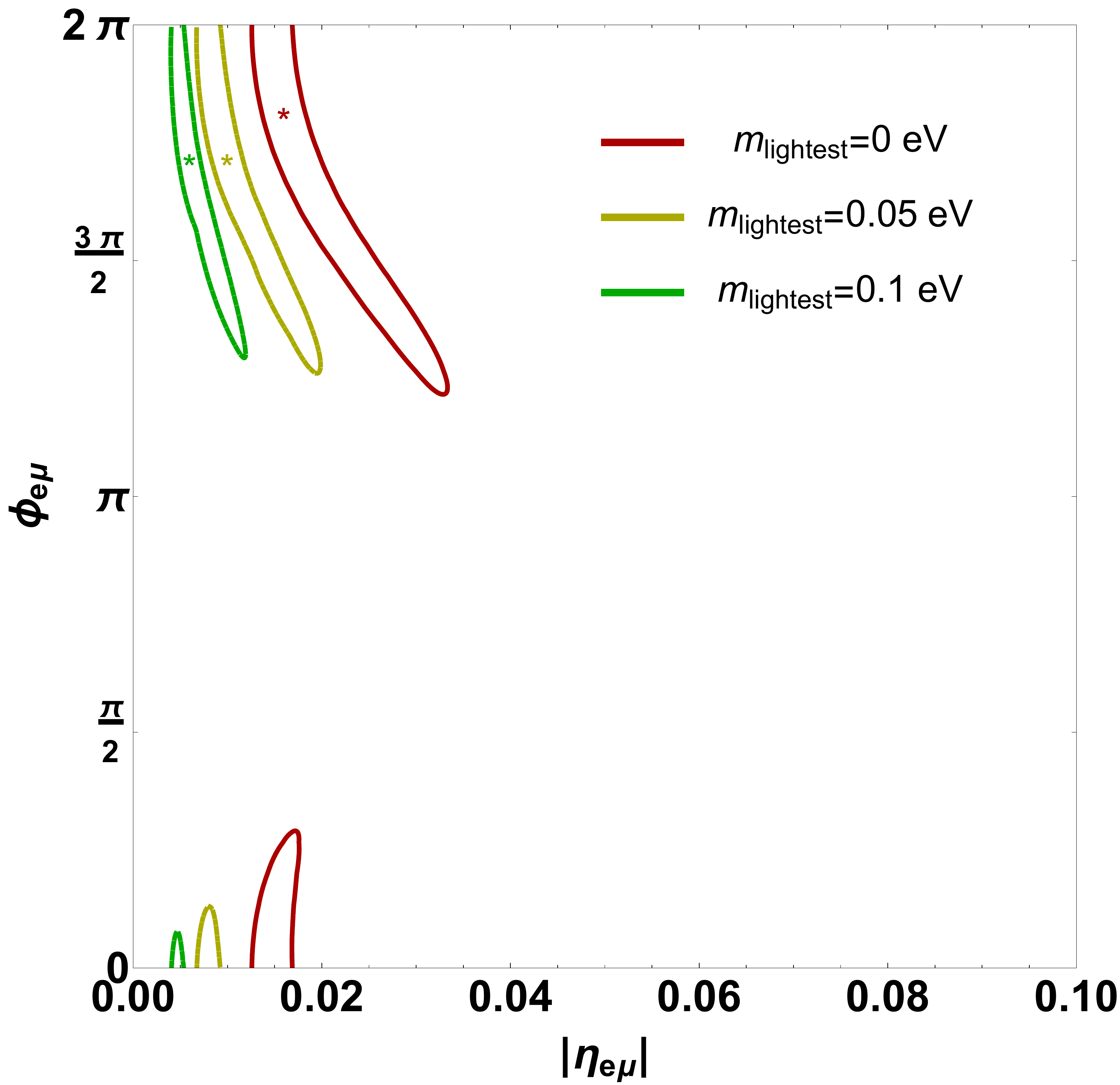}
\includegraphics[width=0.32\textwidth]{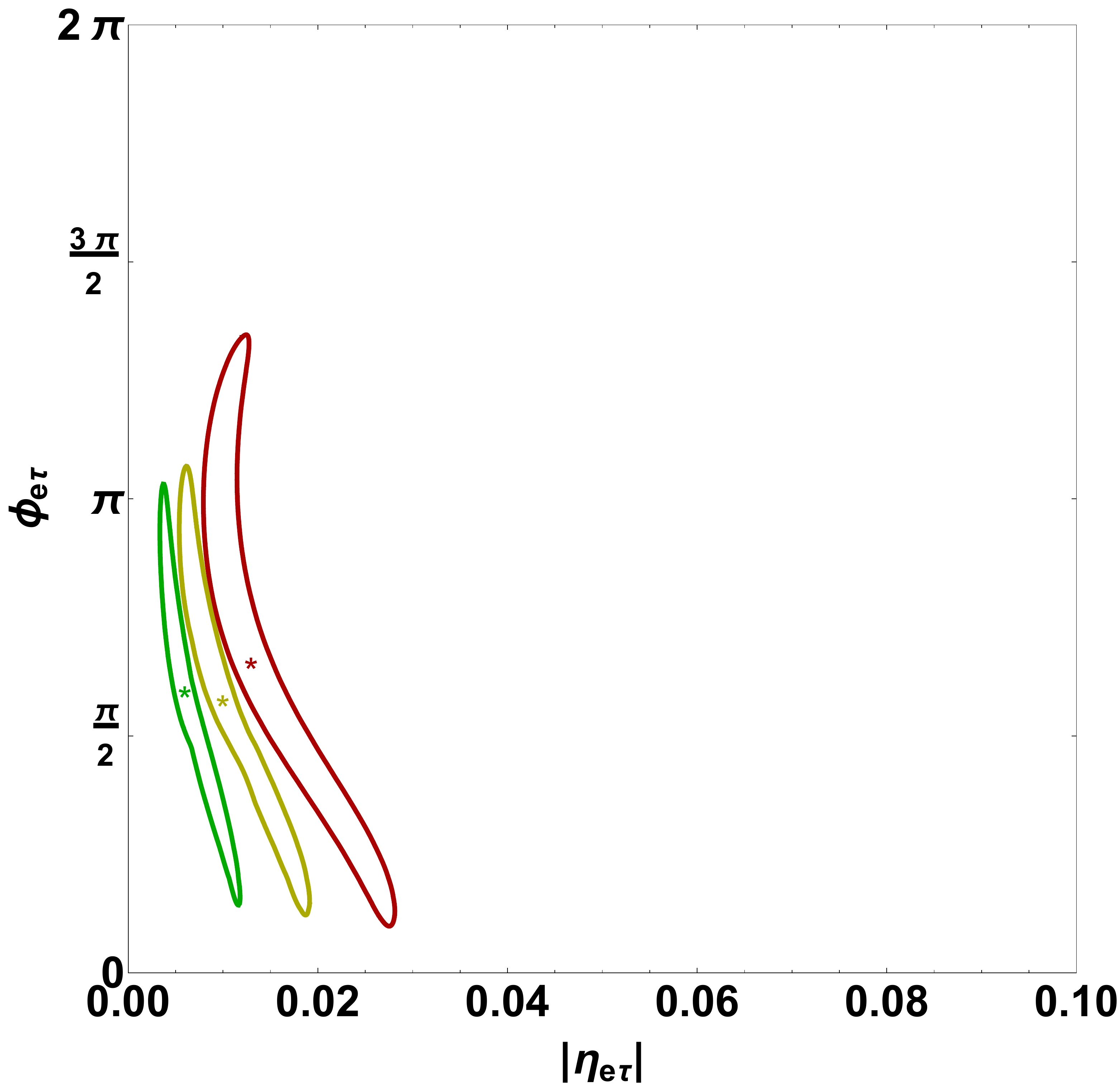}
\includegraphics[width=0.32\textwidth]{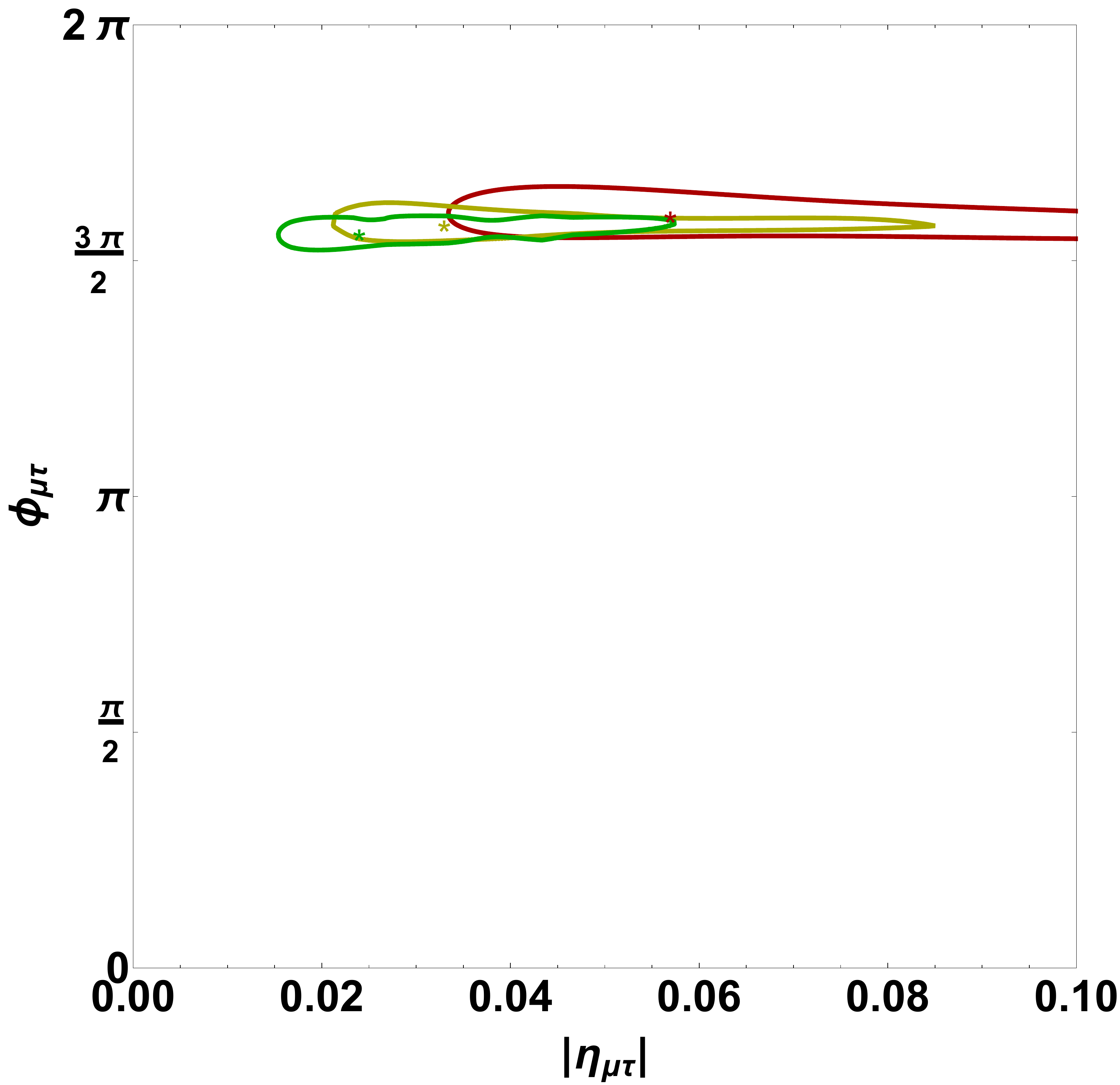}
\caption{68\% contours in DUNE in the $(|\eta_{ij}|-\phi_{ij})$-planes when data are generated using scalar NSI best fits in the IO case considering one NSI parameter at a time. The red curve is obtained with $m_{\rm lightest}=0$, the blue one with $m_{\rm lightest}=0.05$ eV and the green one with $m_{\rm lightest}=0.1$ eV. } 
\label{fig:scalar_mass}
\end{center}
\end{figure}

We have mentioned that one of the most interesting features of the scalar NSI model is that the oscillation probabilities in this case depend on the absolute neutrino mass scale. A non-zero $m_{\rm lightest}$ can slightly change the NOvA-T2K best fits. In particular, increasing the neutrino mass scale, the best fit complex NSI phases are almost the same, while the magnitudes of the $\eta_{ij}$ parameters decrease. The most significant reduction of the best fits can be observed in the IO case, in which, as already pointed out, the phenomenology would allow us to distinguish the scalar NSI parameters from the SM oscillations. In Fig.~\ref{fig:scalar_mass} we show the 68\% contours using the best fits for $m_{\rm lightest}=0,0.05,0.1$ eV. It is clear that the contours shapes are not drastically altered by the decrease of the scalar NSI parameters magnitudes. However, when the lightest neutrino is not massless, DUNE is expected to be able to set upper limits on $\eta_{\mu\tau}$: $|\eta_{\mu\tau}|<0.08 \, (0.06)$ at 68\% CL when $m_{\rm lightest}=0.05 \, (0.1)$ eV.
Combining information from very different densities such as the crust of the Earth and the center of the Sun could allow for a detection of the absolute neutrino mass scale.

\subsection{Sterile neutrinos}

\begin{figure}
\begin{center}
\includegraphics[width=0.4\textwidth]{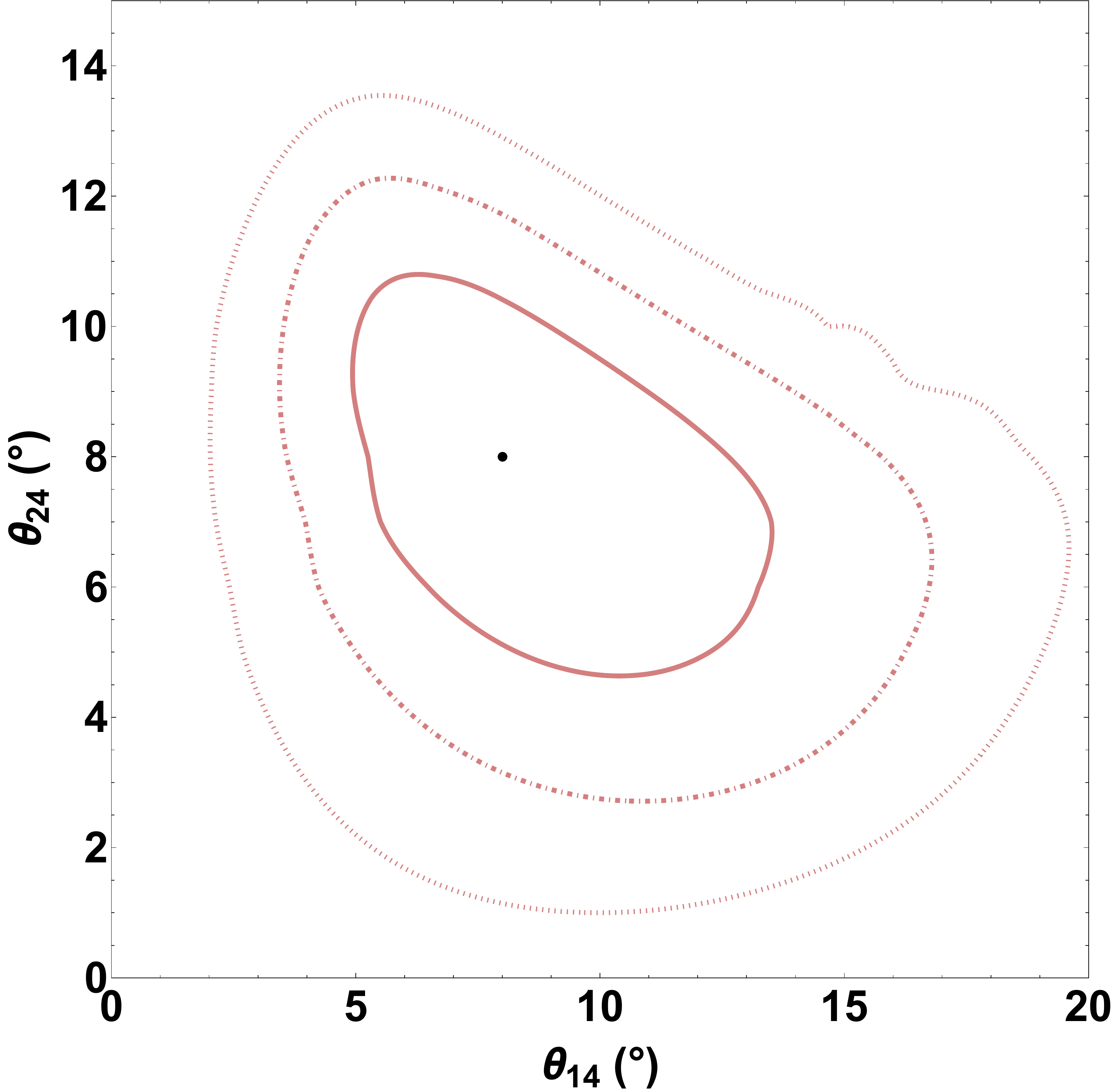}
\includegraphics[width=0.4\textwidth]{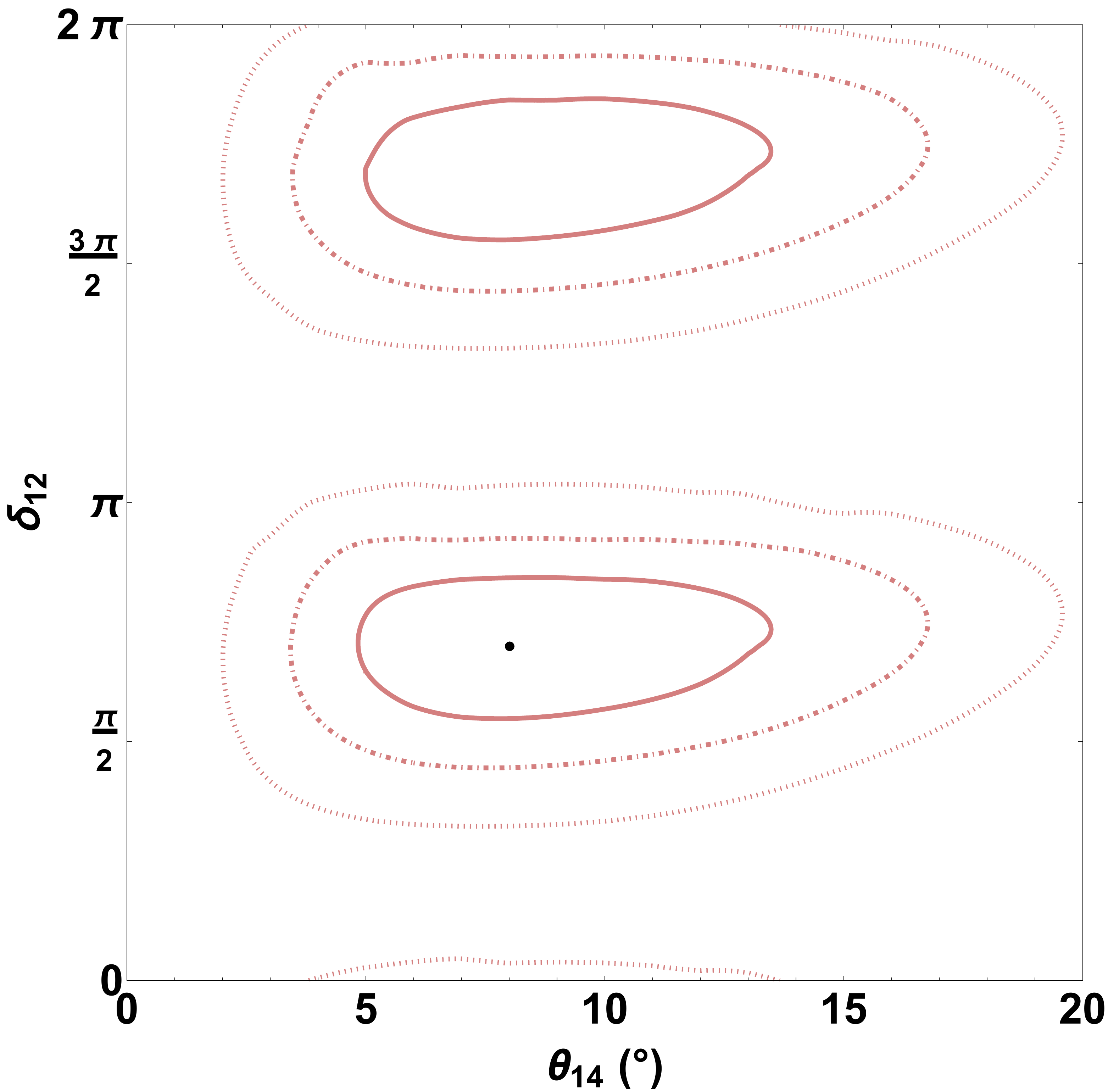} 
\includegraphics[width=0.4\textwidth]{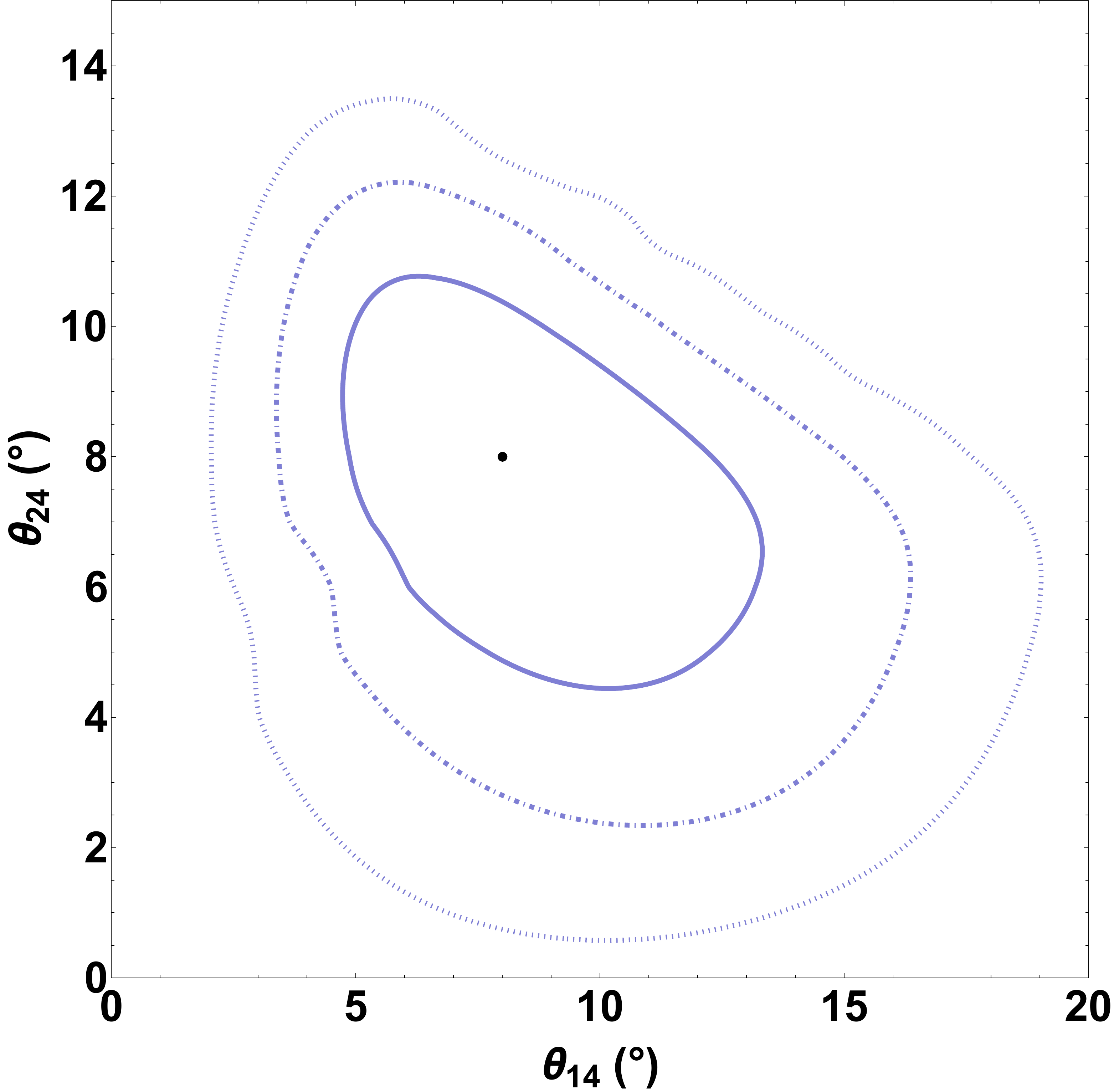}
\includegraphics[width=0.4\textwidth]{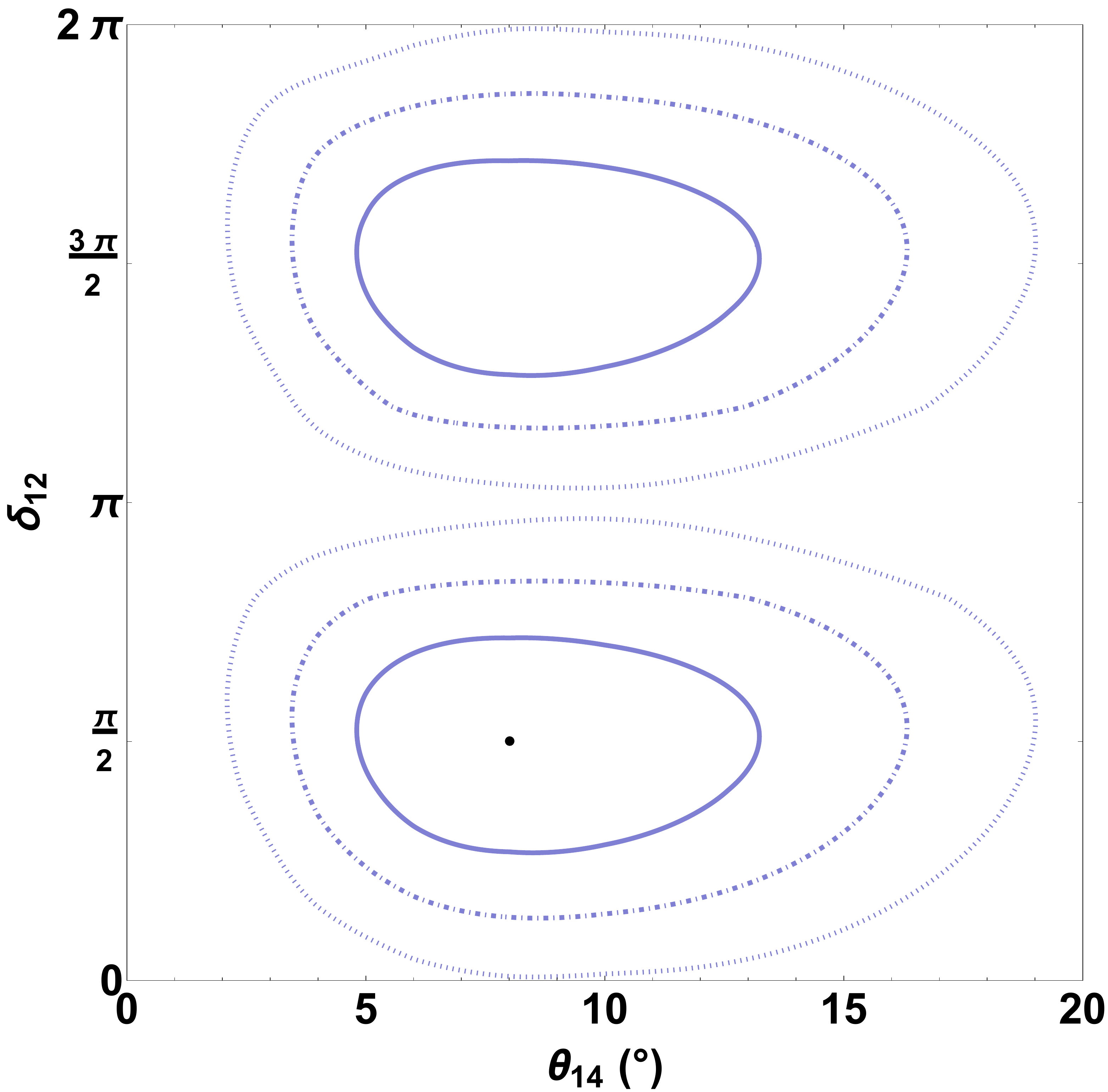}
\caption{68\% (solid lines), 95\% (dashed lines) and 99\% (dotted lines) contours in DUNE in the $(\theta_{14}-\theta_{24})$ and $(\theta_{14}-\delta_{12})$ planes when data are generated using 3+1 model best fits. The three top (bottom) plots with red (blue) contours have been obtained using NO (IO) best fits.}
\label{fig:3+1}
\end{center}
\end{figure}

In Fig.~\ref{fig:3+1} we show the performances of the DUNE Far Detector in constraining the 3+1 parameters if the best fits are the ones reported in table \ref{tab:nova t2k sterile} from NOvA and T2K data. In the analysis, we marginalize over the undisplayed parameters, using the upper bound $\theta_{i4}<25^\circ$ on the non-standard mixing angles\footnote{
This range of $\theta_{i4}$ is motivated by the existing constraints on sterile neutrino mixing angles ranging from $\theta_{14}\lesssim1^\circ$ to $\theta_{34}\lesssim23^\circ$ \cite{Dentler:2018sju}, although we note that these constraints depend considerably on the experiments included and there are numerous anomalies, some of which seem to point to mixing angles in excess of those constraints.}.
In both IO and NO, in the $(\theta_{24}-\theta_{14})$ planes, it is evident that there is a modest amount of anti-correlation between the two mixing angles, which is not surprising given that they both play a similar role.
Related, DUNE is expected to have a similar sensitivity to both the mixing angles for similar true values.
Since we are taking both sterile mixing angles to be $8^\circ$, the 68\% allowed ranges are [5-12]$^\circ$ for $\theta_{14}$ and [4-10]$^\circ$ for $\theta_{24}$.
When we investigate the DUNE capabilities in measuring the new complex phases, we observe in Fig.~\ref{fig:3+1} that, if $\delta_{12}\sim 0.5\pi$ as suggested by NOvA and T2K data, then there exists two local $\Delta\chi^2$ minima, one around the best fit and one around $\delta_{12}\sim 1.5\pi$ for both the IO and NO case.
Thus DUNE would not be able to determine the sign of $\sin\delta_{12}$, but would be able to determine that a new source of CP violation exists.
We checked that, on the other hand, given the best fits in table \ref{tab:nova t2k sterile}, the DUNE experiment will not be sensitive to $\delta_{13}$ and $\delta_{23}$.

Other studies have also investigated sterile neutrinos at DUNE \cite{Donini:2007yf,Dighe:2007uf,Berryman:2015nua,Krasnov:2019kdc,DUNE:2020fgq}.
Most studies assume the SM or different benchmarks from our study, so a direct comparison is not possible, but \cite{Berryman:2015nua} does consider a similar benchmark point in terms of mixing angles and finds very similar sensitivities.

\section{Differentiating the Models}
\label{sec:diff}
\subsection{Method}
We now get to the primary result of this paper, where we explore the ability for DUNE to correctly differentiate among different benchmark scenarios.
Our methodology is to assume that one of the benchmark new physics points represents reality and then attempt to reconstruct it in a different new physics scenario and determine the test statistic for the model comparison test to show the significance at which the data would prefer the correct answer over an incorrect answer.
We compare each of the three classes of models against each other for the most comprehensive such study to date.

The minimum $\Delta\chi^2$'s are shown in tables \ref{tab:vector}, \ref{tab:scalar}, and \ref{tab:3+1}.
The true scenarios are listed along the left column and then the statistical significance in which the SM can be disfavored is shown in the second column labeled SM.
The remaining columns show at what level a different scenario can be disfavored relative to the true scenario which is denoted with a slash as the $\Delta\chi^2$'s in those cases are exact zero.

For the $\chi^2$ computations we marginalize over all the standard oscillation parameters.
We also marginalize over one NSI parameter, magnitude up to 0.5 and phase, for the vector and scalar NSI columns.
For the sterile columns, labeled 3+1, we marginalize over $\Delta m^2_{41}$ in the range $[10^{-5}-10]$ eV$^2$, and the mixing angles in the range $[0-30]^\circ$.
Naturally the $\Delta\chi^2$ in the new physics scenarios will be less than that of the SM as the new physics scenarios include all the degrees of freedom of the SM plus several new ones.

\subsection{Discussion}
\subsubsection{True Vector NSI}
In the case in which data are generated using the best fit points for vector NSI, for both the NO and IO, the data could not be well fit with the three neutrino standard probabilities ($\chi^2>60$). Moreover, all the models obtained generating data with vector NSI ($\varepsilon$) best fits from the NOvA and T2K data can be distinguished at least $7\sigma$ from the scalar NSI models.
When we try to fit the vector NSI data with other vector NSI models or 3+1 models, the only interesting case (namely the only case in which DUNE would struggle in differentiating the models) is when data are generated using $\varepsilon_{e\mu}$ best fits in the IO case. In this framework, the minima of the $\chi^2$ are 10, 13 and 3 when we fit the data with the $\varepsilon_{e\tau}$, $\varepsilon_{\mu\tau}$ and 3+1 models respectively.
These results are generally compatible with \cite{deGouvea:2015ndi} which fit benchmark NSI points in a sterile neutrino framework.

\subsubsection{True Scalar NSI}
Let us now focus on the case in which data are generated using the best fits related to the scalar NSI model.
It is clear that, for NO, the minimum $\chi^2$ values are very small even in the standard model scenario (when all the new physics parameters are fixed to zero). This is because the fit at the NOvA and T2K data, if performed adding the scalar NSI in NO, gives very small best fit values for the $\eta$ couplings.
For this reason, we expect that DUNE would not be able in this case to distinguish at any confidence level these scenarios with the standard three neutrino framework.
If we fit data with models in which other $\eta$ parameters are used in the fit, we obtain even smaller minima for the $\chi^2$, as expected.
The same happens in the case in which we fit the scalar NSI data with sterile neutrino models.
On the other hand, when we try to fit them with vector NSI models, apart from the $\eta_{e \mu}$ case, the best fit point is always found when $\varepsilon_{ij}=0$.
The best fit for the scalar NSI parameters in the IO case give even more interesting results. 
In this case, the three neutrino framework is completely excluded ($\chi^2>30$ in all three cases). However, we can find minima of the $\chi^2$ much smaller than the standard model ones when we fit data with other scalar NSI models. In particular, if data are generated using the best fits for $\eta_{e\tau}$, DUNE would have very limited discrimination capabilities when we switch on only $\eta_{e \mu}$ or $\eta_{\mu\tau}$ at more than $1.5\sigma$.
On the other hand, when data are generated with $\eta_{e \mu}$ or $\eta_{\mu\tau}$ best fits, the $\chi^2$ related to the fits with the other $\eta$ models are in the range $4.6-6.3$. Thus, DUNE would also have some ability to distinguish different scalar NSI scenarios in the IO if the true values for the $\eta$-s are the ones compatible with the NOvA-T2K tension.
If we try to fit scalar NSI data with single parameter vector NSI models, DUNE would differentiate them at a good confidence level. Only in the $\eta_{\mu\tau}$ case, the vector NSI models could be rejected at less than $5\sigma$. Finally, the IO scalar NSI models generated with $\eta_{e\tau}$ and $\eta_{\mu\tau}$ best fits cannot be distinguished at more than $3.2\sigma$ from the sterile neutrino models. On the other hand, we reach almost $5\sigma$ in the case of the $\eta_{e\mu}$ model.

All these results have been obtained considering only one parameter at-a-time for each model, namely switching on only one off-diagonal scalar or vector parameter for each fit model. We checked that, if we vary all the off-diagonal parameters together, varying their phases in the entire allowed range, the results are not changing drastically. This means that, for instance, if we fit data generated with one of the $\eta$'s best fits using the vector NSI model in which we scan over the whole $(\varepsilon_{e\mu},\varepsilon_{e\tau},\varepsilon_{\mu\tau})$ parameter space, the minimum of the $\chi^2$ is very close to the smaller minima of the $\chi^2$-s obtained using only one parameter at-a-time. This suggests that there are no strong correlations between parameters that could mimic data generated with a different model.

\begin{table}
\centering
\begin{tabular}{l|c|c|c|c|c|c|c|c|}
\cline{2-9} \multicolumn{1}{r|}{$\Delta\chi^2$}
                                                 & \multicolumn{1}{l|}{SM} & \multicolumn{1}{l|}{$\eta_{e\mu}$} & \multicolumn{1}{l|}{$\eta_{e\tau}$} & \multicolumn{1}{l|}{$\eta_{\mu\tau}$} & \multicolumn{1}{l|}{$\varepsilon_{e\mu}$} & \multicolumn{1}{l|}{$\varepsilon_{e\tau}$} & \multicolumn{1}{l|}{$\varepsilon_{\mu\tau}$} & 3+1 \\ \hline
\multicolumn{1}{|l|}{$\varepsilon_{e\mu}$ NO}    & 200                     & 140                                & 140                                 & 170                                   & /                                         & 180                                        & 160                                          & 80  \\ \hline
\multicolumn{1}{|l|}{$\varepsilon_{e\tau}$ NO}   & 60                      & 48                                 & 50                                  & 45                                    & 50                                        & /                                          & 50                                           & 40  \\ \hline
\multicolumn{1}{|l|}{$\varepsilon_{\mu\tau}$ NO} & 200                     & 180                                & 170                                 & 180                                   & 160                                       & 180                                        & /                                            & 80  \\ \hline
\multicolumn{1}{|l|}{$\varepsilon_{e\mu}$ IO}    & 170                     & 80                                 & 75                                  & 90                                    & /                                         & 10                                         & 13                                           & 3   \\ \hline
\multicolumn{1}{|l|}{$\varepsilon_{e\tau}$ IO}   & 70                      & 50                                 & 50                                  & 45                                    & 45                                        & /                                          & 60                                           & 20  \\ \hline
\multicolumn{1}{|l|}{$\varepsilon_{\mu\tau}$ IO} & 500                     & 400                                & 400                                 & 400                                   & 300                                       & 350                                        & /                                           & 160 \\ \hline
\end{tabular}
\caption{\label{tab:vector}
$\Delta \chi^2$ obtained fitting to each case all the models taken into account using as true values the best fits of the vector NSI case.}
\end{table}

\begin{table}
\centering
\begin{tabular}{l|c|c|c|c|c|c|c|c|}
\cline{2-9} \multicolumn{1}{r|}{$\Delta\chi^2$}
                                          & \multicolumn{1}{l|}{SM} & \multicolumn{1}{l|}{$\eta_{e\mu}$} & \multicolumn{1}{l|}{$\eta_{e\tau}$} & \multicolumn{1}{l|}{$\eta_{\mu\tau}$} & \multicolumn{1}{l|}{$\varepsilon_{e\mu}$} & \multicolumn{1}{l|}{$\varepsilon_{e\tau}$} & \multicolumn{1}{l|}{$\varepsilon_{\mu\tau}$} & 3+1  \\ \hline
\multicolumn{1}{|l|}{$\eta_{e\mu}$ NO}    & 0.14                    & /                                  & 0.005                               & 0.088                                 & 0.071                                     & 0.033                                      & 0.055                                        & 0.02 \\ \hline
\multicolumn{1}{|l|}{$\eta_{e\tau}$ NO}   & 0.08                    & 0.003                              & /                                   & 0.041                                 & SM                                        & SM                                         & SM                                           & 0.01 \\ \hline
\multicolumn{1}{|l|}{$\eta_{\mu\tau}$ NO} & 0.60                    & 0.48                               & 0.48                                & /                                     & SM                                        & SM                                         & SM                                           & 0.02 \\ \hline
\multicolumn{1}{|l|}{$\eta_{e\mu}$ IO}    & 100                     & /                                  & 4.7                                 & 6.3                                   & 80                                        & 70                                         & 90                                           & 21   \\ \hline
\multicolumn{1}{|l|}{$\eta_{e\tau}$ IO}   & 60                      & 1.0                                & /                                   & 1.5                                   & 44                                        & 38                                         & 50                                           & 11   \\ \hline
\multicolumn{1}{|l|}{$\eta_{\mu\tau}$ IO} & 30                      & 4.6                                & 4.8                                 & /                                     & 23                                        & 20                                         & 29                                           & 12   \\ \hline
\end{tabular}
\caption{ \label{tab:scalar}Same as table \ref{tab:vector} but with true values set to the best fits of scalar NSI.}
\end{table}

\begin{table}[]
\centering \begin{tabular}{l|c|c|c|c|c|c|c|}
\cline{2-8} \multicolumn{1}{r|}{$\Delta\chi^2$}
                             & \multicolumn{1}{l|}{SM} & \multicolumn{1}{l|}{$\eta_{e\mu}$} & \multicolumn{1}{l|}{$\eta_{e\tau}$} & \multicolumn{1}{l|}{$\eta_{\mu\tau}$} & \multicolumn{1}{l|}{$\varepsilon_{e\mu}$} & \multicolumn{1}{l|}{$\varepsilon_{e\tau}$} & \multicolumn{1}{l|}{$\varepsilon_{\mu\tau}$} \\ \hline
\multicolumn{1}{|l|}{3+1 NO} & 20                      & 8.2                                 & 7.9                                  & 6.7                                    & 5.2                                         & 6.6                                          & 10                                           \\ \hline
\multicolumn{1}{|l|}{3+1 IO} & 20                      & 13                                 & SM                                  & 18                                    & 7.4                                         & 6.2                                          & 9.5                                            \\ \hline
\end{tabular}
\caption{ \label{tab:3+1}Same as table \ref{tab:vector} but with true values set to the best fits of 3+1 sterile neutrino model.}
\end{table}

\subsubsection{True Sterile Neutrino}
Finally, when the data are obtained in the 3+1 framework (see table \ref{tab:3+1}), with the best fits shown in table \ref{tab:nova t2k sterile}, we observe that the SM solution is excluded with $\Delta \chi^2=20$, namely at 4.5$\sigma$.
On the other hand, when the fit is performed with the NSI models, it is clear that in both NO and IO, the $\Delta \chi^2$ is reduced to 5.2 in the vector case. The vector model which can be distinguished more easily from the 3+1 is the one where we turn on $\varepsilon_{\mu\tau}$ in NO ($\Delta \chi^2=10$). This is because $\varepsilon_{\mu\tau}$ in DUNE, as already mentioned, is the one that mostly modifies the disappearance probability. In the scalar case, on the other hand, the $\Delta\chi^2$ are not drastically reduced in respect to the Standard Model case if the mass ordering is inverted. Indeed, if we perform the fit allowing $\eta_{e\mu}$ to vary we reach $\Delta\chi^2=13$, while in the opposite case, when we change the value of $\eta_{e\tau}$, the best fit is achieved when the scalar NSI parameter vanishes, namely in the Standard Model case. This suggests us that in inverted ordering, if the sterile neutrino mixing angles are as small as $8^\circ$ and $\theta_{34}$ is fixed to zero, the phenomenology of 3+1 and scalar NSI models are moving in the opposite direction. On the contrary, in the scalar NSI NO case, the $\Delta\chi^2$ are as low as the vector NSI ones. We checked also in this case that a full marginalization over the NSI parameter spaces do not alter significantly the $\Delta \chi^2$-s.

\section{Conclusions}
DUNE will be the state-of-the-art neutrino oscillation experiment making high precision long-baseline measurements with a modest matter effect, and with a fairly broad energy spectrum.
As has been discussed in the literature, this will enable crucial tests of many new physics scenarios that affect oscillations.
Some common new physics scenarios are light sterile neutrinos, vector non-standard neutrino interactions (NSI), and scalar NSI.
To realistically understand DUNE's sensitivity, we considered a number of benchmark scenarios motivated by existing long-baseline accelerator neutrino oscillation data from NOvA and T2K.
We confirmed in section \ref{sec:results} that, in fact, DUNE does have very good sensitivity to a majority of relevant new physics cases at the edge of current limits.
For example, we showed, for the first time, that DUNE has sensitivity to off-diagonal scalar NSI.

While identifying any kind of additional new physics in the neutrino sector is extremely exciting, in order to confirm it, its nature must be understood.
In order to do that, one must check if multiple different new physics scenarios can describe the same data, or if the data can differentiate among these scenarios.
Our main results, that is the ability of DUNE to distinguish among different new physics frameworks, are reported in tables \ref{tab:vector}, \ref{tab:scalar} and \ref{tab:3+1}.
We find that, in general, DUNE can differentiate among vector NSI, scalar NSI, and sterile neutrinos at modest to excellent significance for these interesting benchmark points.
It is not always possible, however, to easily identify which of the new physics parameters are responsible for the new physics as there are some degeneracies among e.g.~$\varepsilon_{e\mu}$ and $\varepsilon_{e\tau}$ and among all the off-diagonal scalar NSI parameters.
Nonetheless, additional measurements from HK, IceCube, KM3NeT, and other experiments would even further resolve the flavor structure of the new physics due to the presence of more baselines and energies.

\acknowledgments
We thank Julia Gehrlein for helpful comments.
PBD acknowledges support from the US Department of Energy under Grant Contract DE-SC0012704.

\appendix

\section{Approximate Expressions for Neutrino Oscillation Probabilities}
\label{sec:approximate}
In this appendix we report the relevant transition probabilities for both vector and scalar NSI, for the transition channels used in our numerical computation.

The analytical behavior of the probabilities are put in a better readable form expanding up to the second order in some small parameters, namely $\alpha=\Delta m^2_{21}/\Delta m^2_{31}$ and the deviation of the mixing angles from the tri-bimaximal mixing $r$, $s$, $t$, defined as \cite{King:2007pr,Pakvasa:2007zj}:
\begin{equation}
\label{expansion}
s_{13} = \frac{r}{\sqrt{2}}\,, \qquad s_{12} = \frac{1}{\sqrt{3}}(1+s)\,,\qquad s_{23} =  \frac{1}{\sqrt{2}}(1+t)\,.
\end{equation}
From the current global fits on the oscillation data, all previous parameters  $r$, $s$, and $t$ are of ${\cal{O}}(0.1)$ or smaller \cite{Capozzi:2021fjo}. Since at the DUNE baseline the matter effects are expected to be of $\mathcal{O}(10\%)$, we can further expand in $V = a L/( 4E \Delta_{31}) = a / \Delta m^2_{31}$. 

\subsection{Vector NSI}

At first order in $\mathcal{O}(\varepsilon_{\alpha\beta})$ and with the identification $\varepsilon_{\alpha\beta}\to \varepsilon_{\alpha\beta} \,e^{i\phi_{\alpha\beta}}$, the NSI contribution to the electron appearance probability is:
\begin{align}
\nonumber P(\nu_\mu\to\nu_e)^{NSI}={}&\frac{4}{3}V\alpha\Delta_{31}[-\Delta_{31}(\varepsilon_{e\tau}\cos{\phi_{e\tau}}-\varepsilon_{e\mu}\cos{\phi_{e\mu}})\\
&-\sin\Delta_{31}\varepsilon_{e\tau}(\cos\Delta_{31}\cos\phi_{e\tau}-\sin\Delta_{31}\sin\phi_{e\tau})\\
\nonumber
& -\sin\Delta_{31}\varepsilon_{e\mu}(\cos\Delta_{31}\cos\phi_{e\mu}-\sin\Delta_{31}\sin\phi_{e\mu})]\\
\nonumber&+2Vr\sin\Delta_{31}[\Delta_{31}\cos\Delta_{31}(\varepsilon_{e\mu}\cos(\delta-\phi_{e\mu})-\varepsilon_{e\tau}\cos(\delta-\phi_{e\tau}))
\\
\nonumber& +\Delta_{31}\sin\Delta_{31}(\varepsilon_{e\tau}\sin(\delta-\phi_{e\tau})-\varepsilon_{e\mu}\sin(\delta-\phi_{e\mu}))\\
\nonumber
& +\sin\Delta_{31}(\varepsilon_{e\mu}\cos(\delta-\phi_{e\mu})+\varepsilon_{e\tau}\cos(\delta-\phi_{e\tau}))]\,.
\end{align}
We observe that, at the perturbative order taken into account, the probability depends on $\varepsilon_{e\mu}$ and $\varepsilon_{e\tau}$ only. In particular, if $\phi_{e\tau}=\phi_{e\mu}$, the only relevant combinations are $\varepsilon_{e\mu}+\varepsilon_{e\tau}$ and $\varepsilon_{e\mu}-\varepsilon_{e\tau}$. The same happens if $\phi_{e\mu}=\pi+\phi_{e\tau}$.
For the muon disappearance channel, the NSI contribution is:
\begin{align}
\nonumber P(\nu_\mu\to\nu_\mu)^{NSI}={}&V\bigg\{-8\Delta_{31}\varepsilon_{\mu\tau}\cos\phi_{\mu\tau}\cos\Delta_{31}\sin\Delta_{31}+\frac{4}{3} \alpha\Delta_{31}^2[\varepsilon_{e\tau}\cos\phi_{e\tau}\cos^2\Delta_{31} \\
\nonumber & -\varepsilon_{e\mu}\cos\phi_{e\mu}\cos^2\Delta_{31}+4\varepsilon_{\mu\tau}\cos\phi_{\mu\tau}(1-2\sin^2\Delta_{31})]\\
&+4t\Delta_{31}\varepsilon_{\tau\tau}\cos\Delta_{31}\sin\Delta_{31}\label{eq:approx vector dis}\\
\nonumber&+4r\Delta_{31}\varepsilon_{e\mu}\cos(\delta-\phi_{e\mu})\cos\Delta_{31}\sin\Delta_{31}-4a\varepsilon_{\tau\tau}\sin^2\Delta_{31}\\
\nonumber&+\frac{4}{3}\alpha\Delta_{31}[\varepsilon_{e\tau}\cos\phi_{e\tau}\cos\Delta_{31}\sin\Delta_{31}-\varepsilon_{e\mu}\cos\phi_{e\mu}\cos\Delta_{31}\sin\Delta_{31}]\bigg\}\,.
\end{align}
Compared to the previous case, all NSI parameters but $\varepsilon_{ee}$ contribute.
However, the $\varepsilon_{\mu\tau}$ term is not suppressed by any of the small parameters $r$, $s$ and $t$. Thus, this probability is expected to be more sensitive to $\varepsilon_{\mu\tau}$ and its phase.

In light of such formulae, we can try to understand the results in table \ref{tab:vector}. It is clear that, due to the strong dependence of the disappearance probability on $\varepsilon_{\mu\tau}$, if we generate data with this parameter's true value different from zero, we expect very large $\chi^2$. Indeed, it is very difficult to find values for the other parameters which fit well the disappearance data. Moreover, even if the appearance probability has the same dependence on $\varepsilon_{e\mu}$ and $\varepsilon_{e\tau}$, the disappearance probability contains one more term in $\varepsilon_{e\mu}$. For this reason, it is easy to fit data generated with $\varepsilon_{e\tau}\neq0$ with a theory that contains $\varepsilon_{e\mu}$, but the opposite is more complicated. This is clear from table \ref{tab:vector}, since $\chi^2$ in the $\varepsilon_{e\mu}$ rows are bigger in respect to the $\varepsilon_{e\tau}$ ones.

\subsection{Scalar NSI}
In the scalar NSI case, we neglect for simplicity the solar mass splitting, which would elongate in a considerable way the probabilities, making them difficult to understand. For the electron appearance, and with the identification $\eta_{\alpha\beta}\to \eta_{\alpha\beta} \,e^{i\phi_{\alpha\beta}}$,
we have:
\begin{align}
\nonumber P(\nu_\mu\to\nu_e)^{NSI}={}&[\eta_{e\mu}^2+\eta_{e\tau}^2+2r\eta_{e\mu}\cos(\delta+\phi_{12})+2r\eta_{e\tau}\cos(\delta+\phi_{13})]\sin^2\Delta_{13}+\\
\nonumber
&+2\eta_{e\mu}\eta_{e\tau}\cos(\phi_{12}-\phi_{13})\sin^2\Delta_{13}+\\
& -2V[\eta_{e\mu}^2+\eta_{e\tau}^2+2r\eta_{e\mu}\cos(\delta+\phi_{12})+2r\eta_{e\tau}\cos(\delta+\phi_{13})]\times \\ \nonumber &\times[\Delta_{13}\cos\Delta_{13}\sin\Delta_{13}-\sin^2\Delta_{13}]+\\ \nonumber
&-2V\eta_{e\mu}\eta_{e\tau}\cos(\phi_{12}-\phi_{13})\sin^2\Delta_{13}[\Delta_{13}\cos\Delta_{13}\sin\Delta_{13}-\sin^2\Delta_{13}]\,.
\end{align}
It can be noticed that the probability depends only on $\eta_{e\mu}$ and $\eta_{e\tau}$. Both parameters appear in second order terms, proportional to $\eta^2$ or to $r\eta$.

The $\nu_\mu$ disappearance probability, on the other hand, can be written as:

\begin{align}
\nonumber P(\nu_\mu\to\nu_\mu)^{NSI}={}&-\Delta_{13}(\eta_{\mu\mu}+\eta_{\tau\tau}+2\eta_{\mu\tau}\cos\phi_{\mu\tau})\sin2\Delta_{13}+\\ \nonumber
&-\frac{1}{2}\Delta_{13}(\eta_{e\mu}^2+\eta_{e\tau}^2+4\eta_{\mu\tau}^2\sin\phi_{\mu\tau}^2)\sin2\Delta_{13}+ \\ \nonumber
&-4\Delta_{13}^2\eta_{\mu\tau}^2\cos\phi_{\mu\tau}^2\sin2\Delta_{13} + \\ 
& +\frac{1}{2}(\eta_{\mu\mu}^2+\eta_{\tau\tau}^2)[1-(1-\Delta_{13}^2)\cos2\Delta_{13}-\Delta_{13}\sin2\Delta_{13}]+ \\ \nonumber 
& -2\Delta_{13}[r\eta_{e\mu}\cos(\delta-\phi_{e\mu})+r\eta_{e\tau}\cos(\delta+\phi_{e\tau})]\sin2\Delta_{13}+ \\ \nonumber
& +2t(\eta_{\tau\tau}-\eta_{\mu\mu})[-1+\cos2\Delta_{13}+\Delta_{13}\sin2\Delta_{13}]+ \\ \nonumber
& -3\Delta_{13}\eta_{e\mu}\eta_{e\tau}\cos(\phi_{e\mu}-\phi_{e\tau})\sin2\Delta_{13}+ \\ \nonumber
& -2\Delta_{13}\eta_{\mu\mu}\eta_{\mu\tau}\cos\phi_{\mu\tau}(2\Delta_{13}\cos2\Delta_{13}+\sin2\Delta_{13})+\\ \nonumber
& +\eta_{\mu\mu}\eta_{\tau\tau}[-1+(1-2\Delta_{13}^2)\cos2\Delta_{13}+\Delta_{13}\sin2\Delta_{13}]+ \\ \nonumber
& -2\Delta_{13}\eta_{\mu\tau}\eta_{\tau\tau}\cos\phi_{\mu\tau}(2\Delta_{13}\cos2\Delta_{13}+\sin2\Delta_{13})\,.
\label{scalardis}
\end{align}
In this case, all the scalar NSI parameters  but $\eta_{ee}$ appear. $\eta_{\mu\mu}$, $\eta_{\tau\tau}$ and $\eta_{\mu\tau}$ modify the probability at the first order in perturbation theory; we expect them to affect the $\Delta m_{13}^2$ measurement, since they are strongly correlated to $\sin2\Delta_{13}$. Apart from the atmospheric mass splitting, the scalar NSI parameters appear coupled also to $r$ and $t$.

Notice that, in our analytical approach, the probabilities are symmetric under the $\eta_{e\tau}$-$\eta_{e\mu}$ exchange. However, from table \ref{tab:scalar}, it is clear that this symmetry cannot be exact. Indeed, we checked that, when we include in the expansion also the solar mass splitting, the symmetry is broken, as expected. From table \ref{tab:scalar} we can also interestingly see that, even though the disappearance probability strongly depends on $\eta_{\mu\tau}$, since the best fit for $\phi_{\mu\tau}$ is similar to $\pi/2$, the leading term in the probability is suppressed. Thus, the fits to the data generated with $\eta_{\mu\tau}$ are better than the ones where data are generated using the other scalar NSI parameters. 

\subsection{Sterile neutrinos}

In this case, expanding in the small $s_{i4}=\sin\theta_{i4}$ quantities up to the second order and averaging out the fast oscillations driven by $\Delta m_{41}^2$, we obtain
\begin{align}
P_{\nu_\mu\to\nu_e}^{3+1}=0\,,
\end{align}
and
\begin{align}
P_{\nu_\mu\to\nu_\mu}^{3+1}=&-2s_{24}^2\cos^2\Delta_{31}-2V' \Delta_{31}s_{24}s_{34}\cos\delta_{23}\sin2\Delta_{31} \,,
\end{align}
where $V'$ is the NC matter potential, which appears in the probabilities due to the presence of a fourth sterile state.
Notice that given our PMNS parameterization, in the standard three-flavor oscillation probabilities we substitute the usual complex phase $\delta$ with the combination $\delta_{13}-\delta_{12}-\delta_{23}$. It is clear that, at our expansion order, only $\theta_{24}$ modifies the disappearance probabilities, multiplied however by the small (at the atmospheric peak) $\cos^2\Delta_{31}$. This explains why in table \ref{tab:3+1} the Standard Model is excluded only at 4.5$\sigma$ even if the true values for $\theta_{14}$ and $\theta_{24}$ are relatively large (8$^\circ$). Moreover, it is clear that, since $\delta_{12}\sim1.9\pi$ or $0$ in the two best fits, the DUNE sensitivity to the CP violating phase translates directly to the sensitivity to $\delta_{12}$, whose best fits are maximal. Another interesting feature of the disappearance probability is that it is always smaller for $\theta_{24}\neq0$ . This explains why, for instance, a scalar NSI model in NO fits better the 3+1 data than the same model in IO. We can indeed observe that, the third line in eq.~\ref{scalardis} reduces the disappearance probability in NO, while it enhances the probability in IO.

\section{NSI Fits To NOvA, T2K, and Reactor Data}
\label{fitAppendix}
In this section we revisit the fits to NOvA and T2K data presented in \cite{Denton:2020uda} for vector NSI and provide the first fits to NOvA and T2K data in the presence of scalar NSI.
The details of the fit are described there, but we will briefly summarize the inputs here.

The fit includes neutrino and anti-neutrino appearance data from both NOvA and T2K assuming a single energy bin.
The fit also includes disappearance data from both experiments where it is then effectively mapped onto a Hamiltonian to derive the relevant constraints.
To constrain the remaining parameters, information from Daya Bay \cite{DayaBay:2018yms} was used for $\theta_{13}$ and $\Delta m^2_{31}$ for both mass orderings and information from KamLAND \cite{KamLAND:2013rgu} was used for the solar parameters $\theta_{12}$ and $\Delta m^2_{21}$, along with the fact that $\Delta m^2_{21}>0$ from SNO, Borexino, and SuperK.
A fit was then performed to the NSI parameters while minimizing the test statistic over $\Delta m^2_{31}$, $\theta_{23}$, and $\delta$; the other three oscillation parameters were determined to have no appreciable effect.

\subsection{Vector NSI}
In \cite{Denton:2020uda} a fit to NOvA and T2K data was performed in the context of vector NSI.
We reproduce the main result in table \ref{tab:nova t2k vector}.

\begin{table}
\centering
\begin{tabular}{c|c||c|c|c|c}
MO&NSI&$|\eps_{\alpha\beta}|$&$\phi_{\alpha\beta}/\pi$&$\delta/\pi$&$\Delta\chi^2$\\\hline
\multirow{3}{*}{NO}&$\eps_{e\mu}$&0.19&1.50&1.46&4.44\\
&$\eps_{e\tau}$&0.28&1.60&1.46&3.65\\
&$\eps_{\mu\tau}$&0.35&0.60&1.83&0.90\\\hline
\multirow{3}{*}{IO}&$\eps_{e\mu}$&0.04&1.50&1.52&0.23\\
&$\eps_{e\tau}$&0.15&1.46&1.59&0.69\\
&$\eps_{\mu\tau}$&0.17&0.14&1.51&1.03
\end{tabular}
\caption{Best fit values to NOvA and T2K data and $\Delta\chi^2=\chi^2_{\rm SM}-\chi^2_{\rm NSI}$ for a fixed MO considering one complex vector NSI parameter at a time. (For the SM, $\chi^2_{\rm NO}-\chi^2_{\rm IO}=2.3\,$.)}
\label{tab:nova t2k vector}
\end{table}

\subsection{Scalar NSI}
\label{sec:nova t2k scalar}
\begin{figure}
\centering
\includegraphics[width=0.49\textwidth]{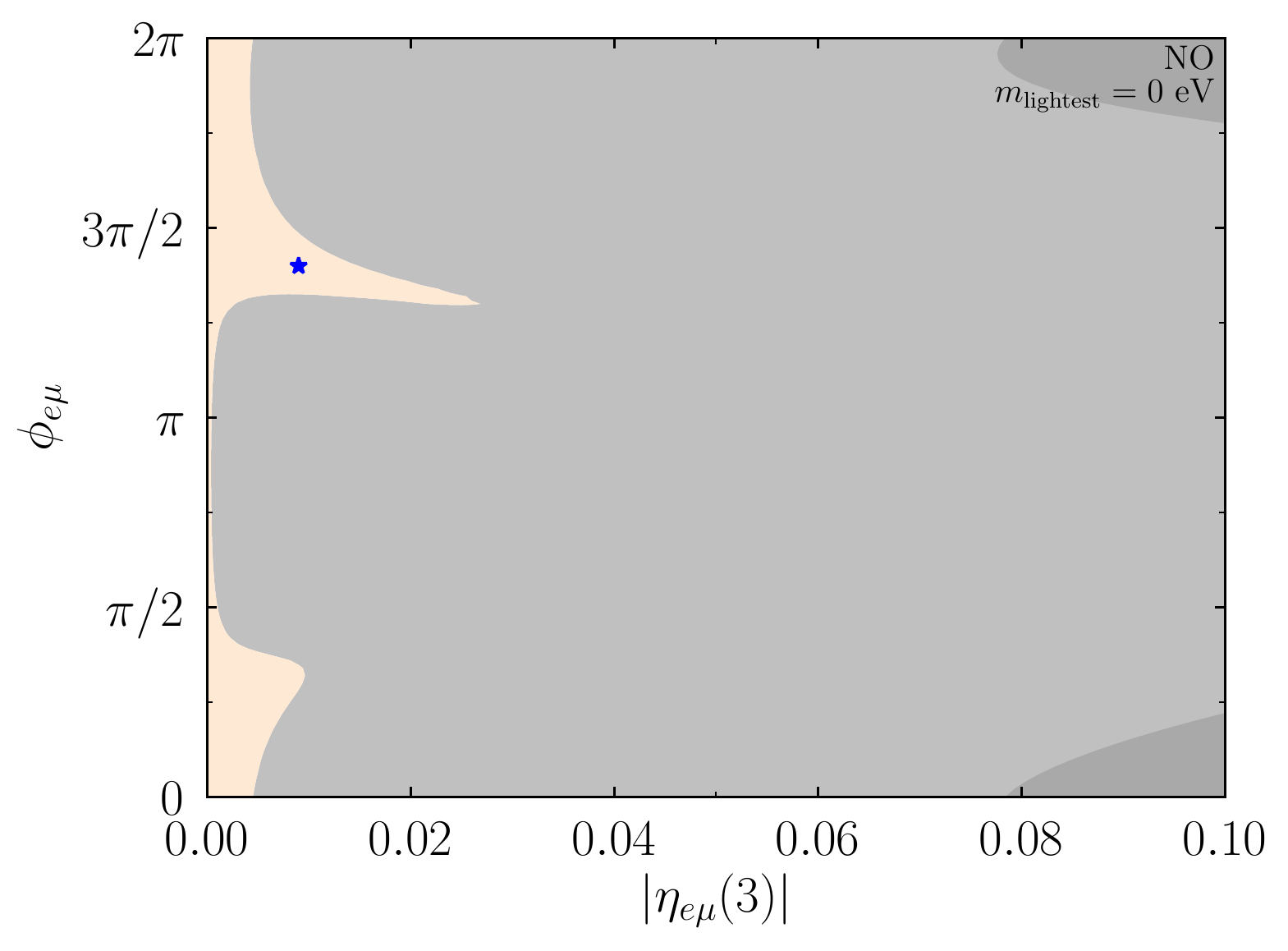}
\includegraphics[width=0.49\textwidth]{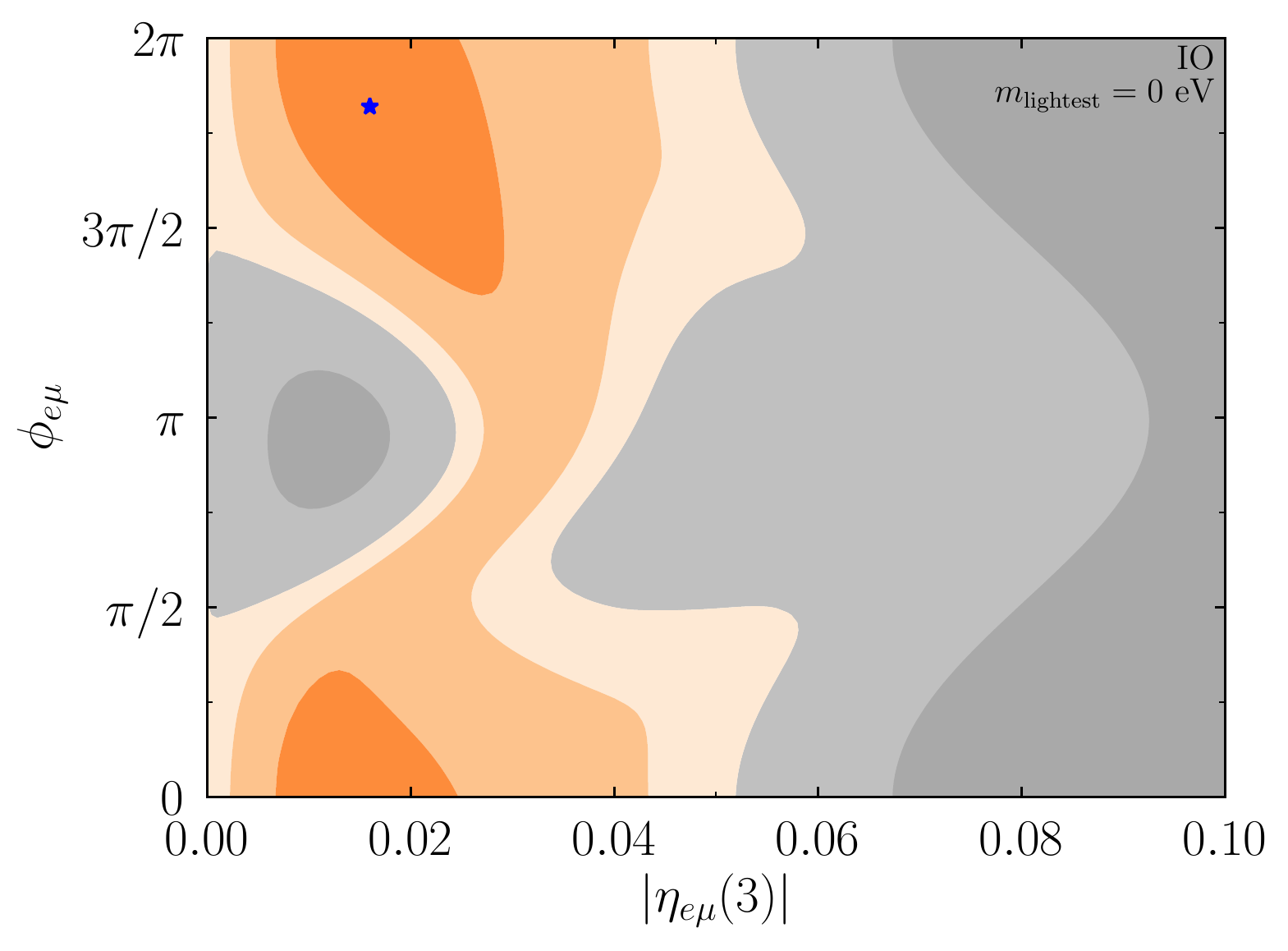}
\includegraphics[width=0.49\textwidth]{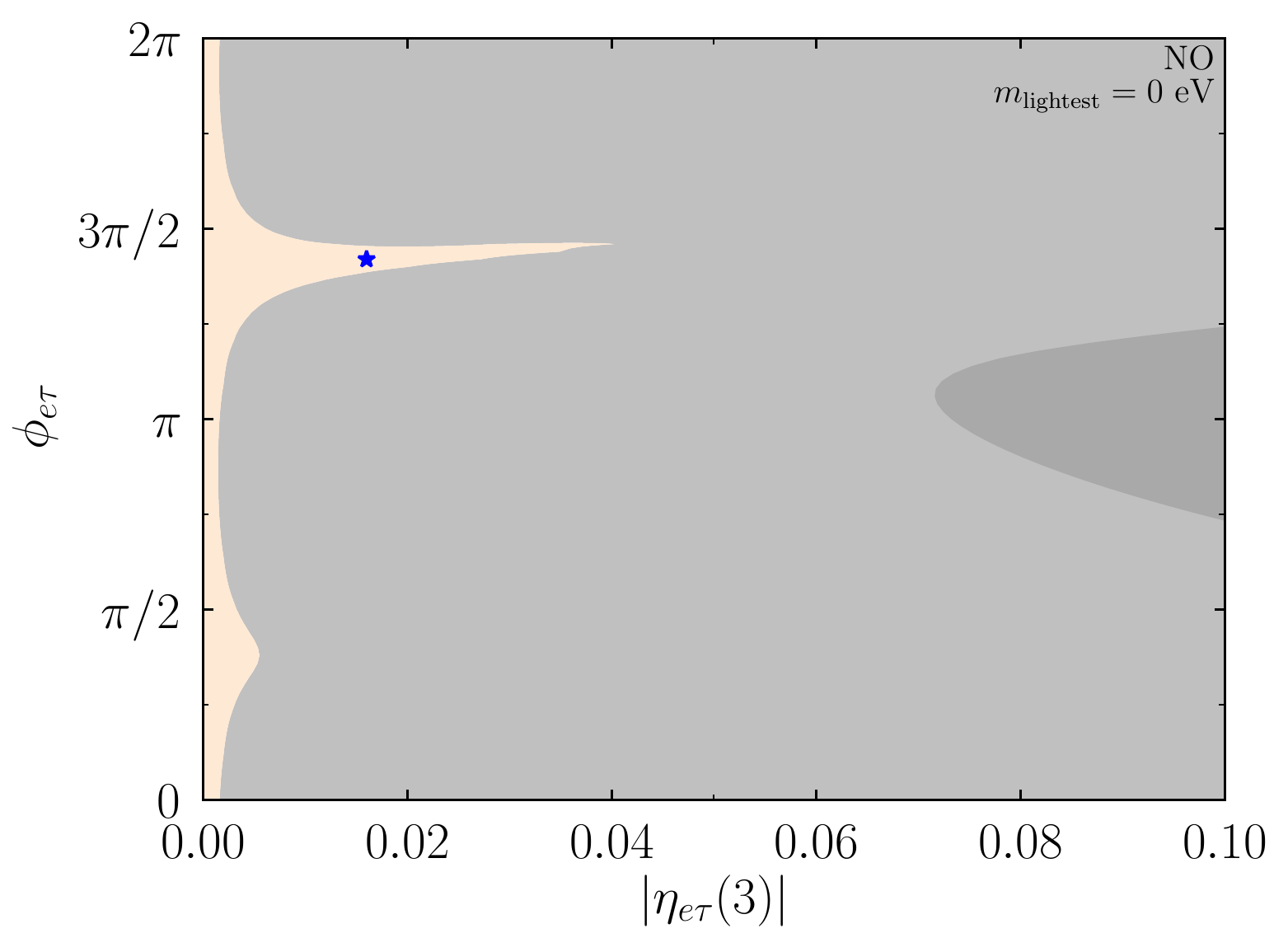}
\includegraphics[width=0.49\textwidth]{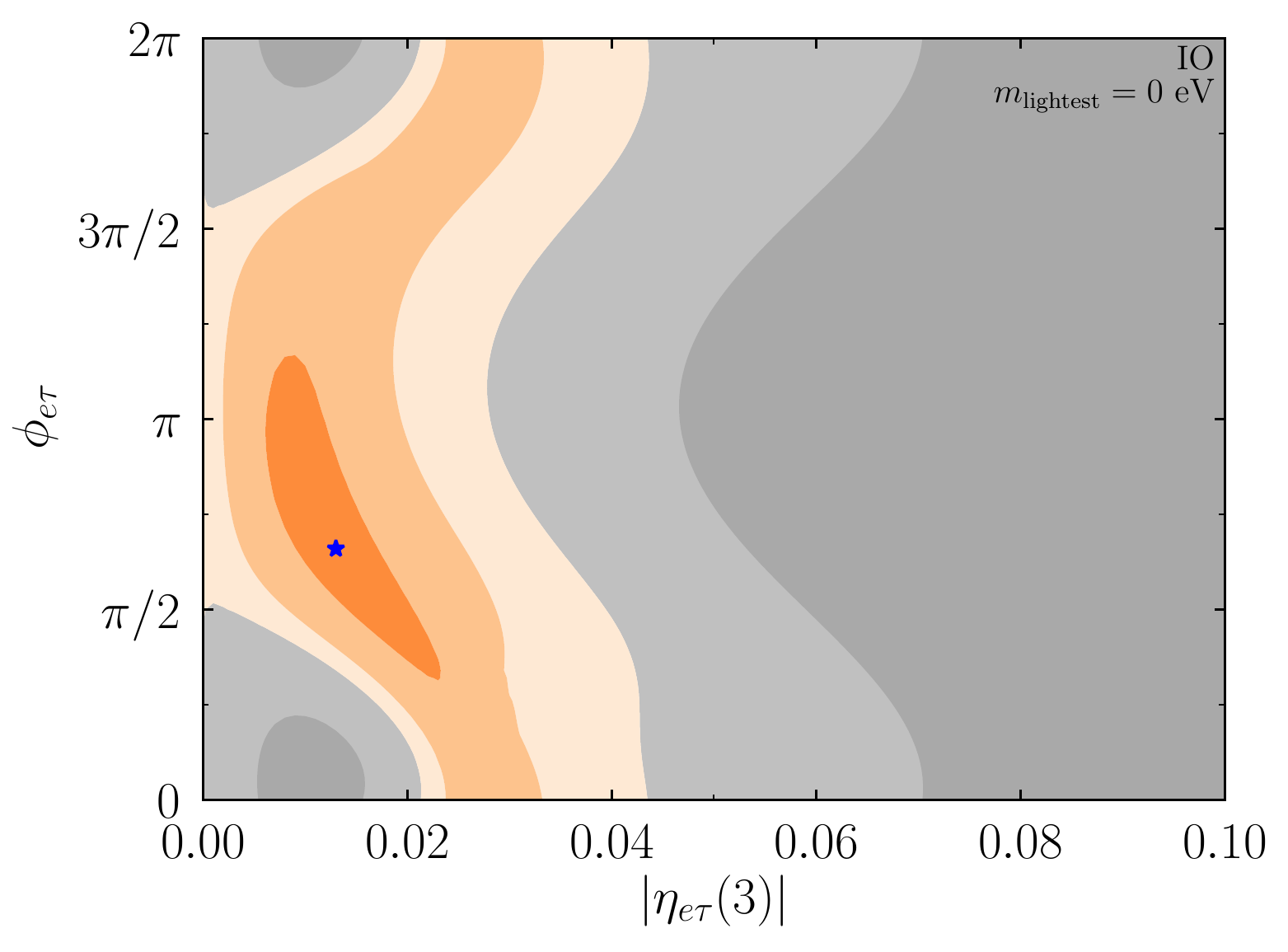}
\includegraphics[width=0.49\textwidth]{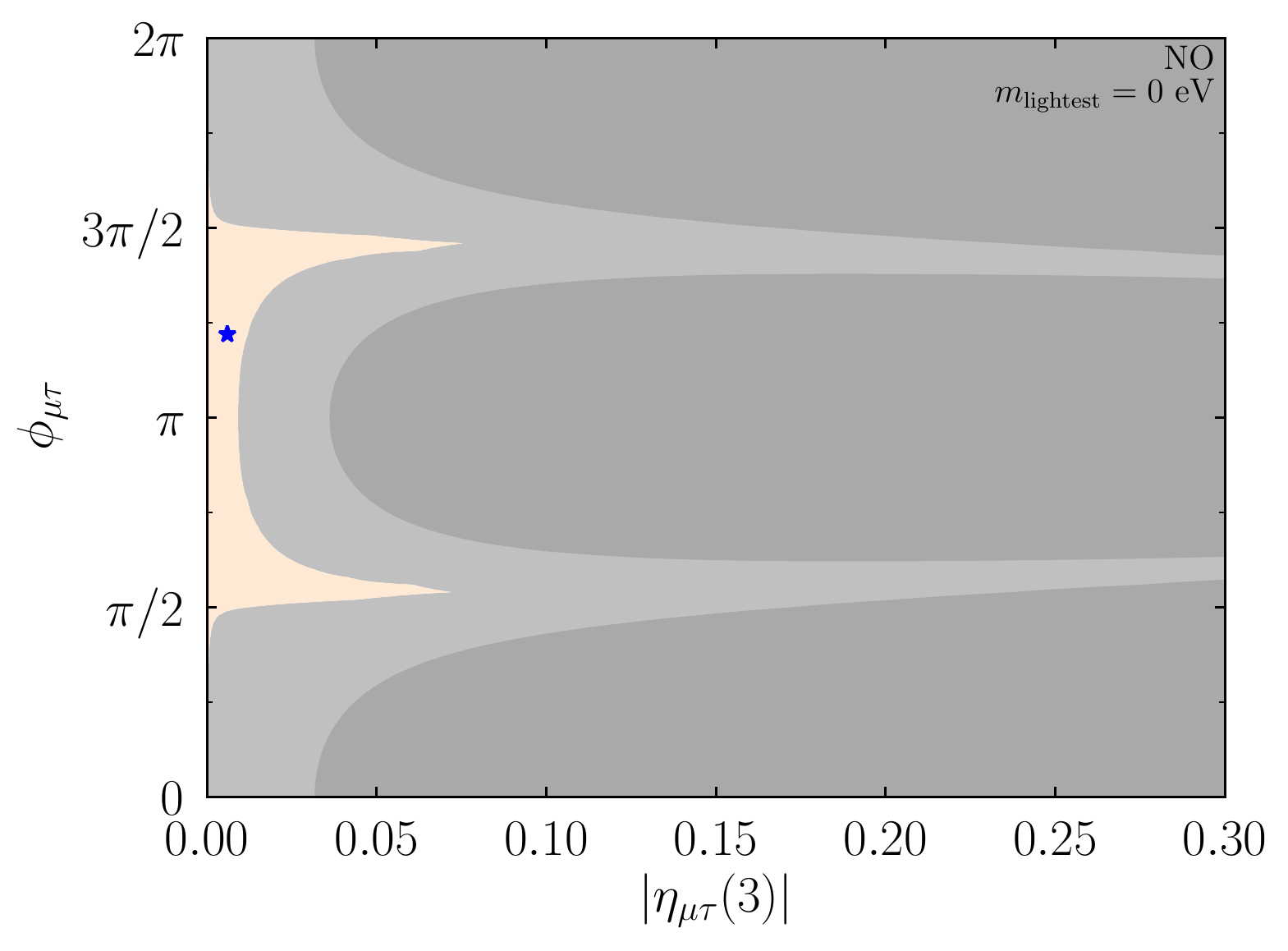}
\includegraphics[width=0.49\textwidth]{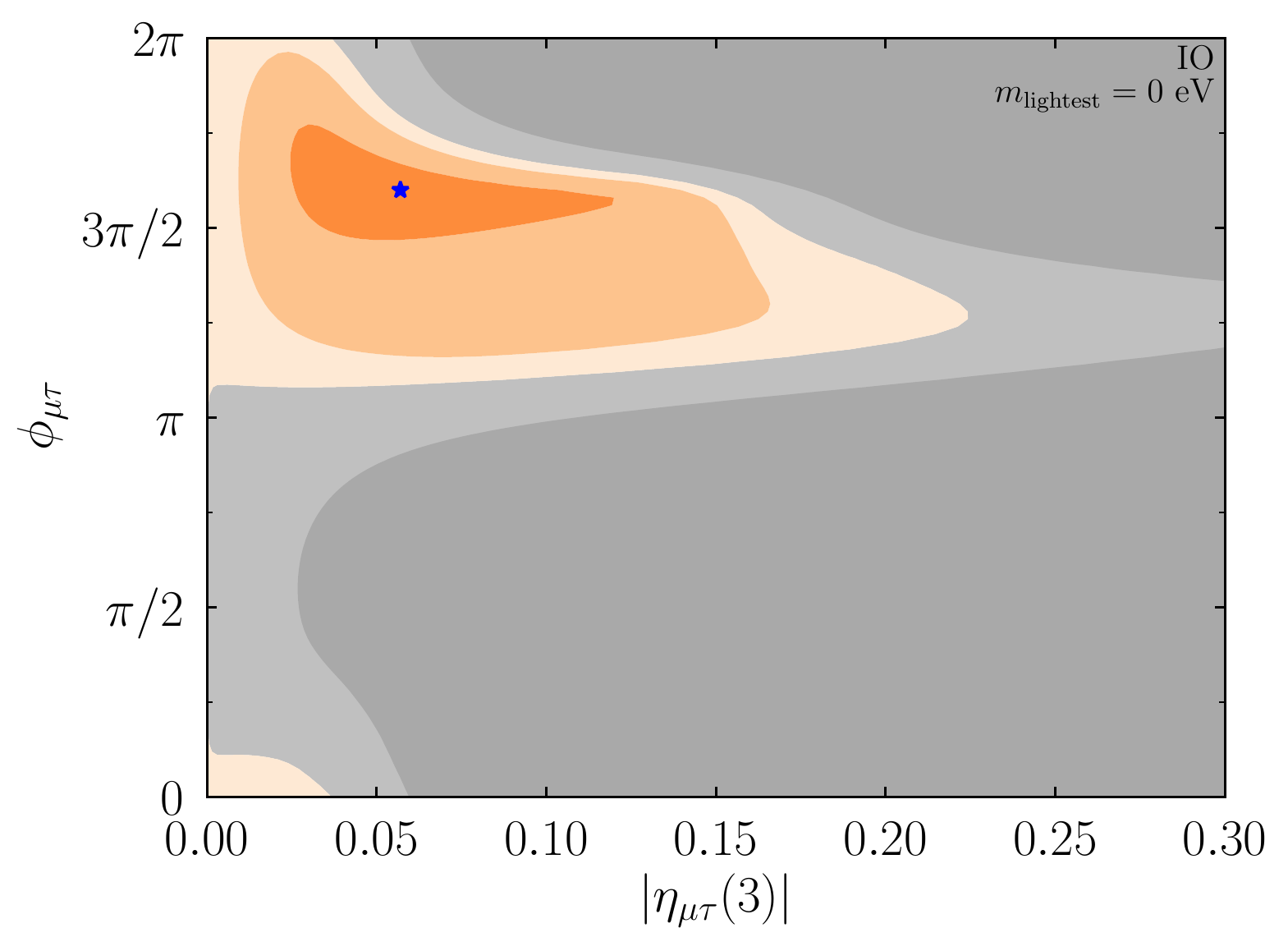}
\caption{The preferred and disfavored regions of scalar NSI given by NOvA and T2K data with information from Daya Bay and KamLAND data as well.
$\Delta m^2_{31}$, $\theta_{23}$, and $\delta$ are minimized over and $\theta_{13}$, $\theta_{12}$, and $\Delta m^2_{21}$ are fixed to the best fit values from Daya Bay and KamLAND.
The top, middle, and bottom rows correspond to $\eta_{e\mu}(3)$, $\eta_{e\tau}(3)$, and $\eta_{\mu\tau}(3)$ respectively, and the left and right columns correspond to the NO and IO respectively.
The mass of the lightest neutrino is taken to be zero here.
The $(3)$ refers to the fact that the scalar NSI parameters are plotted as rescaled to a density of 3 g/cm$^3$.
The blue stars are the best fit points, the light gray regions are slightly disfavored and the dark gray regions are disfavored at 68\% CL.
The successive orange colors represent integer units of $\Delta\chi^2$.}
\label{fig:nova t2k scalar}
\end{figure}

We redo the analysis of NOvA and T2K data as described above in the context of scalar NSI.
The results are reported in Fig.~\ref{fig:nova t2k scalar} and table \ref{tab:nova t2k scalar}.

\begin{table}
\centering
\begin{tabular}{c|c||c|c|c|c}
\multicolumn{6}{c}{$m_{\rm lightest}=0$ eV}\\\hline
MO&NSI&$|\eta_{\alpha\beta}(3)|$&$\phi_{\alpha\beta}/\pi$&$\delta/\pi$&$\Delta\chi^2$\\\hline
\multirow{3}{*}{NO}&$\eta_{e\mu}(3)$&0.009&1.40&1.17&0.04\\
&$\eta_{e\tau}(3)$&0.016&1.42&1.10&0.02\\
&$\eta_{\mu\tau}(3)$&0.006&1.22&1.11&0.08\\\hline
\multirow{3}{*}{IO}&$\eta_{e\mu}(3)$&0.016&1.82&1.86&2.33\\
&$\eta_{e\tau}(3)$&0.013&0.66&1.89&2.20\\
&$\eta_{\mu\tau}(3)$&0.057&1.60&1.85&2.33\\\multicolumn{2}{c}{}\\
\multicolumn{6}{c}{$m_{\rm lightest}=0.05$ eV}\\\hline
MO&NSI&$|\eta_{\alpha\beta}(3)|$&$\phi_{\alpha\beta}/\pi$&$\delta/\pi$&$\Delta\chi^2$\\\hline
\multirow{3}{*}{NO}&$\eta_{e\mu}(3)$&0.002&1.66&1.18&0.10\\
&$\eta_{e\tau}(3)$&0.003&0.62&1.13&0.08\\
&$\eta_{\mu\tau}(3)$&0.009&0.56&1.17&0.06\\\hline
\multirow{3}{*}{IO}&$\eta_{e\mu}(3)$&0.010&1.72&1.88&2.21\\
&$\eta_{e\tau}(3)$&0.010&0.58&1.90&2.18\\
&$\eta_{\mu\tau}(3)$&0.033&1.58&1.79&2.36\\\multicolumn{2}{c}{}\\
\multicolumn{6}{c}{$m_{\rm lightest}=0.10$ eV}\\\hline
MO&NSI&$|\eta_{\alpha\beta}(3)|$&$\phi_{\alpha\beta}/\pi$&$\delta/\pi$&$\Delta\chi^2$\\\hline
\multirow{3}{*}{NO}&$\eta_{e\mu}(3)$&0.001&1.74&1.17&0.12\\
&$\eta_{e\tau}(3)$&0.002&0.64&1.14&0.11\\
&$\eta_{\mu\tau}(3)$&0.006&0.56&1.19&0.06\\\hline
\multirow{3}{*}{IO}&$\eta_{e\mu}(3)$&0.006&1.72&1.86&2.20\\
&$\eta_{e\tau}(3)$&0.006&0.60&1.88&2.19\\
&$\eta_{\mu\tau}(3)$&0.024&1.56&1.83&2.36
\end{tabular}
\caption{Best fit values to NOvA and T2K data and $\Delta\chi^2=\chi^2_{\rm SM}-\chi^2_{\rm NSI}$ for a fixed MO considering one complex scalar NSI parameter at a time, rescaled to what it would be for a density of 3 g/cm$^3$ for various values of $m_{\rm lightest}$. (For the SM, $\chi^2_{\rm NO}-\chi^2_{\rm IO}=2.3\,$.)}
\label{tab:nova t2k scalar}
\end{table}

We make some comments on these results.
We note that the IO is always preferred over the NO, but at even lower significance than for vector NSI.
In addition, we also recall that SuperK's atmospheric data prefers the NO \cite{Esteban:2020cvm,Kelly:2020fkv,Denton:2020uda} and will be affected by scalar NSI in quite a different fashion to long-baseline data \cite{Ge:2018uhz,Medhi:2021wxj}.
Thus we find that scalar NSI is not a satisfactory improvement to the slight NOvA and T2K tension, but we can still regard the best fit points as valuable benchmarks moving forward.

Unlike in vector NSI, for scalar NSI the absolute neutrino mass scale plays a role.
In general the effect of $m_{\rm lightest}$, even to values quite a bit larger than allowed by cosmology \cite{DiValentino:2021hoh} where the upper limit is $\sim$few$\times10^{-2}$ eV, on the results is quite small.
That said, in some cases we see that the best fit value changes by a fair amount as $m_{\rm lightest}$ changes;
this is due to the existence of multiple quasi-degenerate local minima that slightly change as $m_{\rm lightest}$ changes.

\bibliographystyle{JHEP}
\bibliography{dune_scenarios}

\providecommand{\href}[2]{#2}\begingroup\raggedright\begin{thebibliography}{100}

\bibitem{Ayres:2007tu}
{\scshape NOvA} collaboration, \emph{{The NOvA Technical Design Report}}, .

\bibitem{Abe:2011ks}
{\scshape T2K} collaboration, \emph{{The T2K Experiment}},
  \href{https://doi.org/10.1016/j.nima.2011.06.067}{\emph{Nucl. Instrum. Meth.
  A} {\bfseries 659} (2011) 106}
  [\href{https://arxiv.org/abs/1106.1238}{{\ttfamily 1106.1238}}].

\bibitem{Abe:2015zbg}
{\scshape Hyper-Kamiokande Proto-} collaboration, \emph{{Physics potential of a
  long-baseline neutrino oscillation experiment using a J-PARC neutrino beam
  and Hyper-Kamiokande}},
  \href{https://doi.org/10.1093/ptep/ptv061}{\emph{PTEP} {\bfseries 2015}
  (2015) 053C02} [\href{https://arxiv.org/abs/1502.05199}{{\ttfamily
  1502.05199}}].

\bibitem{Acciarri:2015uup}
{\scshape DUNE} collaboration, \emph{{Long-Baseline Neutrino Facility (LBNF)
  and Deep Underground Neutrino Experiment (DUNE)}: {Conceptual Design Report,
  Volume 2: The Physics Program for DUNE at LBNF}},
  \href{https://arxiv.org/abs/1512.06148}{{\ttfamily 1512.06148}}.

\bibitem{Denton:2020uda}
P.B.~Denton, J.~Gehrlein and R.~Pestes, \emph{{$CP$ -Violating Neutrino
  Nonstandard Interactions in Long-Baseline-Accelerator Data}},
  \href{https://doi.org/10.1103/PhysRevLett.126.051801}{\emph{Phys. Rev. Lett.}
  {\bfseries 126} (2021) 051801}
  [\href{https://arxiv.org/abs/2008.01110}{{\ttfamily 2008.01110}}].

\bibitem{Chatterjee:2020kkm}
S.S.~Chatterjee and A.~Palazzo, \emph{{Nonstandard Neutrino Interactions as a
  Solution to the $NO\nu A$ and T2K Discrepancy}},
  \href{https://doi.org/10.1103/PhysRevLett.126.051802}{\emph{Phys. Rev. Lett.}
  {\bfseries 126} (2021) 051802}
  [\href{https://arxiv.org/abs/2008.04161}{{\ttfamily 2008.04161}}].

\bibitem{Chatterjee:2020yak}
S.S.~Chatterjee and A.~Palazzo, \emph{{Interpretation of NO$\nu$A and T2K data
  in the presence of a light sterile neutrino}},
  \href{https://arxiv.org/abs/2005.10338}{{\ttfamily 2005.10338}}.

\bibitem{Miranda:2019ynh}
L.S.~Miranda, P.~Pasquini, U.~Rahaman and S.~Razzaque, \emph{{Searching for
  non-unitary neutrino oscillations in the present T2K and NO$\nu $A data}},
  \href{https://doi.org/10.1140/epjc/s10052-021-09227-0}{\emph{Eur. Phys. J. C}
  {\bfseries 81} (2021) 444}
  [\href{https://arxiv.org/abs/1911.09398}{{\ttfamily 1911.09398}}].

\bibitem{Forero:2021azc}
D.V.~Forero, C.~Giunti, C.A.~Ternes and M.~Tortola, \emph{{Nonunitary neutrino
  mixing in short and long-baseline experiments}},
  \href{https://doi.org/10.1103/PhysRevD.104.075030}{\emph{Phys. Rev. D}
  {\bfseries 104} (2021) 075030}
  [\href{https://arxiv.org/abs/2103.01998}{{\ttfamily 2103.01998}}].

\bibitem{deGouvea:2022kma}
A.~de~Gouv\^ea, G.~Jusino~S\'anchez and K.J.~Kelly, \emph{{Very Light Sterile
  Neutrinos at NOvA and T2K}},
  \href{https://arxiv.org/abs/2204.09130}{{\ttfamily 2204.09130}}.

\bibitem{Majhi:2022wyp}
R.~Majhi, D.K.~Singha, K.N.~Deepthi and R.~Mohanta, \emph{{Vector leptoquark
  $U_3$ and CP violation at T2K, NOvA experiments}},
  \href{https://arxiv.org/abs/2205.04269}{{\ttfamily 2205.04269}}.

\bibitem{Arguelles:2019xgp}
C.A.~Arg\"uelles et~al., \emph{{New opportunities at the next-generation
  neutrino experiments I: BSM neutrino physics and dark matter}},
  \href{https://doi.org/10.1088/1361-6633/ab9d12}{\emph{Rept. Prog. Phys.}
  {\bfseries 83} (2020) 124201}
  [\href{https://arxiv.org/abs/1907.08311}{{\ttfamily 1907.08311}}].

\bibitem{Arguelles:2022xxa}
C.A.~Arg\"uelles et~al., \emph{{Snowmass White Paper: Beyond the Standard Model
  effects on Neutrino Flavor}},  in \emph{{2022 Snowmass Summer Study}}, 3,
  2022 [\href{https://arxiv.org/abs/2203.10811}{{\ttfamily 2203.10811}}].

\bibitem{Berryman:2014qha}
J.M.~Berryman, A.~de~Gouvea and D.~Hernandez, \emph{{Solar Neutrinos and the
  Decaying Neutrino Hypothesis}},
  \href{https://doi.org/10.1103/PhysRevD.92.073003}{\emph{Phys. Rev. D}
  {\bfseries 92} (2015) 073003}
  [\href{https://arxiv.org/abs/1411.0308}{{\ttfamily 1411.0308}}].

\bibitem{Picoreti:2015ika}
R.~Picoreti, M.M.~Guzzo, P.C.~de~Holanda and O.L.G.~Peres, \emph{{Neutrino
  Decay and Solar Neutrino Seasonal Effect}},
  \href{https://doi.org/10.1016/j.physletb.2016.08.007}{\emph{Phys. Lett. B}
  {\bfseries 761} (2016) 70}
  [\href{https://arxiv.org/abs/1506.08158}{{\ttfamily 1506.08158}}].

\bibitem{SNO:2018pvg}
{\scshape SNO} collaboration, \emph{{Constraints on Neutrino Lifetime from the
  Sudbury Neutrino Observatory}},
  \href{https://doi.org/10.1103/PhysRevD.99.032013}{\emph{Phys. Rev. D}
  {\bfseries 99} (2019) 032013}
  [\href{https://arxiv.org/abs/1812.01088}{{\ttfamily 1812.01088}}].

\bibitem{Gonzalez-Garcia:2008mgl}
M.C.~Gonzalez-Garcia and M.~Maltoni, \emph{{Status of Oscillation plus Decay of
  Atmospheric and Long-Baseline Neutrinos}},
  \href{https://doi.org/10.1016/j.physletb.2008.04.041}{\emph{Phys. Lett. B}
  {\bfseries 663} (2008) 405}
  [\href{https://arxiv.org/abs/0802.3699}{{\ttfamily 0802.3699}}].

\bibitem{Gomes:2014yua}
R.A.~Gomes, A.L.G.~Gomes and O.L.G.~Peres, \emph{{Constraints on neutrino decay
  lifetime using long-baseline charged and neutral current data}},
  \href{https://doi.org/10.1016/j.physletb.2014.12.014}{\emph{Phys. Lett. B}
  {\bfseries 740} (2015) 345}
  [\href{https://arxiv.org/abs/1407.5640}{{\ttfamily 1407.5640}}].

\bibitem{Abrahao:2015rba}
T.~Abrah\~ao, H.~Minakata, H.~Nunokawa and A.A.~Quiroga, \emph{{Constraint on
  Neutrino Decay with Medium-Baseline Reactor Neutrino Oscillation
  Experiments}}, \href{https://doi.org/10.1007/JHEP11(2015)001}{\emph{JHEP}
  {\bfseries 11} (2015) 001}
  [\href{https://arxiv.org/abs/1506.02314}{{\ttfamily 1506.02314}}].

\bibitem{Pagliaroli:2016zab}
G.~Pagliaroli, N.~Di~Marco and M.~Mannarelli, \emph{{Enhanced tau neutrino
  appearance through invisible decay}},
  \href{https://doi.org/10.1103/PhysRevD.93.113011}{\emph{Phys. Rev. D}
  {\bfseries 93} (2016) 113011}
  [\href{https://arxiv.org/abs/1603.08696}{{\ttfamily 1603.08696}}].

\bibitem{Coloma:2017zpg}
P.~Coloma and O.L.G.~Peres, \emph{{Visible neutrino decay at DUNE}},
  \href{https://arxiv.org/abs/1705.03599}{{\ttfamily 1705.03599}}.

\bibitem{Choubey:2017dyu}
S.~Choubey, S.~Goswami and D.~Pramanik, \emph{{A study of invisible neutrino
  decay at DUNE and its effects on $\theta_{23}$ measurement}},
  \href{https://doi.org/10.1007/JHEP02(2018)055}{\emph{JHEP} {\bfseries 02}
  (2018) 055} [\href{https://arxiv.org/abs/1705.05820}{{\ttfamily
  1705.05820}}].

\bibitem{Gago:2017zzy}
A.M.~Gago, R.A.~Gomes, A.L.G.~Gomes, J.~Jones-Perez and O.L.G.~Peres,
  \emph{{Visible neutrino decay in the light of appearance and disappearance
  long baseline experiments}},
  \href{https://doi.org/10.1007/JHEP11(2017)022}{\emph{JHEP} {\bfseries 11}
  (2017) 022} [\href{https://arxiv.org/abs/1705.03074}{{\ttfamily
  1705.03074}}].

\bibitem{Choubey:2017eyg}
S.~Choubey, S.~Goswami, C.~Gupta, S.M.~Lakshmi and T.~Thakore,
  \emph{{Sensitivity to neutrino decay with atmospheric neutrinos at the
  INO-ICAL detector}},
  \href{https://doi.org/10.1103/PhysRevD.97.033005}{\emph{Phys. Rev. D}
  {\bfseries 97} (2018) 033005}
  [\href{https://arxiv.org/abs/1709.10376}{{\ttfamily 1709.10376}}].

\bibitem{Denton:2018aml}
P.B.~Denton and I.~Tamborra, \emph{{Invisible Neutrino Decay Could Resolve
  IceCube\textquoteright{}s Track and Cascade Tension}},
  \href{https://doi.org/10.1103/PhysRevLett.121.121802}{\emph{Phys. Rev. Lett.}
  {\bfseries 121} (2018) 121802}
  [\href{https://arxiv.org/abs/1805.05950}{{\ttfamily 1805.05950}}].

\bibitem{deSalas:2018kri}
P.F.~de~Salas, S.~Pastor, C.A.~Ternes, T.~Thakore and M.~T\'ortola,
  \emph{{Constraining the invisible neutrino decay with KM3NeT-ORCA}},
  \href{https://doi.org/10.1016/j.physletb.2018.12.066}{\emph{Phys. Lett. B}
  {\bfseries 789} (2019) 472}
  [\href{https://arxiv.org/abs/1810.10916}{{\ttfamily 1810.10916}}].

\bibitem{Ascencio-Sosa:2018lbk}
M.V.~Ascencio-Sosa, A.M.~Calatayud-Cadenillas, A.M.~Gago and J.~Jones-P\'erez,
  \emph{{Matter effects in neutrino visible decay at future long-baseline
  experiments}},
  \href{https://doi.org/10.1140/epjc/s10052-018-6276-0}{\emph{Eur. Phys. J. C}
  {\bfseries 78} (2018) 809}
  [\href{https://arxiv.org/abs/1805.03279}{{\ttfamily 1805.03279}}].

\bibitem{Choubey:2018cfz}
S.~Choubey, D.~Dutta and D.~Pramanik, \emph{{Invisible neutrino decay in the
  light of NOvA and T2K data}},
  \href{https://doi.org/10.1007/JHEP08(2018)141}{\emph{JHEP} {\bfseries 08}
  (2018) 141} [\href{https://arxiv.org/abs/1805.01848}{{\ttfamily
  1805.01848}}].

\bibitem{Funcke:2019grs}
L.~Funcke, G.~Raffelt and E.~Vitagliano, \emph{{Distinguishing Dirac and
  Majorana neutrinos by their decays via Nambu-Goldstone bosons in the
  gravitational-anomaly model of neutrino masses}},
  \href{https://doi.org/10.1103/PhysRevD.101.015025}{\emph{Phys. Rev. D}
  {\bfseries 101} (2020) 015025}
  [\href{https://arxiv.org/abs/1905.01264}{{\ttfamily 1905.01264}}].

\bibitem{Abdullahi:2020rge}
A.~Abdullahi and P.B.~Denton, \emph{{Visible Decay of Astrophysical Neutrinos
  at IceCube}}, \href{https://doi.org/10.1103/PhysRevD.102.023018}{\emph{Phys.
  Rev. D} {\bfseries 102} (2020) 023018}
  [\href{https://arxiv.org/abs/2005.07200}{{\ttfamily 2005.07200}}].

\bibitem{Ghoshal:2020hyo}
A.~Ghoshal, A.~Giarnetti and D.~Meloni, \emph{{Neutrino Invisible Decay at
  DUNE: a multi-channel analysis}},
  \href{https://doi.org/10.1088/1361-6471/abdfab}{\emph{J. Phys. G} {\bfseries
  48} (2021) 055004} [\href{https://arxiv.org/abs/2003.09012}{{\ttfamily
  2003.09012}}].

\bibitem{Porto-Silva:2020gma}
Y.P.~Porto-Silva, S.~Prakash, O.L.G.~Peres, H.~Nunokawa and H.~Minakata,
  \emph{{Constraining visible neutrino decay at KamLAND and JUNO}},
  \href{https://doi.org/10.1140/epjc/s10052-020-08573-9}{\emph{Eur. Phys. J. C}
  {\bfseries 80} (2020) 999}
  [\href{https://arxiv.org/abs/2002.12134}{{\ttfamily 2002.12134}}].

\bibitem{Choubey:2020dhw}
S.~Choubey, M.~Ghosh, D.~Kempe and T.~Ohlsson, \emph{{Exploring invisible
  neutrino decay at ESSnuSB}},
  \href{https://doi.org/10.1007/JHEP05(2021)133}{\emph{JHEP} {\bfseries 05}
  (2021) 133} [\href{https://arxiv.org/abs/2010.16334}{{\ttfamily
  2010.16334}}].

\bibitem{Picoreti:2021yct}
R.~Picoreti, D.~Pramanik, P.C.~de~Holanda and O.L.G.~Peres, \emph{{Updating
  \ensuremath{\nu}3 lifetime from solar antineutrino spectra}},
  \href{https://doi.org/10.1103/PhysRevD.106.015025}{\emph{Phys. Rev. D}
  {\bfseries 106} (2022) 015025}
  [\href{https://arxiv.org/abs/2109.13272}{{\ttfamily 2109.13272}}].

\bibitem{Kostelecky:2003cr}
V.A.~Kostelecky and M.~Mewes, \emph{{Lorentz and CPT violation in neutrinos}},
  \href{https://doi.org/10.1103/PhysRevD.69.016005}{\emph{Phys. Rev. D}
  {\bfseries 69} (2004) 016005}
  [\href{https://arxiv.org/abs/hep-ph/0309025}{{\ttfamily hep-ph/0309025}}].

\bibitem{LSND:2005oop}
{\scshape LSND} collaboration, \emph{{Tests of Lorentz violation in anti-nu(mu)
  ---\ensuremath{>} anti-nu(e) oscillations}},
  \href{https://doi.org/10.1103/PhysRevD.72.076004}{\emph{Phys. Rev. D}
  {\bfseries 72} (2005) 076004}
  [\href{https://arxiv.org/abs/hep-ex/0506067}{{\ttfamily hep-ex/0506067}}].

\bibitem{MINOS:2008fnv}
{\scshape MINOS} collaboration, \emph{{Testing Lorentz Invariance and CPT
  Conservation with NuMI Neutrinos in the MINOS Near Detector}},
  \href{https://doi.org/10.1103/PhysRevLett.101.151601}{\emph{Phys. Rev. Lett.}
  {\bfseries 101} (2008) 151601}
  [\href{https://arxiv.org/abs/0806.4945}{{\ttfamily 0806.4945}}].

\bibitem{MINOS:2010kat}
{\scshape MINOS} collaboration, \emph{{A Search for Lorentz Invariance and CPT
  Violation with the MINOS Far Detector}},
  \href{https://doi.org/10.1103/PhysRevLett.105.151601}{\emph{Phys. Rev. Lett.}
  {\bfseries 105} (2010) 151601}
  [\href{https://arxiv.org/abs/1007.2791}{{\ttfamily 1007.2791}}].

\bibitem{IceCube:2010fyu}
{\scshape IceCube} collaboration, \emph{{Search for a Lorentz-violating
  sidereal signal with atmospheric neutrinos in IceCube}},
  \href{https://doi.org/10.1103/PhysRevD.82.112003}{\emph{Phys. Rev. D}
  {\bfseries 82} (2010) 112003}
  [\href{https://arxiv.org/abs/1010.4096}{{\ttfamily 1010.4096}}].

\bibitem{Kostelecky:2011gq}
A.~Kostelecky and M.~Mewes, \emph{{Neutrinos with Lorentz-violating operators
  of arbitrary dimension}},
  \href{https://doi.org/10.1103/PhysRevD.85.096005}{\emph{Phys. Rev. D}
  {\bfseries 85} (2012) 096005}
  [\href{https://arxiv.org/abs/1112.6395}{{\ttfamily 1112.6395}}].

\bibitem{MiniBooNE:2011pix}
{\scshape MiniBooNE} collaboration, \emph{{Test of Lorentz and CPT violation
  with Short Baseline Neutrino Oscillation Excesses}},
  \href{https://doi.org/10.1016/j.physletb.2012.12.020}{\emph{Phys. Lett. B}
  {\bfseries 718} (2013) 1303}
  [\href{https://arxiv.org/abs/1109.3480}{{\ttfamily 1109.3480}}].

\bibitem{DoubleChooz:2012eiq}
{\scshape Double Chooz} collaboration, \emph{{First Test of Lorentz Violation
  with a Reactor-based Antineutrino Experiment}},
  \href{https://doi.org/10.1103/PhysRevD.86.112009}{\emph{Phys. Rev. D}
  {\bfseries 86} (2012) 112009}
  [\href{https://arxiv.org/abs/1209.5810}{{\ttfamily 1209.5810}}].

\bibitem{MINOS:2012ozn}
{\scshape MINOS} collaboration, \emph{{Search for Lorentz invariance and CPT
  violation with muon antineutrinos in the MINOS Near Detector}},
  \href{https://doi.org/10.1103/PhysRevD.85.031101}{\emph{Phys. Rev. D}
  {\bfseries 85} (2012) 031101}
  [\href{https://arxiv.org/abs/1201.2631}{{\ttfamily 1201.2631}}].

\bibitem{Diaz:2013iba}
J.S.~D\'\i{}az, T.~Katori, J.~Spitz and J.M.~Conrad, \emph{{Search for
  neutrino-antineutrino oscillations with a reactor experiment}},
  \href{https://doi.org/10.1016/j.physletb.2013.10.058}{\emph{Phys. Lett. B}
  {\bfseries 727} (2013) 412}
  [\href{https://arxiv.org/abs/1307.5789}{{\ttfamily 1307.5789}}].

\bibitem{Rebel:2013vc}
B.~Rebel and S.~Mufson, \emph{{The Search for Neutrino-Antineutrino Mixing
  Resulting from Lorentz Invariance Violation using neutrino interactions in
  MINOS}},
  \href{https://doi.org/10.1016/j.astropartphys.2013.07.006}{\emph{Astropart.
  Phys.} {\bfseries 48} (2013) 78}
  [\href{https://arxiv.org/abs/1301.4684}{{\ttfamily 1301.4684}}].

\bibitem{Super-Kamiokande:2014exs}
{\scshape Super-Kamiokande} collaboration, \emph{{Test of Lorentz invariance
  with atmospheric neutrinos}},
  \href{https://doi.org/10.1103/PhysRevD.91.052003}{\emph{Phys. Rev. D}
  {\bfseries 91} (2015) 052003}
  [\href{https://arxiv.org/abs/1410.4267}{{\ttfamily 1410.4267}}].

\bibitem{EXO-200:2016hbz}
{\scshape EXO-200} collaboration, \emph{{First Search for Lorentz and CPT
  Violation in Double Beta Decay with EXO-200}},
  \href{https://doi.org/10.1103/PhysRevD.93.072001}{\emph{Phys. Rev. D}
  {\bfseries 93} (2016) 072001}
  [\href{https://arxiv.org/abs/1601.07266}{{\ttfamily 1601.07266}}].

\bibitem{IceCube:2017qyp}
{\scshape IceCube} collaboration, \emph{{Neutrino Interferometry for
  High-Precision Tests of Lorentz Symmetry with IceCube}},
  \href{https://doi.org/10.1038/s41567-018-0172-2}{\emph{Nature Phys.}
  {\bfseries 14} (2018) 961}
  [\href{https://arxiv.org/abs/1709.03434}{{\ttfamily 1709.03434}}].

\bibitem{T2K:2017ega}
{\scshape T2K} collaboration, \emph{{Search for Lorentz and CPT violation using
  sidereal time dependence of neutrino flavor transitions over a short
  baseline}}, \href{https://doi.org/10.1103/PhysRevD.95.111101}{\emph{Phys.
  Rev. D} {\bfseries 95} (2017) 111101}
  [\href{https://arxiv.org/abs/1703.01361}{{\ttfamily 1703.01361}}].

\bibitem{DayaBay:2018fsh}
{\scshape Daya Bay} collaboration, \emph{{Search for a time-varying electron
  antineutrino signal at Daya Bay}},
  \href{https://doi.org/10.1103/PhysRevD.98.092013}{\emph{Phys. Rev. D}
  {\bfseries 98} (2018) 092013}
  [\href{https://arxiv.org/abs/1809.04660}{{\ttfamily 1809.04660}}].

\bibitem{Reynoso:2016hjr}
M.M.~Reynoso and O.A.~Sampayo, \emph{{Propagation of high-energy neutrinos in a
  background of ultralight scalar dark matter}},
  \href{https://doi.org/10.1016/j.astropartphys.2016.05.004}{\emph{Astropart.
  Phys.} {\bfseries 82} (2016) 10}
  [\href{https://arxiv.org/abs/1605.09671}{{\ttfamily 1605.09671}}].

\bibitem{Berlin:2016woy}
A.~Berlin, \emph{{Neutrino Oscillations as a Probe of Light Scalar Dark
  Matter}}, \href{https://doi.org/10.1103/PhysRevLett.117.231801}{\emph{Phys.
  Rev. Lett.} {\bfseries 117} (2016) 231801}
  [\href{https://arxiv.org/abs/1608.01307}{{\ttfamily 1608.01307}}].

\bibitem{Krnjaic:2017zlz}
G.~Krnjaic, P.A.N.~Machado and L.~Necib, \emph{{Distorted neutrino oscillations
  from time varying cosmic fields}},
  \href{https://doi.org/10.1103/PhysRevD.97.075017}{\emph{Phys. Rev. D}
  {\bfseries 97} (2018) 075017}
  [\href{https://arxiv.org/abs/1705.06740}{{\ttfamily 1705.06740}}].

\bibitem{Brdar:2017kbt}
V.~Brdar, J.~Kopp, J.~Liu, P.~Prass and X.-P.~Wang, \emph{{Fuzzy dark matter
  and nonstandard neutrino interactions}},
  \href{https://doi.org/10.1103/PhysRevD.97.043001}{\emph{Phys. Rev. D}
  {\bfseries 97} (2018) 043001}
  [\href{https://arxiv.org/abs/1705.09455}{{\ttfamily 1705.09455}}].

\bibitem{Liao:2018byh}
J.~Liao, D.~Marfatia and K.~Whisnant, \emph{{Light scalar dark matter at
  neutrino oscillation experiments}},
  \href{https://doi.org/10.1007/JHEP04(2018)136}{\emph{JHEP} {\bfseries 04}
  (2018) 136} [\href{https://arxiv.org/abs/1803.01773}{{\ttfamily
  1803.01773}}].

\bibitem{Capozzi:2018bps}
F.~Capozzi, I.M.~Shoemaker and L.~Vecchi, \emph{{Neutrino Oscillations in Dark
  Backgrounds}},
  \href{https://doi.org/10.1088/1475-7516/2018/07/004}{\emph{JCAP} {\bfseries
  07} (2018) 004} [\href{https://arxiv.org/abs/1804.05117}{{\ttfamily
  1804.05117}}].

\bibitem{Huang:2018cwo}
G.-Y.~Huang and N.~Nath, \emph{{Neutrinophilic Axion-Like Dark Matter}},
  \href{https://doi.org/10.1140/epjc/s10052-018-6391-y}{\emph{Eur. Phys. J. C}
  {\bfseries 78} (2018) 922}
  [\href{https://arxiv.org/abs/1809.01111}{{\ttfamily 1809.01111}}].

\bibitem{Farzan:2019yvo}
Y.~Farzan, \emph{{Ultra-light scalar saving the 3 + 1 neutrino scheme from the
  cosmological bounds}},
  \href{https://doi.org/10.1016/j.physletb.2019.134911}{\emph{Phys. Lett. B}
  {\bfseries 797} (2019) 134911}
  [\href{https://arxiv.org/abs/1907.04271}{{\ttfamily 1907.04271}}].

\bibitem{Cline:2019seo}
J.M.~Cline, \emph{{Viable secret neutrino interactions with ultralight dark
  matter}}, \href{https://doi.org/10.1016/j.physletb.2019.135182}{\emph{Phys.
  Lett. B} {\bfseries 802} (2020) 135182}
  [\href{https://arxiv.org/abs/1908.02278}{{\ttfamily 1908.02278}}].

\bibitem{Dev:2020kgz}
A.~Dev, P.A.N.~Machado and P.~Mart\'\i{}nez-Mirav\'e, \emph{{Signatures of
  ultralight dark matter in neutrino oscillation experiments}},
  \href{https://doi.org/10.1007/JHEP01(2021)094}{\emph{JHEP} {\bfseries 01}
  (2021) 094} [\href{https://arxiv.org/abs/2007.03590}{{\ttfamily
  2007.03590}}].

\bibitem{Huang:2021kam}
G.-y.~Huang and N.~Nath, \emph{{Neutrino meets ultralight dark matter:
  0\ensuremath{\nu}\ensuremath{\beta}\ensuremath{\beta} decay and cosmology}},
  \href{https://doi.org/10.1088/1475-7516/2022/05/034}{\emph{JCAP} {\bfseries
  05} (2022) 034} [\href{https://arxiv.org/abs/2111.08732}{{\ttfamily
  2111.08732}}].

\bibitem{Losada:2021bxx}
M.~Losada, Y.~Nir, G.~Perez and Y.~Shpilman, \emph{{Probing scalar dark matter
  oscillations with neutrino oscillations}},
  \href{https://doi.org/10.1007/JHEP04(2022)030}{\emph{JHEP} {\bfseries 04}
  (2022) 030} [\href{https://arxiv.org/abs/2107.10865}{{\ttfamily
  2107.10865}}].

\bibitem{Chun:2021ief}
E.J.~Chun, \emph{{Neutrino Transition in Dark Matter}},
  \href{https://arxiv.org/abs/2112.05057}{{\ttfamily 2112.05057}}.

\bibitem{Dev:2022bae}
A.~Dev, G.~Krnjaic, P.~Machado and H.~Ramani, \emph{{Constraining Feeble
  Neutrino Interactions with Ultralight Dark Matter}},
  \href{https://arxiv.org/abs/2205.06821}{{\ttfamily 2205.06821}}.

\bibitem{Liu:1997km}
Y.~Liu, L.-z.~Hu and M.-L.~Ge, \emph{{The Effect of quantum mechanics violation
  on neutrino oscillation}},
  \href{https://doi.org/10.1103/PhysRevD.56.6648}{\emph{Phys. Rev. D}
  {\bfseries 56} (1997) 6648}.

\bibitem{Chang:1998ea}
C.-H.~Chang, W.-S.~Dai, X.-Q.~Li, Y.~Liu, F.-C.~Ma and Z.-j.~Tao,
  \emph{{Possible effects of quantum mechanics violation induced by certain
  quantum gravity on neutrino oscillations}},
  \href{https://doi.org/10.1103/PhysRevD.60.033006}{\emph{Phys. Rev. D}
  {\bfseries 60} (1999) 033006}
  [\href{https://arxiv.org/abs/hep-ph/9809371}{{\ttfamily hep-ph/9809371}}].

\bibitem{Benatti:2000ph}
F.~Benatti and R.~Floreanini, \emph{{Open system approach to neutrino
  oscillations}},
  \href{https://doi.org/10.1088/1126-6708/2000/02/032}{\emph{JHEP} {\bfseries
  02} (2000) 032} [\href{https://arxiv.org/abs/hep-ph/0002221}{{\ttfamily
  hep-ph/0002221}}].

\bibitem{Lisi:2000zt}
E.~Lisi, A.~Marrone and D.~Montanino, \emph{{Probing possible decoherence
  effects in atmospheric neutrino oscillations}},
  \href{https://doi.org/10.1103/PhysRevLett.85.1166}{\emph{Phys. Rev. Lett.}
  {\bfseries 85} (2000) 1166}
  [\href{https://arxiv.org/abs/hep-ph/0002053}{{\ttfamily hep-ph/0002053}}].

\bibitem{Super-Kamiokande:2004orf}
{\scshape Super-Kamiokande} collaboration, \emph{{Evidence for an oscillatory
  signature in atmospheric neutrino oscillation}},
  \href{https://doi.org/10.1103/PhysRevLett.93.101801}{\emph{Phys. Rev. Lett.}
  {\bfseries 93} (2004) 101801}
  [\href{https://arxiv.org/abs/hep-ex/0404034}{{\ttfamily hep-ex/0404034}}].

\bibitem{BalieiroGomes:2016ykp}
G.~Balieiro~Gomes, M.M.~Guzzo, P.C.~de~Holanda and R.L.N.~Oliveira,
  \emph{{Parameter Limits for Neutrino Oscillation with Decoherence in
  KamLAND}}, \href{https://doi.org/10.1103/PhysRevD.95.113005}{\emph{Phys. Rev.
  D} {\bfseries 95} (2017) 113005}
  [\href{https://arxiv.org/abs/1603.04126}{{\ttfamily 1603.04126}}].

\bibitem{BalieiroGomes:2018gtd}
G.~Balieiro~Gomes, D.V.~Forero, M.M.~Guzzo, P.C.~De~Holanda and
  R.L.N.~Oliveira, \emph{{Quantum Decoherence Effects in Neutrino Oscillations
  at DUNE}}, \href{https://doi.org/10.1103/PhysRevD.100.055023}{\emph{Phys.
  Rev. D} {\bfseries 100} (2019) 055023}
  [\href{https://arxiv.org/abs/1805.09818}{{\ttfamily 1805.09818}}].

\bibitem{Stuttard:2021uyw}
T.~Stuttard, \emph{{Neutrino signals of lightcone fluctuations resulting from
  fluctuating spacetime}},
  \href{https://doi.org/10.1103/PhysRevD.104.056007}{\emph{Phys. Rev. D}
  {\bfseries 104} (2021) 056007}
  [\href{https://arxiv.org/abs/2103.15313}{{\ttfamily 2103.15313}}].

\bibitem{Hellmann:2021jyz}
D.~Hellmann, H.~P\"as and E.~Rani, \emph{{Searching new particles at neutrino
  telescopes with quantum-gravitational decoherence}},
  \href{https://doi.org/10.1103/PhysRevD.105.055007}{\emph{Phys. Rev. D}
  {\bfseries 105} (2022) 055007}
  [\href{https://arxiv.org/abs/2103.11984}{{\ttfamily 2103.11984}}].

\bibitem{Blennow:2016jkn}
M.~Blennow, P.~Coloma, E.~Fernandez-Martinez, J.~Hernandez-Garcia and
  J.~Lopez-Pavon, \emph{{Non-Unitarity, sterile neutrinos, and Non-Standard
  neutrino Interactions}},
  \href{https://doi.org/10.1007/JHEP04(2017)153}{\emph{JHEP} {\bfseries 04}
  (2017) 153} [\href{https://arxiv.org/abs/1609.08637}{{\ttfamily
  1609.08637}}].

\bibitem{Parke:2015goa}
S.~Parke and M.~Ross-Lonergan, \emph{{Unitarity and the three flavor neutrino
  mixing matrix}},
  \href{https://doi.org/10.1103/PhysRevD.93.113009}{\emph{Phys. Rev. D}
  {\bfseries 93} (2016) 113009}
  [\href{https://arxiv.org/abs/1508.05095}{{\ttfamily 1508.05095}}].

\bibitem{Denton:2021mso}
P.B.~Denton and J.~Gehrlein, \emph{{New oscillation and scattering constraints
  on the tau row matrix elements without assuming unitarity}},
  \href{https://doi.org/10.1007/JHEP06(2022)135}{\emph{JHEP} {\bfseries 06}
  (2022) 135} [\href{https://arxiv.org/abs/2109.14575}{{\ttfamily
  2109.14575}}].

\bibitem{Wolfenstein:1977ue}
L.~Wolfenstein, \emph{{Neutrino Oscillations in Matter}},
  \href{https://doi.org/10.1103/PhysRevD.17.2369}{\emph{Phys. Rev. D}
  {\bfseries 17} (1978) 2369}.

\bibitem{deGouvea:2015ndi}
A.~de~Gouvêa and K.J.~Kelly, \emph{{Non-standard Neutrino Interactions at
  DUNE}}, \href{https://doi.org/10.1016/j.nuclphysb.2016.03.013}{\emph{Nucl.
  Phys. B} {\bfseries 908} (2016) 318}
  [\href{https://arxiv.org/abs/1511.05562}{{\ttfamily 1511.05562}}].

\bibitem{Proceedings:2019qno}
\emph{{Neutrino Non-Standard Interactions: A Status Report}}, vol.~2, 2019.
\newblock 10.21468/SciPostPhysProc.2.001.

\bibitem{Bakhti:2020fde}
P.~Bakhti and M.~Rajaee, \emph{{Sensitivities of future reactor and
  long-baseline neutrino experiments to NSI}},
  \href{https://doi.org/10.1103/PhysRevD.103.075003}{\emph{Phys. Rev. D}
  {\bfseries 103} (2021) 075003}
  [\href{https://arxiv.org/abs/2010.12849}{{\ttfamily 2010.12849}}].

\bibitem{DUNE:2020fgq}
{\scshape DUNE} collaboration, \emph{{Prospects for beyond the Standard Model
  physics searches at the Deep Underground Neutrino Experiment}},
  \href{https://doi.org/10.1140/epjc/s10052-021-09007-w}{\emph{Eur. Phys. J. C}
  {\bfseries 81} (2021) 322}
  [\href{https://arxiv.org/abs/2008.12769}{{\ttfamily 2008.12769}}].

\bibitem{Chatterjee:2021wac}
S.S.~Chatterjee, P.S.B.~Dev and P.A.N.~Machado, \emph{{Impact of improved
  energy resolution on DUNE sensitivity to neutrino non-standard
  interactions}}, \href{https://doi.org/10.1007/JHEP08(2021)163}{\emph{JHEP}
  {\bfseries 08} (2021) 163}
  [\href{https://arxiv.org/abs/2106.04597}{{\ttfamily 2106.04597}}].

\bibitem{Giarnetti:2021wur}
A.~Giarnetti and D.~Meloni, \emph{{New Sources of Leptonic CP Violation at the
  DUNE Neutrino Experiment}},
  \href{https://doi.org/10.3390/universe7070240}{\emph{Universe} {\bfseries 7}
  (2021) 240} [\href{https://arxiv.org/abs/2106.00030}{{\ttfamily
  2106.00030}}].

\bibitem{Donini:2007yf}
A.~Donini, M.~Maltoni, D.~Meloni, P.~Migliozzi and F.~Terranova, \emph{{3+1
  sterile neutrinos at the CNGS}},
  \href{https://doi.org/10.1088/1126-6708/2007/12/013}{\emph{JHEP} {\bfseries
  12} (2007) 013} [\href{https://arxiv.org/abs/0704.0388}{{\ttfamily
  0704.0388}}].

\bibitem{Dighe:2007uf}
A.~Dighe and S.~Ray, \emph{{Signatures of heavy sterile neutrinos at long
  baseline experiments}},
  \href{https://doi.org/10.1103/PhysRevD.76.113001}{\emph{Phys. Rev. D}
  {\bfseries 76} (2007) 113001}
  [\href{https://arxiv.org/abs/0709.0383}{{\ttfamily 0709.0383}}].

\bibitem{Berryman:2015nua}
J.M.~Berryman, A.~de~Gouv\^ea, K.J.~Kelly and A.~Kobach, \emph{{Sterile
  neutrino at the Deep Underground Neutrino Experiment}},
  \href{https://doi.org/10.1103/PhysRevD.92.073012}{\emph{Phys. Rev. D}
  {\bfseries 92} (2015) 073012}
  [\href{https://arxiv.org/abs/1507.03986}{{\ttfamily 1507.03986}}].

\bibitem{Acero:2022wqg}
M.A.~Acero et~al., \emph{{White Paper on Light Sterile Neutrino Searches and
  Related Phenomenology}},  \href{https://arxiv.org/abs/2203.07323}{{\ttfamily
  2203.07323}}.

\bibitem{Ge:2018uhz}
S.-F.~Ge and S.J.~Parke, \emph{{Scalar Nonstandard Interactions in Neutrino
  Oscillation}},
  \href{https://doi.org/10.1103/PhysRevLett.122.211801}{\emph{Phys. Rev. Lett.}
  {\bfseries 122} (2019) 211801}
  [\href{https://arxiv.org/abs/1812.08376}{{\ttfamily 1812.08376}}].

\bibitem{LSND:2001aii}
{\scshape LSND} collaboration, \emph{{Evidence for neutrino oscillations from
  the observation of $\bar{\nu}_e$ appearance in a $\bar{\nu}_\mu$ beam}},
  \href{https://doi.org/10.1103/PhysRevD.64.112007}{\emph{Phys. Rev. D}
  {\bfseries 64} (2001) 112007}
  [\href{https://arxiv.org/abs/hep-ex/0104049}{{\ttfamily hep-ex/0104049}}].

\bibitem{T2K:2014xvp}
{\scshape T2K} collaboration, \emph{{Search for short baseline $\nu_e$
  disappearance with the T2K near detector}},
  \href{https://doi.org/10.1103/PhysRevD.91.051102}{\emph{Phys. Rev. D}
  {\bfseries 91} (2015) 051102}
  [\href{https://arxiv.org/abs/1410.8811}{{\ttfamily 1410.8811}}].

\bibitem{Giunti:2010zu}
C.~Giunti and M.~Laveder, \emph{{Statistical Significance of the Gallium
  Anomaly}}, \href{https://doi.org/10.1103/PhysRevC.83.065504}{\emph{Phys. Rev.
  C} {\bfseries 83} (2011) 065504}
  [\href{https://arxiv.org/abs/1006.3244}{{\ttfamily 1006.3244}}].

\bibitem{Mention:2011rk}
G.~Mention, M.~Fechner, T.~Lasserre, T.A.~Mueller, D.~Lhuillier, M.~Cribier
  et~al., \emph{{The Reactor Antineutrino Anomaly}},
  \href{https://doi.org/10.1103/PhysRevD.83.073006}{\emph{Phys. Rev. D}
  {\bfseries 83} (2011) 073006}
  [\href{https://arxiv.org/abs/1101.2755}{{\ttfamily 1101.2755}}].

\bibitem{Barinov:2021asz}
V.V.~Barinov et~al., \emph{{Results from the Baksan Experiment on Sterile
  Transitions (BEST)}},
  \href{https://doi.org/10.1103/PhysRevLett.128.232501}{\emph{Phys. Rev. Lett.}
  {\bfseries 128} (2022) 232501}
  [\href{https://arxiv.org/abs/2109.11482}{{\ttfamily 2109.11482}}].

\bibitem{MiniBooNE:2020pnu}
{\scshape MiniBooNE} collaboration, \emph{{Updated MiniBooNE neutrino
  oscillation results with increased data and new background studies}},
  \href{https://doi.org/10.1103/PhysRevD.103.052002}{\emph{Phys. Rev. D}
  {\bfseries 103} (2021) 052002}
  [\href{https://arxiv.org/abs/2006.16883}{{\ttfamily 2006.16883}}].

\bibitem{Denton:2021czb}
P.B.~Denton, \emph{{Sterile Neutrino Search with MicroBooNE\textquoteright{}s
  Electron Neutrino Disappearance Data}},
  \href{https://doi.org/10.1103/PhysRevLett.129.061801}{\emph{Phys. Rev. Lett.}
  {\bfseries 129} (2022) 061801}
  [\href{https://arxiv.org/abs/2111.05793}{{\ttfamily 2111.05793}}].

\bibitem{Pontecorvo:1957cp}
B.~Pontecorvo, \emph{{Mesonium and anti-mesonium}}, {\emph{Sov. Phys. JETP}
  {\bfseries 6} (1957) 429}.

\bibitem{Maki:1962mu}
Z.~Maki, M.~Nakagawa and S.~Sakata, \emph{{Remarks on the unified model of
  elementary particles}}, \href{https://doi.org/10.1143/PTP.28.870}{\emph{Prog.
  Theor. Phys.} {\bfseries 28} (1962) 870}.

\bibitem{Coloma:2015kiu}
P.~Coloma, \emph{{Non-Standard Interactions in propagation at the Deep
  Underground Neutrino Experiment}},
  \href{https://doi.org/10.1007/JHEP03(2016)016}{\emph{JHEP} {\bfseries 03}
  (2016) 016} [\href{https://arxiv.org/abs/1511.06357}{{\ttfamily
  1511.06357}}].

\bibitem{Esteban:2020cvm}
I.~Esteban, M.C.~Gonzalez-Garcia, M.~Maltoni, T.~Schwetz and A.~Zhou,
  \emph{{The fate of hints: updated global analysis of three-flavor neutrino
  oscillations}}, \href{https://doi.org/10.1007/JHEP09(2020)178}{\emph{JHEP}
  {\bfseries 09} (2020) 178}
  [\href{https://arxiv.org/abs/2007.14792}{{\ttfamily 2007.14792}}].

\bibitem{Coloma:2017egw}
P.~Coloma, P.B.~Denton, M.~Gonzalez-Garcia, M.~Maltoni and T.~Schwetz,
  \emph{{Curtailing the Dark Side in Non-Standard Neutrino Interactions}},
  \href{https://doi.org/10.1007/JHEP04(2017)116}{\emph{JHEP} {\bfseries 04}
  (2017) 116} [\href{https://arxiv.org/abs/1701.04828}{{\ttfamily
  1701.04828}}].

\bibitem{Liao:2017uzy}
J.~Liao and D.~Marfatia, \emph{{COHERENT constraints on nonstandard neutrino
  interactions}},
  \href{https://doi.org/10.1016/j.physletb.2017.10.046}{\emph{Phys. Lett. B}
  {\bfseries 775} (2017) 54}
  [\href{https://arxiv.org/abs/1708.04255}{{\ttfamily 1708.04255}}].

\bibitem{Farzan:2018gtr}
Y.~Farzan, M.~Lindner, W.~Rodejohann and X.-J.~Xu, \emph{{Probing neutrino
  coupling to a light scalar with coherent neutrino scattering}},
  \href{https://doi.org/10.1007/JHEP05(2018)066}{\emph{JHEP} {\bfseries 05}
  (2018) 066} [\href{https://arxiv.org/abs/1802.05171}{{\ttfamily
  1802.05171}}].

\bibitem{Denton:2018xmq}
P.B.~Denton, Y.~Farzan and I.M.~Shoemaker, \emph{{Testing large non-standard
  neutrino interactions with arbitrary mediator mass after COHERENT data}},
  \href{https://doi.org/10.1007/JHEP07(2018)037}{\emph{JHEP} {\bfseries 07}
  (2018) 037} [\href{https://arxiv.org/abs/1804.03660}{{\ttfamily
  1804.03660}}].

\bibitem{Denton:2020hop}
P.B.~Denton and J.~Gehrlein, \emph{{A Statistical Analysis of the COHERENT Data
  and Applications to New Physics}},
  \href{https://doi.org/10.1007/JHEP04(2021)266}{\emph{JHEP} {\bfseries 04}
  (2021) 266} [\href{https://arxiv.org/abs/2008.06062}{{\ttfamily
  2008.06062}}].

\bibitem{Denton:2022nol}
P.B.~Denton and J.~Gehrlein, \emph{{New reactor data improves robustness of
  neutrino mass ordering determination}},
  \href{https://doi.org/10.1103/PhysRevD.106.015022}{\emph{Phys. Rev. D}
  {\bfseries 106} (2022) 015022}
  [\href{https://arxiv.org/abs/2204.09060}{{\ttfamily 2204.09060}}].

\bibitem{Joshipura:2003jh}
A.S.~Joshipura and S.~Mohanty, \emph{{Constraints on flavor dependent long
  range forces from atmospheric neutrino observations at super-Kamiokande}},
  \href{https://doi.org/10.1016/j.physletb.2004.01.057}{\emph{Phys. Lett. B}
  {\bfseries 584} (2004) 103}
  [\href{https://arxiv.org/abs/hep-ph/0310210}{{\ttfamily hep-ph/0310210}}].

\bibitem{Grifols:2003gy}
J.A.~Grifols and E.~Masso, \emph{{Neutrino oscillations in the sun probe long
  range leptonic forces}},
  \href{https://doi.org/10.1016/j.physletb.2003.10.078}{\emph{Phys. Lett. B}
  {\bfseries 579} (2004) 123}
  [\href{https://arxiv.org/abs/hep-ph/0311141}{{\ttfamily hep-ph/0311141}}].

\bibitem{Gonzalez-Garcia:2006vic}
M.C.~Gonzalez-Garcia, P.C.~de~Holanda, E.~Masso and R.~Zukanovich~Funchal,
  \emph{{Probing long-range leptonic forces with solar and reactor neutrinos}},
  \href{https://doi.org/10.1088/1475-7516/2007/01/005}{\emph{JCAP} {\bfseries
  01} (2007) 005} [\href{https://arxiv.org/abs/hep-ph/0609094}{{\ttfamily
  hep-ph/0609094}}].

\bibitem{Bandyopadhyay:2006uh}
A.~Bandyopadhyay, A.~Dighe and A.S.~Joshipura, \emph{{Constraints on
  flavor-dependent long range forces from solar neutrinos and KamLAND}},
  \href{https://doi.org/10.1103/PhysRevD.75.093005}{\emph{Phys. Rev. D}
  {\bfseries 75} (2007) 093005}
  [\href{https://arxiv.org/abs/hep-ph/0610263}{{\ttfamily hep-ph/0610263}}].

\bibitem{Samanta:2010zh}
A.~Samanta, \emph{{Long-range Forces : Atmospheric Neutrino Oscillation at a
  magnetized Detector}},
  \href{https://doi.org/10.1088/1475-7516/2011/09/010}{\emph{JCAP} {\bfseries
  09} (2011) 010} [\href{https://arxiv.org/abs/1001.5344}{{\ttfamily
  1001.5344}}].

\bibitem{Davoudiasl:2011sz}
H.~Davoudiasl, H.-S.~Lee and W.J.~Marciano, \emph{{Long-Range Lepton Flavor
  Interactions and Neutrino Oscillations}},
  \href{https://doi.org/10.1103/PhysRevD.84.013009}{\emph{Phys. Rev. D}
  {\bfseries 84} (2011) 013009}
  [\href{https://arxiv.org/abs/1102.5352}{{\ttfamily 1102.5352}}].

\bibitem{Wise:2018rnb}
M.B.~Wise and Y.~Zhang, \emph{{Lepton Flavorful Fifth Force and Depth-dependent
  Neutrino Matter Interactions}},
  \href{https://doi.org/10.1007/JHEP06(2018)053}{\emph{JHEP} {\bfseries 06}
  (2018) 053} [\href{https://arxiv.org/abs/1803.00591}{{\ttfamily
  1803.00591}}].

\bibitem{Smirnov:2019cae}
A.Y.~Smirnov and X.-J.~Xu, \emph{{Wolfenstein potentials for neutrinos induced
  by ultra-light mediators}},
  \href{https://doi.org/10.1007/JHEP12(2019)046}{\emph{JHEP} {\bfseries 12}
  (2019) 046} [\href{https://arxiv.org/abs/1909.07505}{{\ttfamily
  1909.07505}}].

\bibitem{Coloma:2020gfv}
P.~Coloma, M.C.~Gonzalez-Garcia and M.~Maltoni, \emph{{Neutrino oscillation
  constraints on U(1)' models: from non-standard interactions to long-range
  forces}}, \href{https://doi.org/10.1007/JHEP01(2021)114}{\emph{JHEP}
  {\bfseries 01} (2021) 114}
  [\href{https://arxiv.org/abs/2009.14220}{{\ttfamily 2009.14220}}].

\bibitem{Miranda:2004nb}
O.~Miranda, M.~Tortola and J.~Valle, \emph{{Are solar neutrino oscillations
  robust?}}, \href{https://doi.org/10.1088/1126-6708/2006/10/008}{\emph{JHEP}
  {\bfseries 10} (2006) 008}
  [\href{https://arxiv.org/abs/hep-ph/0406280}{{\ttfamily hep-ph/0406280}}].

\bibitem{Escrihuela:2009up}
F.J.~Escrihuela, O.G.~Miranda, M.A.~Tortola and J.W.F.~Valle,
  \emph{{Constraining nonstandard neutrino-quark interactions with solar,
  reactor and accelerator data}},
  \href{https://doi.org/10.1103/PhysRevD.80.129908}{\emph{Phys. Rev. D}
  {\bfseries 80} (2009) 105009}
  [\href{https://arxiv.org/abs/0907.2630}{{\ttfamily 0907.2630}}].

\bibitem{Gonzalez-Garcia:2013usa}
M.C.~Gonzalez-Garcia and M.~Maltoni, \emph{{Determination of matter potential
  from global analysis of neutrino oscillation data}},
  \href{https://doi.org/10.1007/JHEP09(2013)152}{\emph{JHEP} {\bfseries 09}
  (2013) 152} [\href{https://arxiv.org/abs/1307.3092}{{\ttfamily 1307.3092}}].

\bibitem{Bakhti:2014pva}
P.~Bakhti and Y.~Farzan, \emph{{Shedding light on LMA-Dark solar neutrino
  solution by medium baseline reactor experiments: JUNO and RENO-50}},
  \href{https://doi.org/10.1007/JHEP07(2014)064}{\emph{JHEP} {\bfseries 07}
  (2014) 064} [\href{https://arxiv.org/abs/1403.0744}{{\ttfamily 1403.0744}}].

\bibitem{Coloma:2016gei}
P.~Coloma and T.~Schwetz, \emph{{Generalized mass ordering degeneracy in
  neutrino oscillation experiments}},
  \href{https://doi.org/10.1103/PhysRevD.94.055005}{\emph{Phys. Rev. D}
  {\bfseries 94} (2016) 055005}
  [\href{https://arxiv.org/abs/1604.05772}{{\ttfamily 1604.05772}}].

\bibitem{Farzan:2017xzy}
Y.~Farzan and M.~Tortola, \emph{{Neutrino oscillations and Non-Standard
  Interactions}}, \href{https://doi.org/10.3389/fphy.2018.00010}{\emph{Front.
  in Phys.} {\bfseries 6} (2018) 10}
  [\href{https://arxiv.org/abs/1710.09360}{{\ttfamily 1710.09360}}].

\bibitem{Denton:2021vtf}
P.B.~Denton and S.J.~Parke, \emph{{Parameter symmetries of neutrino
  oscillations in vacuum, matter, and approximation schemes}},
  \href{https://doi.org/10.1103/PhysRevD.105.013002}{\emph{Phys. Rev. D}
  {\bfseries 105} (2022) 013002}
  [\href{https://arxiv.org/abs/2106.12436}{{\ttfamily 2106.12436}}].

\bibitem{Denton:2020exu}
P.B.~Denton, \emph{{A Return To Neutrino Normalcy}},
  \href{https://arxiv.org/abs/2003.04319}{{\ttfamily 2003.04319}}.

\bibitem{Esteban:2018ppq}
I.~Esteban, M.C.~Gonzalez-Garcia, M.~Maltoni, I.~Martinez-Soler and J.~Salvado,
  \emph{{Updated constraints on non-standard interactions from global analysis
  of oscillation data}},
  \href{https://doi.org/10.1007/JHEP08(2018)180}{\emph{JHEP} {\bfseries 08}
  (2018) 180} [\href{https://arxiv.org/abs/1805.04530}{{\ttfamily
  1805.04530}}].

\bibitem{Forero:2016ghr}
D.V.~Forero and W.-C.~Huang, \emph{{Sizable NSI from the SU(2)$_{L}$ scalar
  doublet-singlet mixing and the implications in DUNE}},
  \href{https://doi.org/10.1007/JHEP03(2017)018}{\emph{JHEP} {\bfseries 03}
  (2017) 018} [\href{https://arxiv.org/abs/1608.04719}{{\ttfamily
  1608.04719}}].

\bibitem{Denton:2018dqq}
P.B.~Denton, Y.~Farzan and I.M.~Shoemaker, \emph{{Activating the fourth
  neutrino of the 3+1 scheme}},
  \href{https://doi.org/10.1103/PhysRevD.99.035003}{\emph{Phys. Rev. D}
  {\bfseries 99} (2019) 035003}
  [\href{https://arxiv.org/abs/1811.01310}{{\ttfamily 1811.01310}}].

\bibitem{Dey:2018yht}
U.K.~Dey, N.~Nath and S.~Sadhukhan, \emph{{Non-Standard Neutrino Interactions
  in a Modified $\nu$2HDM}},
  \href{https://doi.org/10.1103/PhysRevD.98.055004}{\emph{Phys. Rev. D}
  {\bfseries 98} (2018) 055004}
  [\href{https://arxiv.org/abs/1804.05808}{{\ttfamily 1804.05808}}].

\bibitem{Babu:2017olk}
K.~Babu, A.~Friedland, P.~Machado and I.~Mocioiu, \emph{{Flavor Gauge Models
  Below the Fermi Scale}},
  \href{https://doi.org/10.1007/JHEP12(2017)096}{\emph{JHEP} {\bfseries 12}
  (2017) 096} [\href{https://arxiv.org/abs/1705.01822}{{\ttfamily
  1705.01822}}].

\bibitem{Farzan:2016wym}
Y.~Farzan and J.~Heeck, \emph{{Neutrinophilic nonstandard interactions}},
  \href{https://doi.org/10.1103/PhysRevD.94.053010}{\emph{Phys. Rev. D}
  {\bfseries 94} (2016) 053010}
  [\href{https://arxiv.org/abs/1607.07616}{{\ttfamily 1607.07616}}].

\bibitem{Farzan:2015hkd}
Y.~Farzan and I.M.~Shoemaker, \emph{{Lepton Flavor Violating Non-Standard
  Interactions via Light Mediators}},
  \href{https://doi.org/10.1007/JHEP07(2016)033}{\emph{JHEP} {\bfseries 07}
  (2016) 033} [\href{https://arxiv.org/abs/1512.09147}{{\ttfamily
  1512.09147}}].

\bibitem{Farzan:2015doa}
Y.~Farzan, \emph{{A model for large non-standard interactions of neutrinos
  leading to the LMA-Dark solution}},
  \href{https://doi.org/10.1016/j.physletb.2015.07.015}{\emph{Phys. Lett. B}
  {\bfseries 748} (2015) 311}
  [\href{https://arxiv.org/abs/1505.06906}{{\ttfamily 1505.06906}}].

\bibitem{Babu:2019mfe}
K.~Babu, P.B.~Dev, S.~Jana and A.~Thapa, \emph{{Non-Standard Interactions in
  Radiative Neutrino Mass Models}},
  \href{https://doi.org/10.1007/JHEP03(2020)006}{\emph{JHEP} {\bfseries 03}
  (2020) 006} [\href{https://arxiv.org/abs/1907.09498}{{\ttfamily
  1907.09498}}].

\bibitem{Babu:2019iml}
K.~Babu, G.~Chauhan and P.~Bhupal~Dev, \emph{{Neutrino Non-Standard
  Interactions via Light Scalars in the Earth, Sun, Supernovae and the Early
  Universe}}, \href{https://doi.org/10.1103/PhysRevD.101.095029}{\emph{Phys.
  Rev. D} {\bfseries 101} (2020) 095029}
  [\href{https://arxiv.org/abs/1912.13488}{{\ttfamily 1912.13488}}].

\bibitem{Venzor:2020ova}
J.~Venzor, A.~P\'erez-Lorenzana and J.~De-Santiago, \emph{{Bounds on
  neutrino-scalar nonstandard interactions from big bang nucleosynthesis}},
  \href{https://doi.org/10.1103/PhysRevD.103.043534}{\emph{Phys. Rev. D}
  {\bfseries 103} (2021) 043534}
  [\href{https://arxiv.org/abs/2009.08104}{{\ttfamily 2009.08104}}].

\bibitem{Medhi:2021wxj}
A.~Medhi, D.~Dutta and M.M.~Devi, \emph{{Exploring the effects of scalar non
  standard interactions on the CP violation sensitivity at DUNE}},
  \href{https://doi.org/10.1007/JHEP06(2022)129}{\emph{JHEP} {\bfseries 06}
  (2022) 129} [\href{https://arxiv.org/abs/2111.12943}{{\ttfamily
  2111.12943}}].

\bibitem{Medhi:2022qmu}
A.~Medhi, M.M.~Devi and D.~Dutta, \emph{{Imprints of scalar NSI on the
  CP-violation sensitivity using synergy among DUNE, T2HK and T2HKK}},
  \href{https://arxiv.org/abs/2209.05287}{{\ttfamily 2209.05287}}.

\bibitem{Sarkar:2022ujy}
T.~Sarkar, \emph{{Effect of non-unitary neutrino mixing in Lorentz violation
  and dark NSI}},  \href{https://arxiv.org/abs/2209.10233}{{\ttfamily
  2209.10233}}.

\bibitem{Dutta:2022fdt}
B.~Dutta, S.~Ghosh, T.~Li, A.~Thompson and A.~Verma, \emph{{Non-standard
  neutrino interactions in light mediator models at reactor experiments}},
  \href{https://arxiv.org/abs/2209.13566}{{\ttfamily 2209.13566}}.

\bibitem{Kreisch:2019yzn}
C.D.~Kreisch, F.-Y.~Cyr-Racine and O.~Dor\'e, \emph{{Neutrino puzzle:
  Anomalies, interactions, and cosmological tensions}},
  \href{https://doi.org/10.1103/PhysRevD.101.123505}{\emph{Phys. Rev. D}
  {\bfseries 101} (2020) 123505}
  [\href{https://arxiv.org/abs/1902.00534}{{\ttfamily 1902.00534}}].

\bibitem{Barenboim:2019tux}
G.~Barenboim, P.B.~Denton and I.M.~Oldengott, \emph{{Constraints on inflation
  with an extended neutrino sector}},
  \href{https://doi.org/10.1103/PhysRevD.99.083515}{\emph{Phys. Rev. D}
  {\bfseries 99} (2019) 083515}
  [\href{https://arxiv.org/abs/1903.02036}{{\ttfamily 1903.02036}}].

\bibitem{Fernandez-Martinez:2007iaa}
E.~Fernandez-Martinez, M.B.~Gavela, J.~Lopez-Pavon and O.~Yasuda,
  \emph{{CP-violation from non-unitary leptonic mixing}},
  \href{https://doi.org/10.1016/j.physletb.2007.03.069}{\emph{Phys. Lett. B}
  {\bfseries 649} (2007) 427}
  [\href{https://arxiv.org/abs/hep-ph/0703098}{{\ttfamily hep-ph/0703098}}].

\bibitem{Babu:2020ivd}
K.S.~Babu, S.~Jana and M.~Lindner, \emph{{Large Neutrino Magnetic Moments in
  the Light of Recent Experiments}},
  \href{https://doi.org/10.1007/JHEP10(2020)040}{\emph{JHEP} {\bfseries 10}
  (2020) 040} [\href{https://arxiv.org/abs/2007.04291}{{\ttfamily
  2007.04291}}].

\bibitem{MINOS:2017cae}
{\scshape MINOS+} collaboration, \emph{{Search for sterile neutrinos in MINOS
  and MINOS+ using a two-detector fit}},
  \href{https://doi.org/10.1103/PhysRevLett.122.091803}{\emph{Phys. Rev. Lett.}
  {\bfseries 122} (2019) 091803}
  [\href{https://arxiv.org/abs/1710.06488}{{\ttfamily 1710.06488}}].

\bibitem{IceCube:2020phf}
{\scshape IceCube} collaboration, \emph{{eV-Scale Sterile Neutrino Search Using
  Eight Years of Atmospheric Muon Neutrino Data from the IceCube Neutrino
  Observatory}},
  \href{https://doi.org/10.1103/PhysRevLett.125.141801}{\emph{Phys. Rev. Lett.}
  {\bfseries 125} (2020) 141801}
  [\href{https://arxiv.org/abs/2005.12942}{{\ttfamily 2005.12942}}].

\bibitem{Hagstotz:2020ukm}
S.~Hagstotz, P.F.~de~Salas, S.~Gariazzo, M.~Gerbino, M.~Lattanzi, S.~Vagnozzi
  et~al., \emph{{Bounds on light sterile neutrino mass and mixing from
  cosmology and laboratory searches}},
  \href{https://doi.org/10.1103/PhysRevD.104.123524}{\emph{Phys. Rev. D}
  {\bfseries 104} (2021) 123524}
  [\href{https://arxiv.org/abs/2003.02289}{{\ttfamily 2003.02289}}].

\bibitem{Dentler:2018sju}
M.~Dentler, A.~Hern\'andez-Cabezudo, J.~Kopp, P.A.N.~Machado, M.~Maltoni,
  I.~Martinez-Soler et~al., \emph{{Updated Global Analysis of Neutrino
  Oscillations in the Presence of eV-Scale Sterile Neutrinos}},
  \href{https://doi.org/10.1007/JHEP08(2018)010}{\emph{JHEP} {\bfseries 08}
  (2018) 010} [\href{https://arxiv.org/abs/1803.10661}{{\ttfamily
  1803.10661}}].

\bibitem{Diaz:2019fwt}
A.~Diaz, C.A.~Arg\"uelles, G.H.~Collin, J.M.~Conrad and M.H.~Shaevitz,
  \emph{{Where Are We With Light Sterile Neutrinos?}},
  \href{https://doi.org/10.1016/j.physrep.2020.08.005}{\emph{Phys. Rept.}
  {\bfseries 884} (2020) 1} [\href{https://arxiv.org/abs/1906.00045}{{\ttfamily
  1906.00045}}].

\bibitem{Boser:2019rta}
S.~B\"oser, C.~Buck, C.~Giunti, J.~Lesgourgues, L.~Ludhova, S.~Mertens et~al.,
  \emph{{Status of Light Sterile Neutrino Searches}},
  \href{https://doi.org/10.1016/j.ppnp.2019.103736}{\emph{Prog. Part. Nucl.
  Phys.} {\bfseries 111} (2020) 103736}
  [\href{https://arxiv.org/abs/1906.01739}{{\ttfamily 1906.01739}}].

\bibitem{Machado:2019oxb}
P.A.~Machado, O.~Palamara and D.W.~Schmitz, \emph{{The Short-Baseline Neutrino
  Program at Fermilab}},
  \href{https://doi.org/10.1146/annurev-nucl-101917-020949}{\emph{Ann. Rev.
  Nucl. Part. Sci.} {\bfseries 69} (2019) 363}
  [\href{https://arxiv.org/abs/1903.04608}{{\ttfamily 1903.04608}}].

\bibitem{Berryman:2020agd}
J.M.~Berryman and P.~Huber, \emph{{Sterile Neutrinos and the Global Reactor
  Antineutrino Dataset}},
  \href{https://doi.org/10.1007/JHEP01(2021)167}{\emph{JHEP} {\bfseries 01}
  (2021) 167} [\href{https://arxiv.org/abs/2005.01756}{{\ttfamily
  2005.01756}}].

\bibitem{Dasgupta:2021ies}
B.~Dasgupta and J.~Kopp, \emph{{Sterile Neutrinos}},
  \href{https://doi.org/10.1016/j.physrep.2021.06.002}{\emph{Phys. Rept.}
  {\bfseries 928} (2021) 1} [\href{https://arxiv.org/abs/2106.05913}{{\ttfamily
  2106.05913}}].

\bibitem{Meloni:2010zr}
D.~Meloni, J.~Tang and W.~Winter, \emph{{Sterile neutrinos beyond LSND at the
  Neutrino Factory}},
  \href{https://doi.org/10.1103/PhysRevD.82.093008}{\emph{Phys. Rev. D}
  {\bfseries 82} (2010) 093008}
  [\href{https://arxiv.org/abs/1007.2419}{{\ttfamily 1007.2419}}].

\bibitem{Bhattacharya:2011ee}
B.~Bhattacharya, A.M.~Thalapillil and C.E.M.~Wagner, \emph{{Implications of
  sterile neutrinos for medium/long-baseline neutrino experiments and the
  determination of $\theta_{13}$}},
  \href{https://doi.org/10.1103/PhysRevD.85.073004}{\emph{Phys. Rev. D}
  {\bfseries 85} (2012) 073004}
  [\href{https://arxiv.org/abs/1111.4225}{{\ttfamily 1111.4225}}].

\bibitem{Hollander:2014iha}
D.~Hollander and I.~Mocioiu, \emph{{Minimal 3+2 sterile neutrino model at
  LBNE}}, \href{https://doi.org/10.1103/PhysRevD.91.013002}{\emph{Phys. Rev. D}
  {\bfseries 91} (2015) 013002}
  [\href{https://arxiv.org/abs/1408.1749}{{\ttfamily 1408.1749}}].

\bibitem{Klop:2014ima}
N.~Klop and A.~Palazzo, \emph{{Imprints of CP violation induced by sterile
  neutrinos in T2K data}},
  \href{https://doi.org/10.1103/PhysRevD.91.073017}{\emph{Phys. Rev. D}
  {\bfseries 91} (2015) 073017}
  [\href{https://arxiv.org/abs/1412.7524}{{\ttfamily 1412.7524}}].

\bibitem{Gandhi:2015xza}
R.~Gandhi, B.~Kayser, M.~Masud and S.~Prakash, \emph{{The impact of sterile
  neutrinos on CP measurements at long baselines}},
  \href{https://doi.org/10.1007/JHEP11(2015)039}{\emph{JHEP} {\bfseries 11}
  (2015) 039} [\href{https://arxiv.org/abs/1508.06275}{{\ttfamily
  1508.06275}}].

\bibitem{Palazzo:2015gja}
A.~Palazzo, \emph{{3-flavor and 4-flavor implications of the latest T2K and
  NO$\nu$A electron (anti-)neutrino appearance results}},
  \href{https://doi.org/10.1016/j.physletb.2016.03.061}{\emph{Phys. Lett. B}
  {\bfseries 757} (2016) 142}
  [\href{https://arxiv.org/abs/1509.03148}{{\ttfamily 1509.03148}}].

\bibitem{Agarwalla:2016mrc}
S.K.~Agarwalla, S.S.~Chatterjee, A.~Dasgupta and A.~Palazzo, \emph{{Discovery
  Potential of T2K and NOvA in the Presence of a Light Sterile Neutrino}},
  \href{https://doi.org/10.1007/JHEP02(2016)111}{\emph{JHEP} {\bfseries 02}
  (2016) 111} [\href{https://arxiv.org/abs/1601.05995}{{\ttfamily
  1601.05995}}].

\bibitem{Agarwalla:2016xxa}
S.K.~Agarwalla, S.S.~Chatterjee and A.~Palazzo, \emph{{Physics Reach of DUNE
  with a Light Sterile Neutrino}},
  \href{https://doi.org/10.1007/JHEP09(2016)016}{\emph{JHEP} {\bfseries 09}
  (2016) 016} [\href{https://arxiv.org/abs/1603.03759}{{\ttfamily
  1603.03759}}].

\bibitem{Dutta:2016glq}
D.~Dutta, R.~Gandhi, B.~Kayser, M.~Masud and S.~Prakash, \emph{{Capabilities of
  long-baseline experiments in the presence of a sterile neutrino}},
  \href{https://doi.org/10.1007/JHEP11(2016)122}{\emph{JHEP} {\bfseries 11}
  (2016) 122} [\href{https://arxiv.org/abs/1607.02152}{{\ttfamily
  1607.02152}}].

\bibitem{Kelly:2017kch}
K.J.~Kelly, \emph{{Searches for new physics at the Hyper-Kamiokande
  experiment}}, \href{https://doi.org/10.1103/PhysRevD.95.115009}{\emph{Phys.
  Rev. D} {\bfseries 95} (2017) 115009}
  [\href{https://arxiv.org/abs/1703.00448}{{\ttfamily 1703.00448}}].

\bibitem{Coloma:2017ptb}
P.~Coloma, D.V.~Forero and S.J.~Parke, \emph{{DUNE Sensitivities to the Mixing
  between Sterile and Tau Neutrinos}},
  \href{https://doi.org/10.1007/JHEP07(2018)079}{\emph{JHEP} {\bfseries 07}
  (2018) 079} [\href{https://arxiv.org/abs/1707.05348}{{\ttfamily
  1707.05348}}].

\bibitem{Ghosh:2017atj}
M.~Ghosh, S.~Gupta, Z.M.~Matthews, P.~Sharma and A.G.~Williams, \emph{{Study of
  parameter degeneracy and hierarchy sensitivity of NO$\nu$A in presence of
  sterile neutrino}},
  \href{https://doi.org/10.1103/PhysRevD.96.075018}{\emph{Phys. Rev. D}
  {\bfseries 96} (2017) 075018}
  [\href{https://arxiv.org/abs/1704.04771}{{\ttfamily 1704.04771}}].

\bibitem{Choubey:2017cba}
S.~Choubey, D.~Dutta and D.~Pramanik, \emph{{Imprints of a light Sterile
  Neutrino at DUNE, T2HK and T2HKK}},
  \href{https://doi.org/10.1103/PhysRevD.96.056026}{\emph{Phys. Rev. D}
  {\bfseries 96} (2017) 056026}
  [\href{https://arxiv.org/abs/1704.07269}{{\ttfamily 1704.07269}}].

\bibitem{Choubey:2017ppj}
S.~Choubey, D.~Dutta and D.~Pramanik, \emph{{Measuring the Sterile Neutrino CP
  Phase at DUNE and T2HK}},
  \href{https://doi.org/10.1140/epjc/s10052-018-5816-y}{\emph{Eur. Phys. J. C}
  {\bfseries 78} (2018) 339}
  [\href{https://arxiv.org/abs/1711.07464}{{\ttfamily 1711.07464}}].

\bibitem{Agarwalla:2018nlx}
S.K.~Agarwalla, S.S.~Chatterjee and A.~Palazzo, \emph{{Signatures of a Light
  Sterile Neutrino in T2HK}},
  \href{https://doi.org/10.1007/JHEP04(2018)091}{\emph{JHEP} {\bfseries 04}
  (2018) 091} [\href{https://arxiv.org/abs/1801.04855}{{\ttfamily
  1801.04855}}].

\bibitem{Gupta:2018qsv}
S.~Gupta, Z.M.~Matthews, P.~Sharma and A.G.~Williams, \emph{{The Effect of a
  Light Sterile Neutrino at NO$\nu$A and DUNE}},
  \href{https://doi.org/10.1103/PhysRevD.98.035042}{\emph{Phys. Rev. D}
  {\bfseries 98} (2018) 035042}
  [\href{https://arxiv.org/abs/1804.03361}{{\ttfamily 1804.03361}}].

\bibitem{KumarAgarwalla:2019blx}
S.~Kumar~Agarwalla, S.S.~Chatterjee and A.~Palazzo, \emph{{Physics potential of
  ESS$\nu$SB in the presence of a light sterile neutrino}},
  \href{https://doi.org/10.1007/JHEP12(2019)174}{\emph{JHEP} {\bfseries 12}
  (2019) 174} [\href{https://arxiv.org/abs/1909.13746}{{\ttfamily
  1909.13746}}].

\bibitem{Majhi:2019hdj}
R.~Majhi, C.~Soumya and R.~Mohanta, \emph{{Light sterile neutrinos and their
  implications on currently running long-baseline and neutrinoless double beta
  decay experiments}}, \href{https://doi.org/10.1088/1361-6471/ab9797}{\emph{J.
  Phys. G} {\bfseries 47} (2020) 095002}
  [\href{https://arxiv.org/abs/1911.10952}{{\ttfamily 1911.10952}}].

\bibitem{Reyimuaji:2019wbn}
Y.~Reyimuaji and C.~Liu, \emph{{Prospects of light sterile neutrino searches in
  long-baseline neutrino oscillations}},
  \href{https://doi.org/10.1007/JHEP06(2020)094}{\emph{JHEP} {\bfseries 06}
  (2020) 094} [\href{https://arxiv.org/abs/1911.12524}{{\ttfamily
  1911.12524}}].

\bibitem{Ghosh:2019zvl}
M.~Ghosh, T.~Ohlsson and S.~Rosauro-Alcaraz, \emph{{Sensitivity to light
  sterile neutrinos at ESSnuSB}},
  \href{https://doi.org/10.1007/JHEP03(2020)026}{\emph{JHEP} {\bfseries 03}
  (2020) 026} [\href{https://arxiv.org/abs/1912.10010}{{\ttfamily
  1912.10010}}].

\bibitem{Penedo:2022etl}
J.T.~Penedo and J.a.~Pulido, \emph{{Baseline and other effects for a sterile
  neutrino at DUNE}},  \href{https://arxiv.org/abs/2207.02331}{{\ttfamily
  2207.02331}}.

\bibitem{Rodejohann:2011vc}
W.~Rodejohann and J.W.F.~Valle, \emph{{Symmetrical Parametrizations of the
  Lepton Mixing Matrix}},
  \href{https://doi.org/10.1103/PhysRevD.84.073011}{\emph{Phys. Rev. D}
  {\bfseries 84} (2011) 073011}
  [\href{https://arxiv.org/abs/1108.3484}{{\ttfamily 1108.3484}}].

\bibitem{IceCubeCollaboration:2021euf}
{\scshape (IceCube Collaboration)*, IceCube} collaboration, \emph{{All-flavor
  constraints on nonstandard neutrino interactions and generalized matter
  potential with three years of IceCube DeepCore data}},
  \href{https://doi.org/10.1103/PhysRevD.104.072006}{\emph{Phys. Rev. D}
  {\bfseries 104} (2021) 072006}
  [\href{https://arxiv.org/abs/2106.07755}{{\ttfamily 2106.07755}}].

\bibitem{Kelly:2020fkv}
K.J.~Kelly, P.A.N.~Machado, S.J.~Parke, Y.F.~Perez-Gonzalez and R.Z.~Funchal,
  \emph{{Neutrino mass ordering in light of recent data}},
  \href{https://doi.org/10.1103/PhysRevD.103.013004}{\emph{Phys. Rev. D}
  {\bfseries 103} (2021) 013004}
  [\href{https://arxiv.org/abs/2007.08526}{{\ttfamily 2007.08526}}].

\bibitem{Goldhagen:2021kxe}
K.~Goldhagen, M.~Maltoni, S.E.~Reichard and T.~Schwetz, \emph{{Testing sterile
  neutrino mixing with present and future solar neutrino data}},
  \href{https://doi.org/10.1140/epjc/s10052-022-10052-2}{\emph{Eur. Phys. J. C}
  {\bfseries 82} (2022) 116}
  [\href{https://arxiv.org/abs/2109.14898}{{\ttfamily 2109.14898}}].

\bibitem{DUNE:2020ypp}
{\scshape DUNE} collaboration, \emph{{Deep Underground Neutrino Experiment
  (DUNE), Far Detector Technical Design Report, Volume II: DUNE Physics}},
  \href{https://arxiv.org/abs/2002.03005}{{\ttfamily 2002.03005}}.

\bibitem{Huber:2004ka}
P.~Huber, M.~Lindner and W.~Winter, \emph{{Simulation of long-baseline neutrino
  oscillation experiments with GLoBES (General Long Baseline Experiment
  Simulator)}}, \href{https://doi.org/10.1016/j.cpc.2005.01.003}{\emph{Comput.
  Phys. Commun.} {\bfseries 167} (2005) 195}
  [\href{https://arxiv.org/abs/hep-ph/0407333}{{\ttfamily hep-ph/0407333}}].

\bibitem{DUNE:2021cuw}
{\scshape DUNE} collaboration, \emph{{Experiment Simulation Configurations
  Approximating DUNE TDR}},  \href{https://arxiv.org/abs/2103.04797}{{\ttfamily
  2103.04797}}.

\bibitem{Fogli:2002pt}
G.L.~Fogli, E.~Lisi, A.~Marrone, D.~Montanino and A.~Palazzo, \emph{{Getting
  the most from the statistical analysis of solar neutrino oscillations}},
  \href{https://doi.org/10.1103/PhysRevD.66.053010}{\emph{Phys. Rev. D}
  {\bfseries 66} (2002) 053010}
  [\href{https://arxiv.org/abs/hep-ph/0206162}{{\ttfamily hep-ph/0206162}}].

\bibitem{deSalas:2020pgw}
P.F.~de~Salas, D.V.~Forero, S.~Gariazzo, P.~Mart\'\i{}nez-Mirav\'e, O.~Mena,
  C.A.~Ternes et~al., \emph{{2020 global reassessment of the neutrino
  oscillation picture}},
  \href{https://doi.org/10.1007/JHEP02(2021)071}{\emph{JHEP} {\bfseries 02}
  (2021) 071} [\href{https://arxiv.org/abs/2006.11237}{{\ttfamily
  2006.11237}}].

\bibitem{Capozzi:2021fjo}
F.~Capozzi, E.~Di~Valentino, E.~Lisi, A.~Marrone, A.~Melchiorri and A.~Palazzo,
  \emph{{The unfinished fabric of the three neutrino paradigm}},
  \href{https://arxiv.org/abs/2107.00532}{{\ttfamily 2107.00532}}.

\bibitem{DUNE:2021tad}
{\scshape DUNE} collaboration, \emph{{Deep Underground Neutrino Experiment
  (DUNE) Near Detector Conceptual Design Report}},
  \href{https://doi.org/10.3390/instruments5040031}{\emph{Instruments}
  {\bfseries 5} (2021) 31} [\href{https://arxiv.org/abs/2103.13910}{{\ttfamily
  2103.13910}}].

\bibitem{Super-Kamiokande:2011dam}
{\scshape Super-Kamiokande} collaboration, \emph{{Study of Non-Standard
  Neutrino Interactions with Atmospheric Neutrino Data in Super-Kamiokande I
  and II}}, \href{https://doi.org/10.1103/PhysRevD.84.113008}{\emph{Phys. Rev.
  D} {\bfseries 84} (2011) 113008}
  [\href{https://arxiv.org/abs/1109.1889}{{\ttfamily 1109.1889}}].

\bibitem{IceCube:2022ubv}
{\scshape IceCube} collaboration, \emph{{Strong Constraints on Neutrino
  Nonstandard Interactions from TeV-Scale $\nu_u$ Disappearance at IceCube}},
  \href{https://doi.org/10.1103/PhysRevLett.129.011804}{\emph{Phys. Rev. Lett.}
  {\bfseries 129} (2022) 011804}
  [\href{https://arxiv.org/abs/2201.03566}{{\ttfamily 2201.03566}}].

\bibitem{Esteban:2019lfo}
I.~Esteban, M.C.~Gonzalez-Garcia and M.~Maltoni, \emph{{On the Determination of
  Leptonic CP Violation and Neutrino Mass Ordering in Presence of Non-Standard
  Interactions: Present Status}},
  \href{https://doi.org/10.1007/JHEP06(2019)055}{\emph{JHEP} {\bfseries 06}
  (2019) 055} [\href{https://arxiv.org/abs/1905.05203}{{\ttfamily
  1905.05203}}].

\bibitem{Masud:2018pig}
M.~Masud, S.~Roy and P.~Mehta, \emph{{Correlations and degeneracies among the
  NSI parameters with tunable beams at DUNE}},
  \href{https://doi.org/10.1103/PhysRevD.99.115032}{\emph{Phys. Rev. D}
  {\bfseries 99} (2019) 115032}
  [\href{https://arxiv.org/abs/1812.10290}{{\ttfamily 1812.10290}}].

\bibitem{Krasnov:2019kdc}
I.~Krasnov, \emph{{DUNE prospects in the search for sterile neutrinos}},
  \href{https://doi.org/10.1103/PhysRevD.100.075023}{\emph{Phys. Rev. D}
  {\bfseries 100} (2019) 075023}
  [\href{https://arxiv.org/abs/1902.06099}{{\ttfamily 1902.06099}}].

\bibitem{King:2007pr}
S.F.~King, \emph{{Parametrizing the lepton mixing matrix in terms of deviations
  from tri-bimaximal mixing}},
  \href{https://doi.org/10.1016/j.physletb.2007.10.078}{\emph{Phys. Lett. B}
  {\bfseries 659} (2008) 244}
  [\href{https://arxiv.org/abs/0710.0530}{{\ttfamily 0710.0530}}].

\bibitem{Pakvasa:2007zj}
S.~Pakvasa, W.~Rodejohann and T.J.~Weiler, \emph{{Unitary parametrization of
  perturbations to tribimaximal neutrino mixing}},
  \href{https://doi.org/10.1103/PhysRevLett.100.111801}{\emph{Phys. Rev. Lett.}
  {\bfseries 100} (2008) 111801}
  [\href{https://arxiv.org/abs/0711.0052}{{\ttfamily 0711.0052}}].

\bibitem{DayaBay:2018yms}
{\scshape Daya Bay} collaboration, \emph{{Measurement of the Electron
  Antineutrino Oscillation with 1958 Days of Operation at Daya Bay}},
  \href{https://doi.org/10.1103/PhysRevLett.121.241805}{\emph{Phys. Rev. Lett.}
  {\bfseries 121} (2018) 241805}
  [\href{https://arxiv.org/abs/1809.02261}{{\ttfamily 1809.02261}}].

\bibitem{KamLAND:2013rgu}
{\scshape KamLAND} collaboration, \emph{{Reactor On-Off Antineutrino
  Measurement with KamLAND}},
  \href{https://doi.org/10.1103/PhysRevD.88.033001}{\emph{Phys. Rev. D}
  {\bfseries 88} (2013) 033001}
  [\href{https://arxiv.org/abs/1303.4667}{{\ttfamily 1303.4667}}].

\bibitem{DiValentino:2021hoh}
E.~Di~Valentino, S.~Gariazzo and O.~Mena, \emph{{Most constraining cosmological
  neutrino mass bounds}},
  \href{https://doi.org/10.1103/PhysRevD.104.083504}{\emph{Phys. Rev. D}
  {\bfseries 104} (2021) 083504}
  [\href{https://arxiv.org/abs/2106.15267}{{\ttfamily 2106.15267}}].

\end{thebibliography}\endgroup

\end{document}